\documentclass[a4paper,10pt]{report}
\usepackage{amssymb}
\usepackage{amsmath}
\usepackage{latexsym}
\usepackage{array}
\usepackage{graphicx}
\usepackage{bm}
\usepackage{exscale}

\def\DIS{\displaystyle}

\usepackage{theorem}
\theoremstyle{break}
\newtheorem{Theorem}{Theorem}[section]
\newtheorem{Proposition}{Proposition}[section]

\newtheorem{Definition}{Definition}[section]
\newtheorem{Lemma}{Lemma}[section]
\def\qed{\hfill\hbox{$\Box$}\vspace{10pt}\break}
\def\hugesymbol#1{\mbox{\strut\rlap{\smash{\Huge$#1$}}\quad}}
\def\C{{\mathbb C}}
\def\Z{{\mathbb Z}}
\def\Q{{\mathbb Q}}
\def\F{{\mathbb F}}
\def\R{{\mathbb R}}
\def\P{{\mathbb P}}
\def\A{{\mathbb A}}
\def\II{${}_{\mbox{\scriptsize{II} }}$}
\def\I{${}_{\mbox{\scriptsize{I} }}$}
\def\sp{\mbox{\scriptsize p}}

\begin{document}

\title{Studies on the discrete integrable equations over finite fields}
\author{
by\\
\\
Masataka Kanki
\\
\\
Graduate School of Mathematical Sciences,\\
The University of Tokyo
}
\date{June 2013}

\maketitle

\begin{abstract}
Discrete dynamical systems over finite fields are investigated and their integrability is discussed.
In particular, the discrete Painlev\'{e} equations and the discrete KdV equation are defined over finite fields and their special solutions are obtained. Although the discrete integrable equations we treat in this paper have been studied intensively over the past, their investigation over the finite fields has not been done as thoroughly, partly because of the indeterminacies that appear in defining the systems.
In this paper we introduce two methods to well-define the equations over the finite fields and apply the methods to several classes of discrete integrable equations.
One method is to extend the space of initial conditions through blowing-up at the singular points. In case of discrete Painlev\'{e} equations, we proved that an finite field analog of the Sakai theory can be applied to construct the space of initial conditions \cite{KMTT2}. We discuss the state transition diagram of the time evolution over finite fields.
The other method is to define the equations over the field of $p$-adic numbers and then reduce them to the finite fields.
The mapping whose time evolutions commute with the reduction is said to have a `good reduction'. We generalize good reduction in order to be applied to integrable mappings, in particular, to the discrete Painlev\'{e} equations \cite{KMTT}. This generalized notion is called an `almost good reduction' (AGR). It is proved that AGR is satisfied for most of the discrete and $q$-discrete Painlev\'{e} equations, and therefore can be used as an integrability detector. Moreover, AGR works as an arithmetic analog of the singularity confinement test \cite{Kanki}.
Next, our methods of extending the initial value space are applied to the soliton systems: the discrete KdV equation and one of its generalized forms \cite{KMT}. The solitary wave phenomena over finite fields are studied and their periodicity discussed. The reduction from the extended field and the field of complex numbers to the finite fields is also defined. The reduction property of two-dimensional lattice systems are studied and the generalization of AGR is presented. 
This article is basically a replication of author's thesis (paper submitted: June 2013, degree conferred: September 2013) and draws from several of our published materials.
Also note that some parts are later modified from its original version to keep up with the latest developments.

\texttt{MSC2010: 37K10, 39A13, 37P20, 34M55}

\texttt{Keywords: discrete integrable equation, discrete Painlev\'{e} equation, discrete KdV equation, almost good reduction, singularity confinement, finite field, field of $p$-adic numbers}

\end{abstract}
\setcounter{tocdepth}{1}
\tableofcontents
\chapter{Introduction}
\label{sec1}
\section{Integrable equations}
As an introduction we briefly review the theory of integrable systems and give an overview of the preceding research.
There exist various distinct notions referred to under the name of `integrable' systems in mathematics and mathematical physics. 
However, we can state without accuracy that the differential equations are `integrable' if they are highly symmetric and have sufficiently `many' first integrals (conserved quantities) so that the integration of them is possible.

In the middle of the 19th century, J. Liouville first defined the notion of `exact integrability' of Hamiltonian systems of classical mechanics in terms of Poisson commuting invariants \cite{Arnold,Goriely}.
Let us consider a Hamiltonian $H=H(\mathbf{p},\mathbf{q})$ with $n$-degree of freedom which is analytic in $\mathbf{p}=(p_1,\cdots,p_n),\mathbf{q}=(q_1,\cdots,q_n)\in\mathbb{R}^n$.
The Hamilton equations are
\[
\dot{q_i}=\frac{\partial H}{\partial p_i},\ \dot{p_i}=-\frac{\partial H}{\partial q_i}\ (i=1,2,\cdots, n).
\]
\begin{Definition}
The Hamiltonian $H(\mathbf{p},\mathbf{q})$ is Liouville integrable if there exist $n$ independent analytic first integrals $I_1=H,I_2,\cdots,I_n$ in involution $($i.e. $\left\{I_i,I_j\right\}=0)$.
\end{Definition}

In the late 1960s, the localized solutions of partial differential equations have been found to be understood by viewing these equations as infinite dimensional integrable systems.
These localized solutions are called solitons. The classical example of solitons is a solution of the Korteweg-de Vries equation (KdV equation) which describes shallow water wave phenomena \cite{Bouss1,Bouss2,Korteweg}. The discovery of solitary wave solutions dates back to the 1830s.
In 1834, Scott Russell discovered a solitary wave phenomenon while observing the motion of a boat in a canal. He noticed that the speed of the waves depends on their size, and that these waves will never merge---a large wave overtakes a small one \cite{Russell1,Russell2}. Later in 1895, Korteweg and de Vries proved that these waves can be simulated by the solutions of the following partial differential equation which is now called the KdV equation:
\begin{equation}
\frac{\partial u(x,t)}{\partial t}+6u(x,t)\frac{\partial u(x,t)}{\partial x}+\frac{\partial^3 u(x,t)}{\partial x^3}=0. \label{continuousKdV}
\end{equation}
The KdV equation became increasingly important when it was discovered that the equation can simulate many physical phenomena such as plasma physics and internal waves.
Zabusky and Kruskal found that the KdV equation was the governing equation of the Fermi-Pasta-Ulam lattice equation, and that the solutions of the KdV equation pass through one another and subsequently retain their characteristic form and velocity \cite{ZK1965}.
It has later been discovered that these soliton equations can be understood from a broader perspective.
In 1980s, M. Sato and Y. Sato discovered that wide class of nonlinear integrable equations and their solutions can be treated uniformly by considering them on an infinite dimensional Grassmannian \cite{SatoSato}. This is the notable `Sato theory', in which the Sato equation is a `master' equation that produces an infinite series of nonlinear partial differential equations.  The theory is also called the theory of the KP hierarchies, since one of the simplest equations among those series of equations is the Kadomtsev-Petviashvili equation (KP equation), which describes shallow water waves of dimension two:
\[
\frac{\partial}{\partial x}\left( 4\frac{\partial u}{\partial t} -12 u\frac{\partial u}{\partial x} -\frac{\partial^3 u}{\partial x^3}\right)-3\frac{\partial^2 u}{\partial y^2}=0.
\]
The KdV equation and its soliton solutions are proved to be obtained from the reduction of the KP equation and its soliton solutions.

Next we review another important class of integrable differential equations: the Painlev\'{e} equations.
The Painlev\'{e} equations were originally discovered by P. Painlev\'{e} and B. Gambier as second order ordinary differential equations whose solutions do not have movable singularities other than poles. \cite{Painleve,Gambier,Okamotobook,Okamoto,Okamoto2,Okamoto3,Okamoto4}.
\begin{Proposition}
Let us consider the differential equation
\begin{equation}
\frac{d^2 y}{dx^2}=R\left(x,y,\frac{dy}{dx}\right), \label{contiP}
\end{equation}
where $R(x,y,z)$ is a rational function of $y,z$ whose coefficients are analytic functions of $x$ 
defined on some domain $D\subset\mathbb{C}^3$.
If the equation \eqref{contiP} does not have movable singular points, then it falls into one of the following cases:
\begin{itemize}
\item Linear equations.
\item Equations  of the form
\[
\left(\frac{dx}{dt}\right)^2-4 x^3+g_2 x+g_3=0.
\]
Their solutions are written by the Weierstra\ss\  elliptic function.
\item Solvable equations.
\item One of the six Painlev\'{e} equations (P\I, P\II, P$_{\mbox{\scriptsize{III}}}$, P$_{\mbox{\scriptsize{IV}}}$, P$_{\mbox{\scriptsize{V}}}$, P$_{\mbox{\scriptsize{VI}}}$). We just present the first two of the Painlev\'{e} equations:
\begin{itemize}
\item Painlev\'{e} I equation (P\I)
\[
\frac{d^2 x}{dt^2}=6x^2+t
\]
\item Painlev\'{e} II equation (P\II)
\[
\frac{d^2 x}{dt^2}=2x^3+tx+\alpha
\]
$\cdots$
\end{itemize}
\end{itemize}
\end{Proposition}
In the 1970s, it has been found that the correlation function
of the two-dimensional Ising model are related to the Painlev\'{e} III equation \cite{Wu}, and since then the Painlev\'{e} equations have been investigated eagerly as one of the classes of integrable equations by K. Okamoto and many other researchers.
Also, the Painlev\'{e} equations can be obtained via similarity reduction of some soliton equations.

\section{Discrete integrable equations}
We review some of the topics on the integrability of discretized equations. Roughly speaking, the discrete integrable systems have `many' conserved quantities and soliton solutions. If the discretization is chosen appropriately, the discrete system preserves the essential properties that the corresponding continuous system possesses.

A discrete version of the KP equation is derived via the Miwa transformation from the KP hierarchy.
The Miwa transformation is the following transformation that changes the variables $(x_1,x_2,\cdots)$ to $(m_1,m_2,\cdots)$:
\[
x_n=\sum_{j=1}^{\infty} m_j \frac{1}{n(a_j)^n},
\]
where $n=1,2,\cdots$ and $a_1,a_2,\cdots \in\mathbb{C}\setminus \{0\}$ are distinct constants.
Let us suppose that the variables $m_i$ take only integer values, and consider a function $\tau(m_1,m_2,\cdots)$, which is a Miwa transformation of the $\tau$-function solution $\tau(x_1,x_2,\cdots)$ of the Sato's bilinear identity.
Then we obtain the following bilinear relation for distinct $i,j,k\in\mathbb{Z}$:
\begin{equation}
(a_i-a_j)\tau_{ij}\tau_k+(a_j-a_k)\tau_{jk}\tau_i+(a_k-a_i)\tau_{ki}\tau_j=0. \label{hirotamiwa}
\end{equation}
The equation \eqref{hirotamiwa} is the discrete KP equation, and is also called `Hirota-Miwa equation'.
The discrete KdV equation is obtained by imposing a restriction
\[
\tau(m_1+1,\ m_2+1,\ m_3)=\tau(m_1,\ m_2,\ m_3)
\]
to the Hirota-Miwa equation. This kind of restriction on the independent variables (imposing shift invariance, omitting some of the variables, e.t.c.) to construct simpler classes of equations is called the `reduction'. It gives the following bilinear form of the discrete KdV equation:
\begin{equation}
(1+\delta)\sigma_{n+1}^{t-1}\sigma_n^{t-1}=\delta \sigma_{n+1}^{t-1}\sigma_n^{t+1} + \sigma_n^t \sigma_{n+1}^t. \label{bilineardkdv}
\end{equation}
Here $\sigma_n^t:=\tau(t,0,n)$.
(Note that, in this paper, the word `reduction' is also used to indicate other process: the projection modulo a maximal ideal.)
By introducing a new variable
\[
x_n^t=\frac{\sigma_n^t\sigma_{n+1}^{t-1}}{\sigma_{n+1}^t \sigma_n^{t-1}},
\]
we obtain the discrete KdV equation as the following nonlinear partial difference equation:
\begin{equation}
\frac{1}{x_{n+1}^{t+1}}-\frac{1}{x_n^t}+\frac{\delta}{1+\delta}\left(x_n^{t+1}-x_{n+1}^t \right)=0.
\label{dKdV1}
\end{equation}
In 1990, the discrete versions of the Painlev\'{e} equations have been discovered by A. Ramani, B. Grammaticos and J. Hietarinta \cite{RGH}.
They are considered to be integrable in the sense that they pass the singularity confinement test.
The singularity confinement test judges whether the spontaneously appearing singularities disappear after a few iteration steps of the systems.
For example, let us consider the following mapping related to the discrete Painlev\'{e} I equation:
\[
x_{n+1}=-x_{n-1}+\frac{1}{x_n^2}.
\]
If we evolve the equation from $x_0=u$ and $x_1=0$, then we have
\[
x_1=0,\ x_2=\infty,\ x_3=0,\ x_4=-\infty+\infty,
\]
thus $x_5$ is indeterminate.
However, if we introduce a small positive parameter $\epsilon>0$ and take $x_1=\epsilon$, then
\[
x_2=\epsilon^{-2}-u,\ x_3=-\frac{\epsilon(1-\epsilon^3-2\epsilon^2 u +\epsilon^4 u^2)}{(\epsilon^2 u -1)^2},
\]
\[
x_4=\frac{(\epsilon^2 u-1)(-u-2\epsilon +\epsilon^4+4\epsilon^3 u+3\epsilon^2 u^2-2\epsilon^5 u^2-3\epsilon^4 u^3+\epsilon^6 u^4)}{(1-\epsilon^3-2\epsilon^2 u +\epsilon^4 u^2)^2}.
\]
By taking the limit $\epsilon\to 0$, we obtain the time evolution as follows:
\[
x_1=0,\ x_2=\infty,\ x_3=0,\ x_4=u.
\]
In this case we can see that, by introducing a parameter $\epsilon$ in the initial value, the indeterminacy resulting from the singularity at $x_1=0$ is `confined' within finite time steps and then the initial value $u$ reappears.
Most integrable discrete systems have been proved to pass the test \cite{Grammaticosetal}.
We will treat some of the discrete Painlev\'{e} equations in the following sections.
Note that, although the singularity confinement test is a very powerful tool to detect the integrability of many discrete equations, it is not easy to apply the test to partial difference equations.
In 2014, after this thesis is submitted, the author and his collaborators invented a new integrability criterion called `co-primeness' condition, which can be considered as one type of generalization of the singularity confinement test.
The benefit of the co-primeness is that it is applicable also to the partial difference equations, however, we do not treat this topic in this article and leave the details to other papers.

\section{Ultra-discrete integrable equations} \label{udintegrable}
The ultra-discrete integrable systems are obtained from the discrete integrable ones through a limiting procedure called `ultra-discretization'.
Both the dependent and independent variables of the ultra-discrete systems take discrete values, usually the integers. Therefore they are considered as cellular automata. The cellular automaton is a discrete computational model which consists of a regular grid of cells. Each cell has a finite number of states, corresponding to the value of the independent variable of the ultra-discrete system.
It is studied not only in mathematical physics, but also in many fields in natural and social sciences such as computability theory, theoretical biology and jamology.

\begin{figure}
\centering
\includegraphics[width=9cm,bb=110 500 480 740]{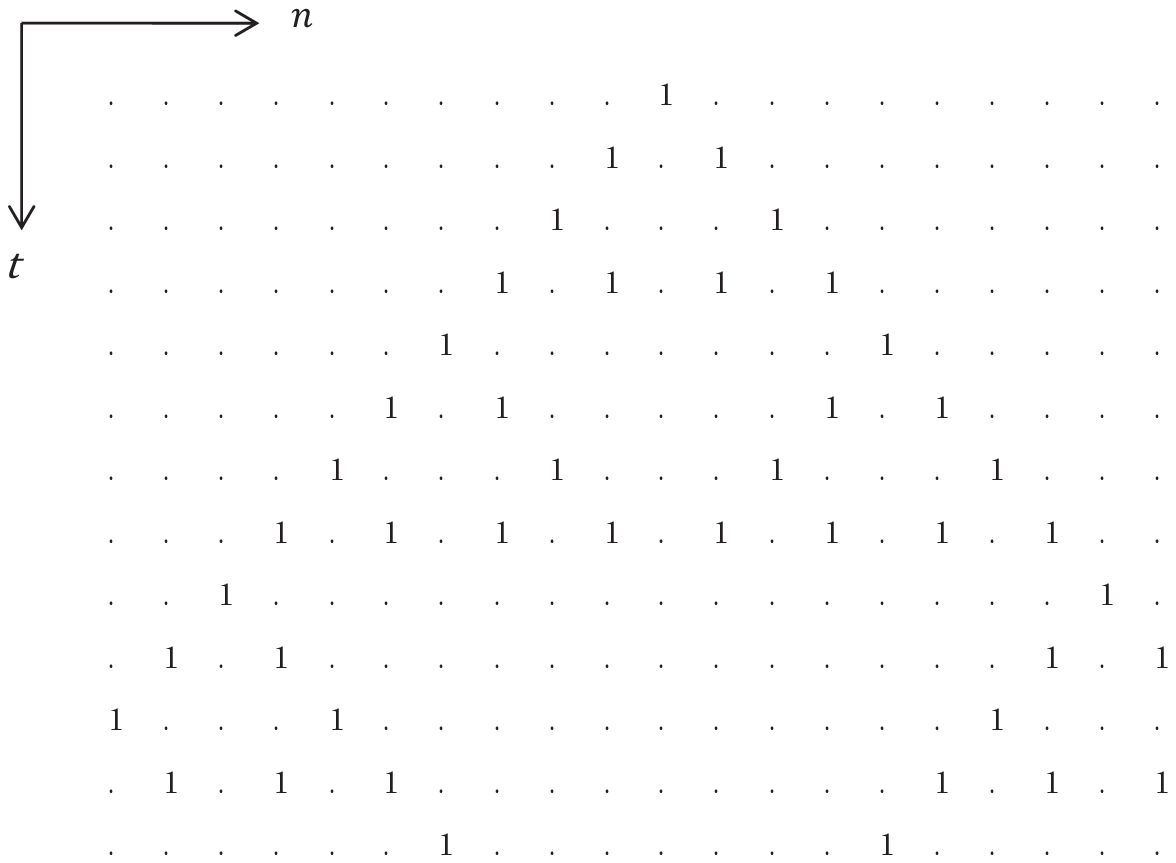}
\caption{The time evolution $x_n^t$ of the ECA with rule $90$, where the dot `.' denotes $x_n^t=0$.}
\label{figuresier}
\end{figure}

One of the most famous cellular automata may be the Elementary Cellular Automata (ECA) \cite{Wolfram}. It is one-dimensional and the time evolution of a cell depends only on its two neighboring cells.
We give an ECA with `rule $90$', which is the `simplest non-trivial' ECA as an example \cite{Wolframetal}. Let the values of one-dimensional cells at time step $t$ be $\{x_n^t\}_{n=-\infty}^\infty$ where each cell satisfies $x_n^t\in\{0,1\}$.
The next step $x_n^{t+1}$ is defined as the exclusive disjunction of $x_{n-1}^t$ and $x_{n+1}^t$, and therefore be expressed as $x_n^{t+1}\equiv x_{n-1}^t+x_{n+1}^t\mod 2$.
The time evolution on a large scale gives the shape of the Sierpi\'{n}ski gasket, a fractal. See the figure \ref{figuresier}.

\begin{figure}
\centering
\includegraphics[width=11cm,bb=100 590 500 740]{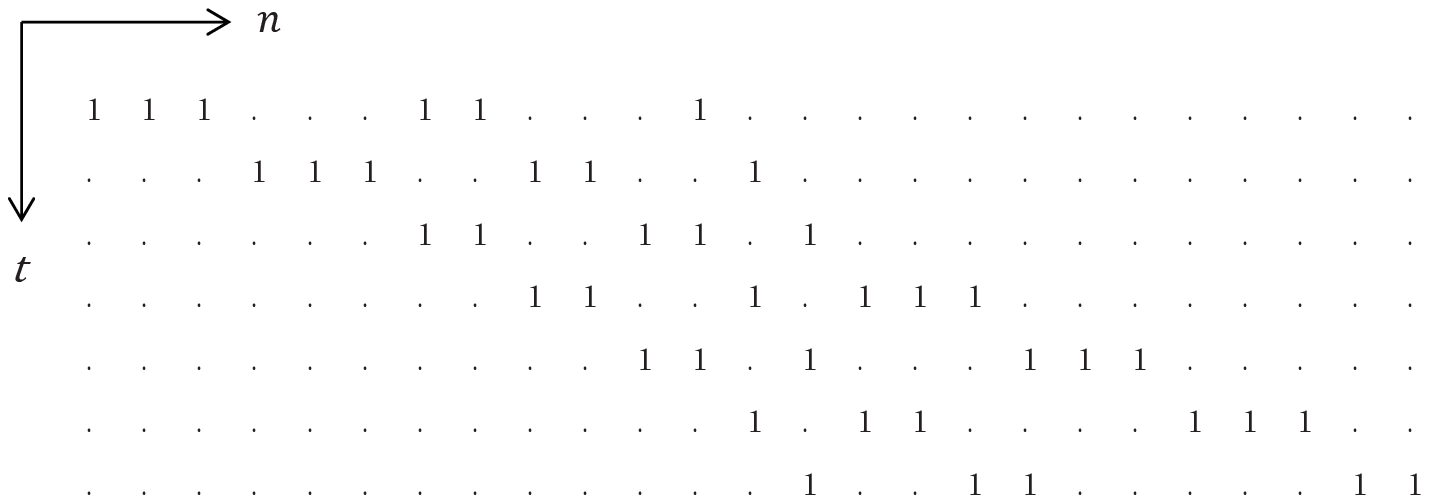}
\caption{The time evolution $U_n^t$ of the BBS, where a dot `.' indicates a zero.}
\label{figurebbs}
\end{figure}

We are also interested in more complex cellular automata whose solutions behave analogous to those of some discrete integrable systems.
We present one way to ultradiscretize the discrete KdV equation and explain how this limiting procedure gives the Box Ball Systems (BBS). The BBS is one of the typical soliton cellular automata discovered by D. Takahashi and J. Satsuma, and has been investigated extensively by T. Tokihiro et. al. \cite{TS1990,T1993,TTMS}. To simplify the process, we use the following lemma:
\begin{Lemma}
Under the boundary condition $\lim_{n\to -\infty}x_n^t=1$, the discrete KdV equation \eqref{dKdV1} takes the following form:
\begin{equation}
x_{n+1}^{t+1}=\left(\delta x_{n+1}^t+(1-\delta)\prod_{k=-\infty}^n\frac{x_k^{t+1}}{x_k^t}\right)^{-1}. \label{dKdVpermanentform}
\end{equation}
\end{Lemma}
To ultradiscretize the equation \eqref{dKdVpermanentform}, we first introduce an auxiliary variable $\varepsilon>0$ and change variables as
\[
x_n^t=e^{U_n^t/\varepsilon},\ \delta=e^{-L/\varepsilon}.
\]
By taking the logarithms of the both sides of \eqref{dKdVpermanentform}, we have
\[
U_{n+1}^{t+1}=-\varepsilon\log\left(e^{(U_{n+1}^t-L)/\varepsilon}+(1-e^{-L/ \varepsilon})\exp\left(\sum_{k=-\infty}^n(U_{k}^{t+1}-U_k^t)/\varepsilon\right)\right).
\]
By taking the limit $\varepsilon\to +0$,
and by using the identities
\[
\lim_{\varepsilon\to +0}\varepsilon\log(\exp(A/\varepsilon)+\exp(B/\varepsilon))=\max(A,B),
\]
and
\[-\max(A,B)=\min(-A,-B),\]
which are true for all $A,B\in\mathbb{R}$, we obtain the evolution equation of the BBS:
\begin{equation}
U_{n+1}^{t+1}=\min\left(L-U_{n+1}^t,\sum_{k=-\infty}^n(U_k^{t}-U_k^{t+1})\right). \label{hakotama1}
\end{equation}
Here the parameter $L$ is called the `capacity of the box'.
In particular, if $L=1$ the BBS \eqref{hakotama1} evolves inside of $\{0,1\}$.
We give an example of the time evolution in the figure \ref{figurebbs}.

This solution corresponds to the three-soliton solution of the (continuous and discrete) KdV equation.
The origin of the name `Box Ball' system is that the solution of BBS can be seen as moving balls $(1$'s$)$ in an infinite array of empty boxes $(0$'s$)$.

Note that taking the ultra-discrete limit is closely related to taking the $p$-adic valuation of the given discrete equations as pointed out by S. Matsutani \cite{Matsutani}. We will look into this approach in more detail at the end of the paper.

Finally let us note on another topic on ultra-discrete equations. The ultra-discrete analogs of the Painlev\'{e} equation have been studied recently. For example, an ultra-discrete version of the $q$-Painlev\'{e} II equation has been obtained through `ultra-discretization with parity variables' \cite{Mimura}, which is a generalized method of ultra-discretization, and its special solution has been obtained. In connection with this theory of extended ultra-discretization procedure, N. Mimura et. al. proposed an ultra-discrete version of the singularity confinement test \cite{Mimura2}.
Another type of singularity confinement test for ultra-discrete equation has been proposed by N. Joshi and S. Lafortune, where the `singularities' for the max-plus equations have been introduced as the non-differentiable points of the piecewise linear functions \cite{Nalini1}. 
 
\section{Arithmetic dynamical systems}
In this section let us summarize the definition of the field of $p$-adic numbers and briefly explain the `good' reduction.
Let $p$ be a prime number. A non-zero rational number $x \in \Q$ ($x \ne 0$) can be written uniquely as $x=p^{v_p(x)} \dfrac{u}{v}$ where $v_p(x), u, v \in \Z$ and $u$ and $v$ are coprime integers neither of which is divisible by $p$.
We call $v_p(x)$ the $p$-adic valuation of $x$.
The $p$-adic norm $|x|_p$ is defined as
\[|x|_p=p^{-v_p(x)}\ (|0|_p=0.)\]
The local field $\Q_p$ is a completion of $\Q$ with respect to the $p$-adic norm. 
It is called the field of $p$-adic numbers, and its subring
\[
\Z_p:=\{x\in \Q_p | \ |x|_p \le 1\ (\leftrightarrow v_p(x)\ge 0)\}
\]
is called the ring of $p$-adic integers \cite{Murty}. 
The $p$-adic norm satisfies a special inequality
\[
|x+y|_p \le \max\{|x|_p,|y|_p \}.
\]
\begin{Definition}
The absolute value $|\cdot|$ of a valued field $K$ is non-archimedean (or also called ultrametric) if the following estimate is satisfied for all $\alpha,\beta\in K$:
\[
|\alpha+\beta | \le \max \{|\alpha|,|\beta| \}.
\]
\end{Definition}
The $p$-adic norm $|\cdot|_p$ of $\mathbb{Q}_p$ is thus non-archimedean, and the field $\mathbb{Q}_p$ is a non-archimedean field. 
Let $\mathfrak{p}=p\Z_p=\left\{x \in \Z_p |\ v_p(x) \ge 1 \right\}$ be the maximal ideal of $\Z_p$.
We define the reduction of $x$ modulo $\mathfrak{p}$ as
\[
\pi:\  \Z_p \ni x \mapsto \pi(x) \in \Z_p/\mathfrak{p} \cong \F_p.
\]
We write $\pi(x)$ as $\tilde{x}$ for simplicity.
Note that the reduction is a ring homomorphism:
\begin{equation}
\widetilde{x \pm y}=\tilde{x} \pm \tilde{y},\quad \widetilde{x \cdot y}=\tilde{x} \cdot \tilde{y},
\quad \widetilde{\left(\frac{x}{y}\right)}=\frac{\tilde{x}}{\tilde{y}}\ (\mbox{for}\ \tilde{y}\ne 0).
\label{prel}
\end{equation}
The element $x\in\mathbb{Z}_p$ is uniquely written as the $p$-adic polynomial series:
\[
x=\sum_{i=0}^\infty x_i p^i,
\]
where each $x_i\in\{0,1,2,\cdots, p-1\}$. The reduction is naturally computed as $\tilde{x}=x_0$.
The reduction map $\pi$ is generalized to $\Q_p^{\times}$:
\begin{equation}\label{padicreductionmap}
\pi:\ \Q_p^{\times}\ni x=p^{v_p(x)} u\ (u\in\Z_p^{\times})\mapsto
\tilde{x}=\left\{
\begin{array}{cl}
0 & (v_p(x)>0)\\
\infty & (v_p(x)<0)\\
\tilde{u} & (v_p(x)=0)
\end{array}
\right. \in \F_p\cup\{\infty\},
\end{equation}
which is no longer homomorphic.
The element $z\in\mathbb{Q}_p \setminus \mathbb{Z}_p$ is uniquely expanded as the Laurent series using $v_p(x)=k<0$:
\[
z=\sum_{i=k}^\infty z_i p^i,
\]
where each $z_i\in\{0,1,2,\cdots, p-1\}$ and $z_k\neq 0$.
In this case, the reduction is $\tilde{z}=\infty$.
For a dynamical system $\phi$ consisting of two rational mappings defined over $(x,y)\in \Q_p^2$:
\[
\phi(x,y)=\left(\frac{\sum_{i,j} d_{ij}x^i y^j}{\sum_{i,j} c_{ij}x^i y^j},\frac{\sum_{i,j} d'_{ij}x^i y^j}{\sum_{i,j} c'_{ij}x^i y^j}\right)\in(\mathbb{Z}_p(x,y))^2,
\]
the `reduced' system
\[
\tilde{\phi}=\pi(\phi)
\]
is defined as the system whose coefficients are all reduced to $\F_p$:
\[
\tilde{\phi}(x,y)=\left(\frac{\sum_{i,j} \tilde{d}_{ij}x^i y^j}{\sum_{i,j} \tilde{c}_{ij}x^i y^j},\frac{\sum_{i,j} \tilde{d}'_{ij}x^i y^j}{\sum_{i,j} \tilde{c}'_{ij}x^i y^j}\right)\in(\mathbb{F}_p (x,y))^2.
\]
We define the notion of `good reduction', which basically means that the time evolution of the system and the reduction modulo $\mathfrak{p}$ commutes.
\begin{Definition}\label{GRdef}
The rational system $\phi$ has a \textit{good reduction} modulo $\mathfrak{p}$ on the domain $\mathcal{D}\subseteq \Z_p^2$ if we have $\widetilde{\phi(x,y)}=\tilde{\phi}(\tilde{x},\tilde{y})$ for any $(x,y) \in \mathcal{D}$. 
\end{Definition}
\begin{figure}
\centering
\includegraphics[width=8cm,bb=160 220 500 420]{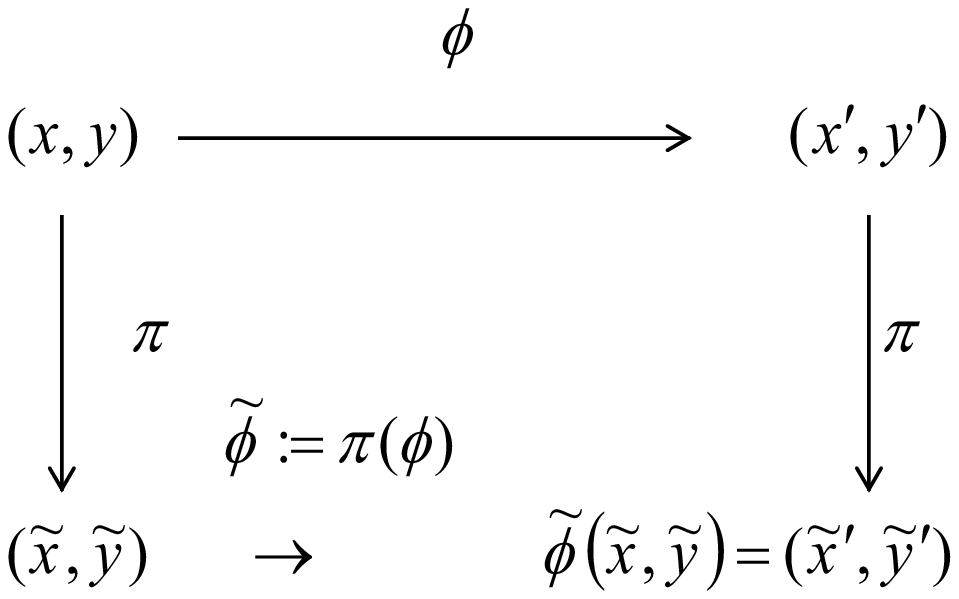}
\caption{Good reduction modulo $\pi$.}
\label{figuregr}
\end{figure}
It is equivalent for the diagram in the figure \ref{figuregr} to be commutative.
Originally, the good reduction was defined for a rational mapping with one variable \cite{Silverman}.
\begin{Definition}[\cite{Silverman}]
A rational map $\phi:\mathbb{P}^1\to\mathbb{P}^1$ defined over the valued field $K$ is said to have good reduction modulo $\mathfrak{p}$ if $\deg (\phi)=\deg (\tilde{\phi})$.
\end{Definition}
A map with a good reduction satisfies the following proposition.
\begin{Proposition}[\cite{Silverman}]
Let $\phi:\mathbb{P}^1\to\mathbb{P}^1$ be a rational map that has good reduction.
Then the map $\phi$ satisfies $\tilde{\phi} (\tilde{P})=\widetilde{\phi(P)}$ for all $P\in\mathbb{P}^1(K)$.
\end{Proposition}
With this property in mind, we define the good reduction for the dynamical systems with two variables as satisfying $\tilde{\phi} (\tilde{P})=\widetilde{\phi(P)}$ (Definition \ref{GRdef}).

\section{Purpose of our research and main results}
The purpose of our research is to define and investigate the discrete integrable equations over finite fields. We wish to study the implication of integrability over finite fields. We also expect to construct cellular automata directly from the discrete systems over finite fields.

In the case of linear discrete equations, for example, we can well-define the equations over finite fields just by changing the field on which the equations are defined to finite fields. However, in the case of nonlinear equations, since the systems are usually formulated by rational functions, the division by $0$ mod $p$ and some indeterminacies such as $0/0$ and $\infty\pm \infty$ frequently appear.
These points makes it difficult for us to well-define the equations over finite fields. Thus there has been few studies on the nonlinear discrete integrable equations defined over finite fields.

There are mainly three approaches to overcome this difficulty.
(a) The first one is to study the equation that does not contain division terms. Santini et. al. studied cellular automata constructed from one type of the Scr\"{o}dinger equations which is free from division
\cite{BSR}.
Bilinear form of the discrete KP and KdV equations \eqref{hirotamiwa}, \eqref{bilineardkdv} have been treated over the finite field $\mathbb{F}_p$ and their soliton solutions over $\mathbb{F}_p$ are obtained \cite{BD,Bialecki,DBK}.
(b) The second one is to restrict the domain of definition of the system so that the indeterminacies do not appear. The discrete Toda equation over finite fields and its graphical structures have been obtained \cite{ytakahashi}. Roberts and Vivaldi studied the integrability over finite fields in terms of the lengths of the periodic orbits \cite{Roberts,Roberts2,Roberts3}.
(c) The third one is to extend the space of initial conditions to make the mapping well-defined.
We investigate this third approach in this paper and try two different schemes.

(c-i) The first scheme is to apply the Sakai's theory on discrete Painlev\'{e} equations to the case of finite domains.
According to the theory developed by K. Okamoto and H. Sakai, the space of initial conditions for the discrete Painlev\'{e} equation becomes a birational surface as we extend the domain $\mathbb{P}^2$ by blowing-up at each singular point \cite{Okamotosp,Sakai}.
We show, in chapter \ref{sec3}, that this procedure is still valid if applied to the finite domain of definition $\mathbb{F}_{p^m}\times \mathbb{F}_{p^m}$. We in particular treated the discrete Painlev\'{e} II equation, and presented the extended domain of initial conditions for $p=3$ and $m=1$. What is more, we have shown that the size of the extended domain we construct is
smaller than that made by the Sakai theory. Since the domain over the finite field has a discrete topology, the extended domain need not to be birational, but needs only to be bijective.

(c-ii) The second scheme of extension is to define the equations over the field of $p$-adic numbers $\mathbb{Q}_p$ and then reduce them to the finite field $\mathbb{F}_p$.
Through this approach, we wish to establish the significance of `integrability' of the systems over finite fields.
For example, if we try to define the discrete Painlev\'{e} equations over the field $\F_p$,
the initial value space is a finite set $\F_p \times \F_p$. Since the system consists of transitions between just $p^2$ points, it is not clear how  we can formulate the integrability of the system from the integrability of the original system defined over $\mathbb{R}$ or $\mathbb{C}$. To resolve this problem we consider a pair of fields $(\Q_p,\ \F_p)$ in chapter \ref{chapter3}. We can say that the system over $\F_p$ is `integrable' if it is integrable over $\Q_p$ and its reduction to the finite field $\F_p$ has an `almost good reduction' property.
We prove that, although the integrable mappings generally do not have a good reduction modulo a prime,
they do have an \textit{almost good reduction} (AGR), which is a generalized notion of good reduction.
We demonstrate that AGR can be used as an integrability detector over finite fields, by proving that
dP\II ,$q$P\I, $q$P\II, $q$P$_{\mbox{\scriptsize III}}$, $q$P$_{\mbox{\scriptsize IV}}$ and $q$P$_{\mbox{\scriptsize V}}$ equations have AGR over appropriate domains.
We also prove that one of the chaotic system, the Hietarinta-Viallet equation, has AGR and conclude that AGR is an arithmetic analog of the singularity confinement method.
We then discuss the relation of our approach to the Diophantine integrability proposed by R. Halburd \cite{Halburd}. We also propose a way to reduce the systems defined over the extended field of $\mathbb{Q}_p$, and then apply the procedure to obtain the cellular automaton from the complex-valued equations.

Lastly, in chapter \ref{sec5}, we apply our methods to the soliton systems, in particular the discrete KdV equation and one of its generalized forms \cite{KMT}.
We present two methods of extension: first one is to use a field of rational functions whose coefficients are in the finite field, and the second one is to use the field of $p$-adic numbers just like we have done in the previous sections. The soliton solutions obtained through both two methods are introduced and their periodicity is discussed. Special types of solitary waves that appear only over the finite fields are presented and their nature is studied. The reduction properties of the two-dimensional lattice systems are discussed.

Let us summarize the main results of this paper. The key definition is definition \ref{AGRdef}, in which almost good reduction property for non-autonomous discrete dynamical systems is formulated.
The main theorems of this paper are the followings:
theorem \ref{thminit} on the space of initial conditions, and theorems
\ref{PropQRT}, \ref{PropdP2}, \ref{PropqP1}, \ref{PropqP2}, \ref{PropqP3}, \ref{PropqP4}, \ref{PropqP5}, \ref{PropHV} on the almost good reduction property for discrete Painlev\'{e} equations.

%
%
\chapter{The space of initial conditions of one-dimensional systems}\label{sec3}
%
%
A discrete Painlev\'{e} equation is a non-autonomous and nonlinear second order ordinary difference equation with several parameters.
When it is defined over a finite field, the dependent variable takes only a finite number of values and its time evolution will attain an indeterminate state in many cases for generic values of the parameters and initial conditions.
\section{Space of initial conditions of dP\II equation}
For example, the discrete Painlev\'{e} II equation (dP\II equation) is defined as 
\begin{equation}
u_{n+1}+u_{n-1}=\frac{z_n u_n+a}{1-u_n^2}\quad (n \in \mathbb{Z}),
\label{dP2equation}
\end{equation}
where $z_n=\delta n + z_0$ and $a, \delta, z_0$ are constant parameters \cite{NP}.
Let $q=p^k$ for a prime $p$ and a positive integer $k \in \Z_+$.
When \eqref{dP2equation} is defined over a finite field $\F_{q}$,
the dependent variable $u_n$ will eventually take values $\pm 1$ for generic parameters and initial values $(u_0,u_1) \in \F_{q}^2$, 
and we cannot proceed to evolve it.
If we extend the domain from $\F_{q}^2$ to $(\P\F_q)^2=(\F_q\cup\{\infty\})^2$, $\P\F_q$ is not a field and we cannot define arithmetic operation in \eqref{dP2equation}. 
To determine its time evolution consistently, we have two choices:
One is to restrict the parameters and the initial values to a smaller domain so that the singularities do not appear.
The other is to extend the domain on which the equation is defined.
In this article, we will adopt the latter approach.
It is convenient to rewrite \eqref{dP2equation} as:
\begin{equation}
\left\{
\begin{array}{cl}
x_{n+1}&=\dfrac{\alpha_n}{1-x_n}+\dfrac{\beta_n}{1+x_n}-y_{n},\\
y_{n+1}&=x_n,
\end{array}
\right.
\label{dP2}
\end{equation}
where $\alpha_n:=\frac{1}{2}(z_n+a),\ \beta_n:=\frac{1}{2}(-z_n+a)$.
Then we can regard \eqref{dP2} as a mapping defined on the domain $\F_q \times \F_q$.
To resolve the indeterminacy at $x_n = \pm 1$, we apply the theory of the state of initial conditions developed by H. Sakai \cite{Sakai}.
First we extend the domain to $\P\F_q \times \P\F_q$, and then blow it up at four points $(x,y)=(\pm 1, \infty), (\infty, \pm 1)$ 
to obtain the space of initial conditions:
\begin{equation}
\tilde{\Omega}^{(n)}:=\mathcal{A}_{(1,\infty)}^{(n)}\cup \mathcal{A}_{(-1,\infty)}^{(n)}\cup \mathcal{A}_{(\infty,1)}^{(n)}\cup \mathcal{A}_{(\infty,-1)}^{(n)},
\label{omega}
\end{equation}
where $\mathcal{A}_{(1,\infty)}^{(n)}$ is the space obtained from the two dimensional affine space $\A^2$ by blowing up twice as
\begin{align*}
\mathcal{A}_{(1,\infty)}^{(n)}&:=\left\{ \left((x-1,y^{-1}),[\xi_1:\eta_1],[u_1:v_1]  \right)\ \Big|\ \right. \\
&\qquad \eta_1 (x-1)=\xi_1 y^{-1},
(\xi_1+\alpha_n \eta_1)v_1=\eta_1(1-x)u_1 \ \Big\} \; \subset \A^2 \times \P \times \P.  
\end{align*}
Similarly, 
\begin{align*}
\mathcal{A}_{(-1,\infty)}^{(n)}&:=\left\{ \left((x+1,y^{-1}),[\xi_2:\eta_2],[u_2:v_2]  \right)\ \Big|\ 
\right.\\
&\qquad \qquad \eta_2 (x+1)=\xi_2 y^{-1},(-\xi_2+\beta_n \eta_2)v_2=\eta_2(1+x)u_2 \ \Big\},\\
\mathcal{A}_{(\infty,1)}^{(n)}&:=\left\{ \left((x^{-1},y-1),[\xi_3:\eta_3],[u_3:v_3]  \right)\ \Big|\ 
\right. \\
& \qquad \qquad \xi_3 (y-1)=\eta_3 x^{-1}, (\eta_3+\alpha_n \xi_3)v_3=\xi_3(1-y)u_3 \ \Big\},\\
\mathcal{A}_{(\infty,-1)}^{(n)}&:=\left\{ \left((x^{-1},y+1),[\xi_4:\eta_4],[u_4:v_4]  \right)\ \Big|\ 
\right. \\
& \qquad \qquad \xi_4 (y+1)=\eta_4 x^{-1}, (-\eta_4+\beta_n \xi_4)v_3=\xi_4(1+y)u_4 \ \Big\}.
\end{align*}  
The bi-rational map \eqref{dP2} is extended to the bijection $\tilde{\phi}_n: \ \tilde{\Omega}^{(n)} \rightarrow \tilde{\Omega}^{(n+1)}$ 
which decomposes as $\tilde{\phi}_n:=\iota_n \circ \tilde{\omega}_n$. 
Here $\iota_n$ is a natural isomorphism which gives $\tilde{\Omega}^{(n)} \cong  \tilde{\Omega}^{(n+1)}$, that is,
on $\mathcal{A}_{(1,\infty)}^{(n)}$ for instance, $\iota_n$ is expressed as 
\begin{align*}
&\left((x-1,y^{-1}),[\xi :\eta ],[u :v ]  \right) \in  \mathcal{A}_{(1,\infty)}^{(n)} \\
&\rightarrow \quad
\left((x-1,y^{-1}),[\xi -\delta/2\cdot\eta:\eta ],[u :v ]  \right) \in  \mathcal{A}_{(1,\infty)}^{(n+1)}.
\end{align*}

The automorphism $\tilde{\omega}_n$ on $\tilde{\Omega}^{(n)}$ is induced from \eqref{dP2} and gives the mapping
\[
\mathcal{A}_{(1, \infty)}^{(n)} \rightarrow \mathcal{A}_{(\infty,1)}^{(n)}, \;
\mathcal{A}_{(\infty,1)}^{(n)} \rightarrow \mathcal{A}_{(-1,\infty)}^{(n)}, \;
\mathcal{A}_{(-1, \infty)}^{(n)} \rightarrow \mathcal{A}_{(\infty,-1)}^{(n)}, \;
\mathcal{A}_{(\infty,-1)}^{(n)} \rightarrow \mathcal{A}_{(1,\infty)}^{(n)}.
\]
Under the map $\mathcal{A}_{(1, \infty)}^{(n)} \rightarrow \mathcal{A}_{(\infty,1)}^{(n)}$,
\begin{align*}
x=1 \ \rightarrow \ E_2^{(\infty,1)} &\qquad u_3=\left(y-\frac{\beta_n}{2}\right)v_3, \\
E_1^{(1,\infty)} \ \rightarrow \ E_1^{(\infty,1)} &\qquad [\xi_1:-\eta_1]=[\alpha_n \xi_3+\eta_3:\xi_3], \\
E_2^{(1,\infty)} \ \rightarrow \ y'=1 &\qquad x'=\frac{u_1}{v_1}+\frac{\beta_n}{2},
\end{align*}
where $(x,y) \in \mathcal{A}_{(1, \infty)}^{(n)}$, $(x',y')\in \mathcal{A}_{(\infty,1)}^{(n)}$, $E_1^{\sp}$ and $E_2^{\sp}$ are the exceptional curves in $\mathcal{A}_{\sp}^{(n)}$ obtained by the first blowing up and the second blowing up respectively at the point p $\in \{(\pm 1, \infty),(\infty,\pm 1)  \}$. 
Similarly under the map $\mathcal{A}_{(\infty,1)}^{(n)} \rightarrow \mathcal{A}_{(-1,\infty)}^{(n)}$,
\begin{align*}
E_1^{(\infty,1)} \ \rightarrow \ E_1^{(-1,\infty)} &\qquad [\xi_3:\eta_3]=[\eta_2:(\beta_n-\alpha_n) \eta_2-\xi_2], \\
E_2^{(\infty,1)} \ \rightarrow \ E_2^{(-1,\infty)} &\qquad [u_3:v_3]=[-\beta_n u_2: \alpha_n v_2].
\end{align*}
The mapping on the other points are defined in a similar manner.
Note that $\tilde{\omega}_n$ is well-defined in the case $\alpha_n=0$ or $\beta_n=0$.
In fact, for $\alpha_n=0$, $E_2^{(1,\infty)}$ and $E_2^{(\infty,1)}$ can be identified with the lines $x=1$ and $y=1$ respectively. 
Therefore we have found that, through the construction of the space of initial conditions, the dP\II equation can be well-defined over finite fields.
However there are some unnecessary elements in the space of initial conditions when we consider a finite field, because we are working on a discrete topology and do not need continuity of the map. 
Let $\tilde{\Omega}^{(n)}$ be the space of initial conditions and $|\tilde{\Omega}^{(n)}|$ be the number of elements of it.
For the dP\II equation, we obtain $|\tilde{\Omega}^{(n)}|=(q+1)^2-4+4(q+1)-4+4(q+1)=q^2+10q+1$, since $\P\F_q$ contains $q+1$ elements.
However an exceptional curve $E_1^{\sp}$ is transferred to another exceptional curve $E_1^{\sp'}$, and $[1:0] \in E_2^{\sp}$ to 
$[1:0] \in E_2^{\sp'}$ or to a point in $E_1^{\sp'}$. Hence we can reduce the space of initial conditions $\tilde{\Omega}^{(n)}$ to the minimal space of initial conditions $\Omega^{(n)}$ which is the minimal subset of $\tilde{\Omega}^{(n)}$ including $\P\F_q\times \P\F_q$, closed under the time evolution.
By subtracting unnecessary elements we find $|\Omega^{(n)}|=(q+1)^2-4+4(q+1)-4=q^2+6q-3$.
In summary, we obtain the following theorem:
\begin{Theorem}\label{thminit}
The domain of the dP\II equation over $\F_q$ can be extended to the minimal domain $\Omega^{(n)}$ on which the time evolution at time step $n$ is well defined. Moreover $|\Omega^{(n)}|=q^2+6q-3$. 
\end{Theorem}

\begin{figure}
\centering
\includegraphics[width=12cm,bb=-152 152 510 662]{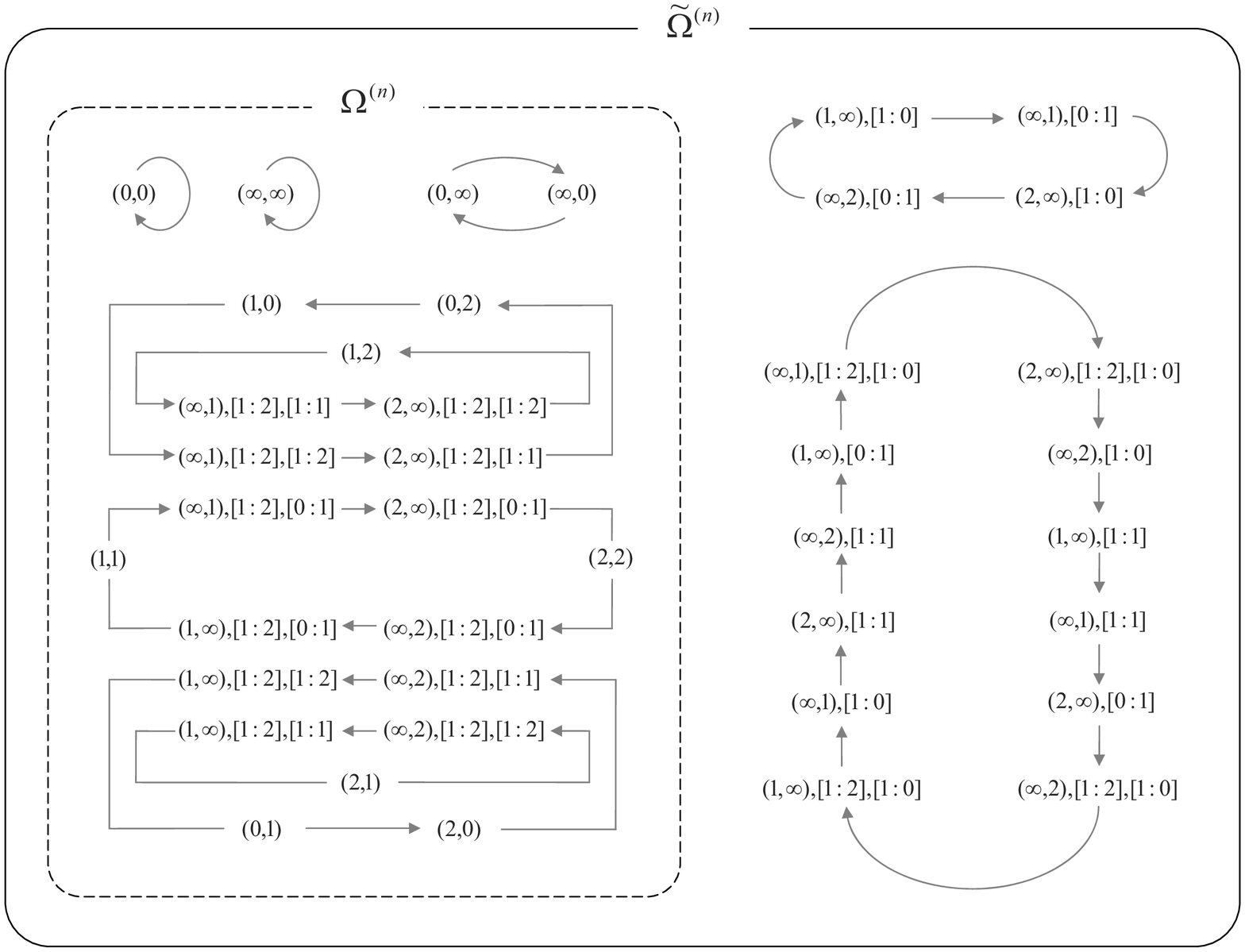}
\caption{The orbit decomposition of the space of initial conditions $\tilde{\Omega}^{(n)}$ and the reduced one $\Omega^{(n)}$ for $q=3$.}
\label{figure1painleve}
\end{figure}
In figure \ref{figure1painleve}, we show a schematic diagram of the map $\tilde{\omega}_n$ on $\tilde{\Omega}^{(n)}$, and its restriction map $\omega_n:=\tilde{\omega}_n|_{\Omega^{(n)}}$ on $\Omega^{(n)}$
with $q=3$, $\alpha_0=1$ and $\beta_0=2$.
We can also say that figure \ref{figure1painleve} is a diagram for the autonomous version of the equation \eqref{dP2} when $\delta=0$.
In the case of $q=3$, we have $|\tilde{\Omega}^{(n)}|=40$ and $|\Omega^{(n)}|=24$. 

The above approach is equally valid for other discrete Painlev\'{e} equations and we can define them over finite fields by constructing isomorphisms on the spaces of initial conditions.
Thus we conclude that a discrete Painlev\'{e} equation can be well defined over a finite field by redefining the initial domain properly. 
Note that, for a general nonlinear equation, explicit construction of the space of initial conditions over a finite field is not so straightforward (see \cite{Takenawa} or a higher order lattice system) and it will not help us to obtain the explicit solutions. We will return to this topic in chapter \ref{sec5}.

\section{Combinatorial construction of the initial value space}
\begin{figure}
\centering
\includegraphics[width=9cm,bb=50 100 400 550]{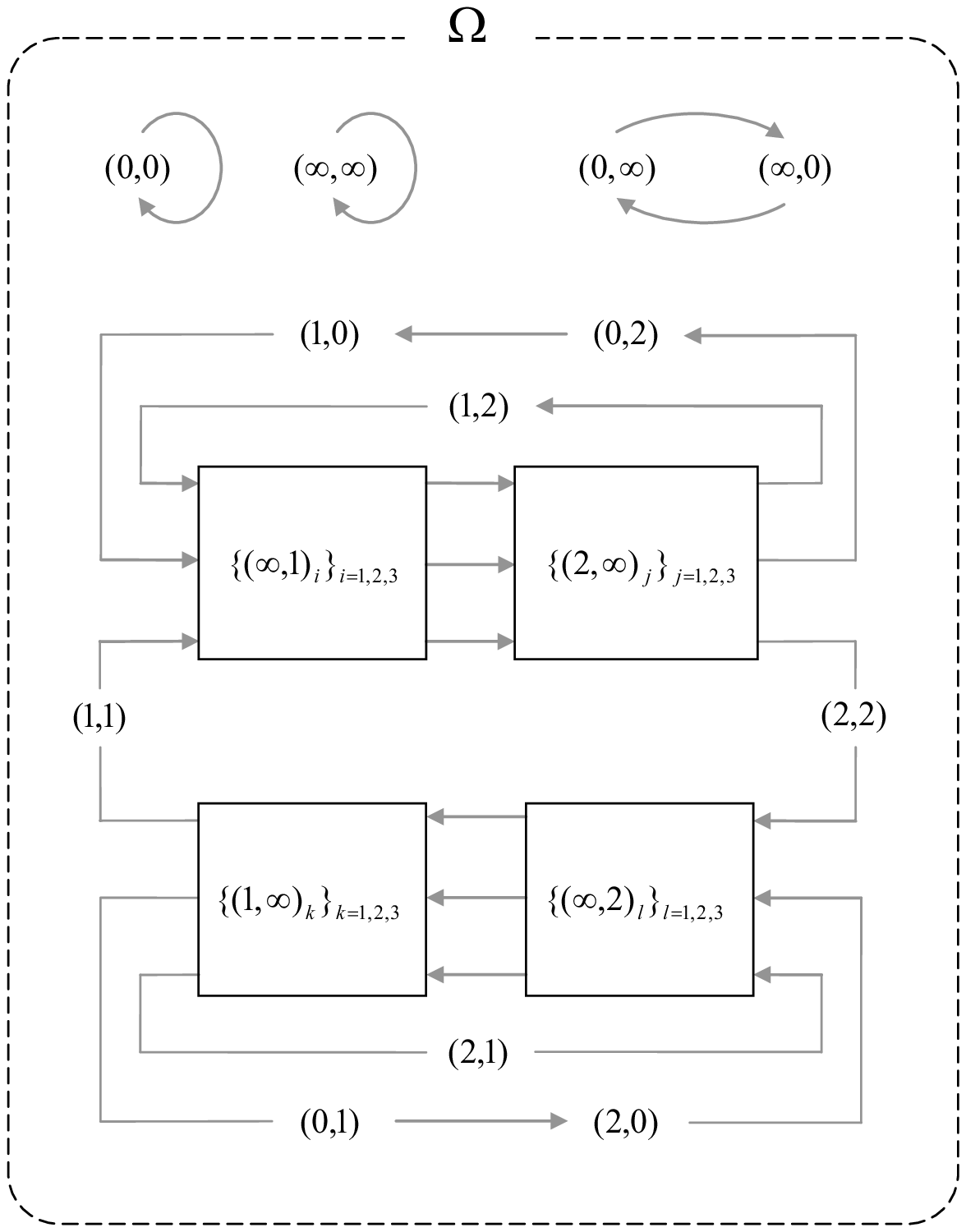}
\caption{The orbit decomposition of the space $\Omega$ for $q=3$ by the mapping $\hat{\phi}$.}
\label{intuitivedP}
\end{figure}

In the previous section, we investigated the space of initial conditions of the dP\II equation by considering a finite field analog of the Sakai theory.
In this section we introduce another method to construct the space of initial conditions over finite fields by directly and intuitively adding finite number of points to the original space $\F_q \times \F_q$.
We take $\alpha_0=1$, $\beta_0=2$, $q=3$ and $\delta=0$ (autonomization) in dP\II equation.
The mapping over $\P\F_3 \times \P\F_3$ is
\begin{equation}
\phi:\ \left\{
\begin{array}{cl}
x'&=-y+\dfrac{1}{1-x}+\dfrac{2}{1+x},\\
y'&=x.
\end{array}
\right.
\end{equation}
First we formally set a division $\dfrac{j}{\infty}\equiv 0$.
We extend the space $\P\F_3 \times \P\F_3$ to $\Omega$, and the map $\phi$ to $\hat{\phi}$, so that
$\hat{\phi}:\Omega\to\Omega$
 is a bijective mapping and that $\hat{\phi}|_{(\P\F_3)^2}=\phi$.
Since,
\[
\phi(1,0)=\phi(1,1)=\phi(1,2)=(\infty,1)
\]
the mapping $\phi$ is not injective. We want $\hat{\phi}$ to be bijective, therefore we set
\begin{equation}
\begin{array}{cl}
\hat{\phi}(1,0)&=(\infty,1)_1\in\Omega,\\
\hat{\phi}(1,1)&=(\infty,1)_2\in\Omega,\\
\hat{\phi}(1,2)&=(\infty,1)_3\in\Omega,
\end{array}
\end{equation}
where $(\infty, 1)_i$, $(i=1,2,3)$ denote distinct points in the extended space $\Omega$.
In the same manner, the point $(\infty,2)\in(\P\F_3)^2$ is divided into three distinct points
$(\infty,2)_1$, $(\infty,2)_2$, $(\infty,2)_3$ in $\Omega$.
Next, since $\phi(\infty,1)=(2,\infty)$ and $\phi(\infty,2)=(1,\infty)$, both the points $(2,\infty)$ and $(1,\infty)$ must be divided into three distinct points in $\Omega$ in order to assure the bijectivity of $\hat{\phi}$.
Lastly, $\phi(1,\infty)$ (and therefore $\hat{\phi}\left((1,\infty)_i \right)$ for $i=1,2,3$) are not well-defined because $x'=-\infty+\frac{1}{0}+1$ is indeterminable. Since we have $y'=1$, we have no choice but to define $\hat{\phi}(1,\infty)_{i}=(j,1)\in\Omega$ in order for the map $\hat{\phi}$ to be well-defined and bijective. Here, $i=1,2,3$ and $j=1,2,3$. Note that $j\neq \infty$ since $(\infty,1)$ has already appeared as the image of the point $(1,0)$.
The same discussion applies to the image of the point $(2,\infty)$.
Summing up the discussions above, we obtain all transitions inside the newly constructed initial value space $\Omega$.
\begin{equation*}
\hat{\phi}:\left\{
\begin{array}{rl|rl}
\Omega&\to \Omega & \Omega & \to \Omega\\ 
(0,0)&\to (0,0) & (2,0)&\to (\infty,2)_1\\
(0,1)&\to (2,0) & (2,1)&\to (\infty,2)_2\\
(0,2)&\to (1,0)&(2,2)&\to (\infty,2)_3\\
(0,\infty)&\to (\infty,0)&(2,\infty)_1&\to (0,2)\\
(\infty,0)&\to (0,\infty)&(2,\infty)_2&\to (1,2)\\
(\infty,\infty)&\to (\infty,\infty)&(2,\infty)_3&\to (2,2)\\
(1,0)&\to (\infty,1)_1&(\infty,1)_{1}&\to (2,\infty)_{i}\\
(1,1)&\to (\infty,1)_2&(\infty,1)_{2}&\to (2,\infty)_{j}\\
(1,2)&\to (\infty,1)_3&(\infty,1)_{3}&\to (2,\infty)_{k}\\
(1,\infty)_1&\to (0,1)&(\infty,2)_{1}&\to (1,\infty)_{l}\\
(1,\infty)_2&\to (1,1)&(\infty,2)_{2}&\to (1,\infty)_{m}\\
(1,\infty)_3&\to (2,1)&(\infty,2)_{3}&\to (1,\infty)_{n}
\end{array}
\right.
\end{equation*}
Here $\{i,j,k\}=\{l,m,n\}=\{1,2,3\}$ and the order of three numbers of each set is not determined with the method in this section. To uniquely determine the correspondences, 
we need to use singularity confinement method.
Note that, apart from the ambiguity above, $\Omega$ exactly corresponds to the space $\Omega^{(n)}$ constructed in the previous section and we have $|\Omega|=|\Omega^{(n)}|=24$.
See the figure \ref{figure1painleve}, and \ref{intuitivedP} for comparison.

%
%
\chapter{One-dimensional systems over a local field and their reduction modulo a prime}\label{chapter3}

We define a generalized notion of good reduction and explain how the notion can be used to detect the integrability of several dynamical systems.
%
%

\section{Almost good reduction}\label{AGRsection}
\begin{Definition}[\cite{KMTT}]\label{AGRdef}
A non-autonomous rational system
\[
\phi_n:\ \Q_p^2 \to (\P \Q_p)^2\  (n \in \Z)
\]
has an almost good reduction modulo $\mathfrak{p}$ on the domain $\mathcal{D}^{(n)}\subseteq \Z_p^2\cap \phi_n^{-1}(\Q_p^2)$, if there
exists a positive integer $m_{\mbox{\rm \scriptsize p};n}$ for any $\mbox{\rm p}=(x,y) \in \mathcal{D}^{(n)}$ and time step $n$ such that
\begin{equation}
\widetilde{\phi_n^{m_{\mbox{\rm \tiny p};n}}(x,y)}=\widetilde{\phi_n^{m_{\mbox{\rm \tiny p};n}}}(\tilde{x},\tilde{y}),
\label{AGR}
\end{equation}
where $\phi_n^m :=\phi_{n+m-1} \circ \phi_{n+m-2} \circ \cdots \circ \phi_n$.
\end{Definition}
\begin{figure}
\centering
\includegraphics[width=12cm, bb=50 200 650 500]{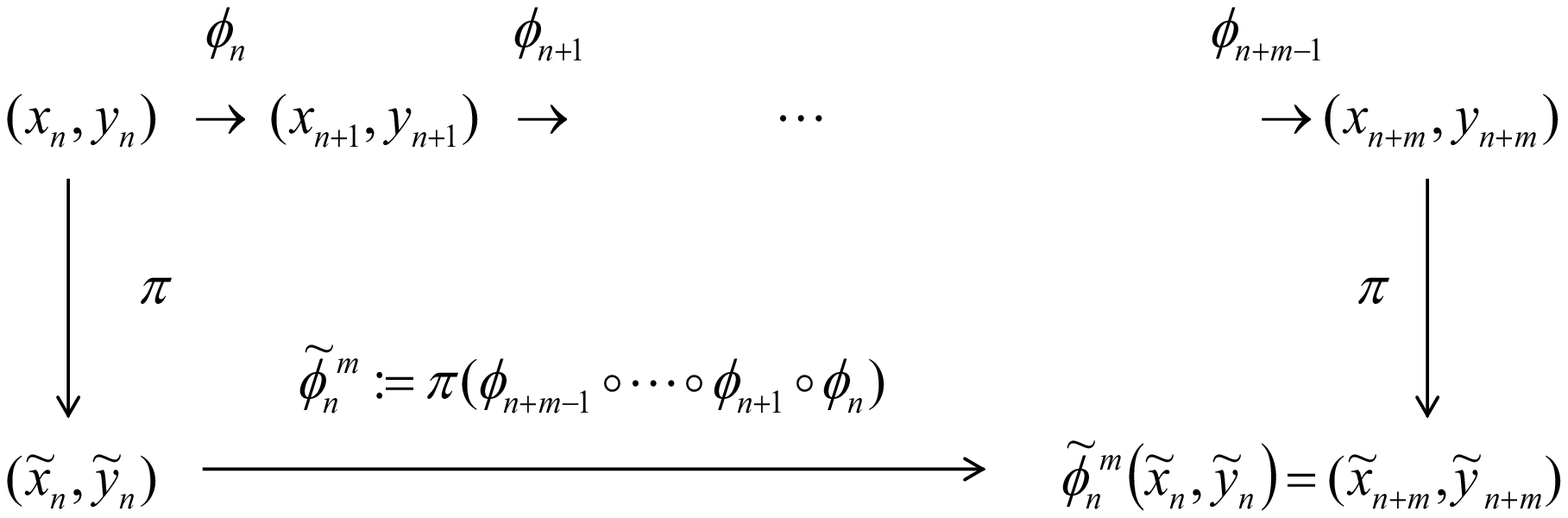}
\caption{Almost good reduction property}
\label{figureagr}
\end{figure}
In this chapter, we take $\mathcal{D}^{(n)}=\Z_p^2\cap \phi_n^{-1}(\Q_p^2)$
and explain that having almost good reduction on $\mathcal{D}^{(n)}$ is equivalent for the mapping to be integrable. If $\mathcal{D}^{(n)}$ does not depend on $n$, we simply write $\mathcal{D}^{(n)}=\mathcal{D}$.
The almost good reduction property is equivalent
for the diagram in figure \ref{figureagr} to be commutative.
Note that if we can take $m_{\mbox{\rm \scriptsize p};n}=1$, the system has a good reduction.
Let us first see the significance of the notion of \textit{almost good reduction}. Let us consider the mapping $\Psi_\gamma$:
\begin{equation}
\left\{
\begin{array}{cl}
x_{n+1}&=\dfrac{ax_n+1}{x_n^\gamma y_n}\\
y_{n+1}&=x_n
\end{array}
\right.,
\label{discretemap}
\end{equation} 
where $a\in\mathbb{Z}_p^{\times}$ ($\leftrightarrow v_p(a)=0$) and $\gamma \in \Z_{\ge 0}$ are parameters. 
The map \eqref{discretemap} is known to be integrable if and only if $\gamma=0,1,2$.
When $\gamma=0,1,2$, \eqref{discretemap} belongs to the QRT family \cite{QRT} and is integrable in the sense that it has a conserved quantity. We also note that when $\gamma=0,1,2$,  \eqref{discretemap} is an autonomous version of the $q$-discrete Painlev\'{e} I equation.
\begin{Theorem}
The rational mapping \eqref{discretemap} with $a\in\mathbb{Z}_p^{\times}$  and $\gamma\in\mathbb{Z}_{\ge 0}$ has an almost good reduction modulo $\mathfrak{p}$ on the domain $\mathcal{D}$ if and only if $\gamma=0,1,2$.
Here $\mathcal{D}=\Z_p^2\cap\Psi_{\gamma}^{-1}(\Q_p^2)$. If $\gamma>0$ then
$\mathcal{D}=\{(x,y) \in \Z^2_p \ |x \ne 0, y \ne 0\}$. If $\gamma=0$ then
$\mathcal{D}=\{(x,y) \in \Z^2_p \ |y \ne 0\}$.
\label{PropQRT}
\end{Theorem}
\textbf{Proof}\;\;
(i) First note that
\[
\widetilde{\Psi_2(x_n,y_n)}=\widetilde{\Psi}_2(\tilde{x}_n,\tilde{y}_n) \qquad \mbox{for $\tilde{x}_n \ne 0, \ \tilde{y}_n \ne 0$},
\]
since $\tilde{x}_{n+1}=\pi\left(\frac{a\tilde{x}_n+1}{\tilde{x}^2_n \tilde{y}_n}\right)$ in this case.

(ii) For $(x_n,y_n)\in\mathcal{D}$ with $\tilde{x}_n=0$ and $\tilde{y}_n \ne 0$, we find that
$\widetilde{\Psi_2^k}(\tilde{x}_n=0,\tilde{y}_n)$ is not defined for $k=1,2$,
however it is defined if $k=3$ and we have
\[
\widetilde{\Psi_2^3(x_n,y_n)}=\widetilde{\Psi_2^3}(\tilde{x}_n=0,\tilde{y}_n)=\left(\dfrac{1}{a^2\tilde{y}_n},0\right).
\]
Since $\tilde{x}_n=0$ is equivalent to $|x|\le p^{-1}\ (\leftrightarrow v_p(x)\ge 1)$, the calculation is done by taking $x_n=p^k\cdot e$ where $k \ge 1$ and $|e|_p=1\ (\leftrightarrow v_p(e)=0 \leftrightarrow e\in\mathbb{Z}_p^{\times})$, and iterating the mapping.

(iii-a) If $\tilde{x}_n \ne 0$ and $\tilde{y}_n=0$ and $v_p(ax_n+b)<v_p(y_n)$ then, by a similar calculation to (ii), we obtain
\[
\widetilde{\Psi_2^5(x_n,y_n)}=\widetilde{\Psi_2^5}(\tilde{x}_n,\tilde{y}_n=0)=\left(0,\dfrac{a^2}{\tilde{x}_n}\right).
\]

(iii-b) If $\tilde{x}_n \ne 0$ and $\tilde{y}_n=0$ and $v_p(ax_n+b)\ge v_p(y_n)$ then,
the apparent singularity is canceled, since $\tilde{x}_{n+2}$ is finite. Then we have,
\[
\widetilde{\Psi_2(x_n,y_n)}=\widetilde{\Psi}_2(\tilde{x}_n,\tilde{y}_n).
\]

(iv) Finally, if $\tilde{x}_n=\tilde{y}_n = 0$, we find that
$\widetilde{\Psi_2^k}(\tilde{x}_n,\tilde{y}_n)$ is not defined for $k=1,2,..,7$,
however
\[
\widetilde{\Psi_2^8(x_n,y_n)}=\widetilde{\Psi_2^8}(\tilde{x}_n=0,\tilde{y}_n=0)=\left(0,0\right) .
\]
In the case (iv), we have an exceptional case just like (iii-b): numerator $ax_{n+i}+1$ may become $0$ modulo $p$ during the time evolution. The singularity is also confined in this exceptional case, because we just arrive at non-infinite values with fewer iterations than (iv) in this case.

Hence the map $\Psi_2$ has almost good reduction modulo $\mathfrak{p}$ on $\mathcal{D}$.
In a similar manner, we find that $\Psi_\gamma$ ($\gamma=0,1$) also has almost good reduction modulo $\mathfrak{p}$ on $\mathcal{D}$. 
On the other hand, for $\gamma \ge 3$ and $\tilde{x}_n=0$, we easily find that
\[
{}^\forall k \in \Z_{\ge 0}, \;\; \widetilde{\Psi_\gamma^{k}(x_n,y_n)} \ne \widetilde{\Psi_\gamma^{k}}(\tilde{x}_n=0,\tilde{y}_n),
\]
since the order of $p$ diverges as we iterate the mapping from $x_n=p^k\cdot e$ where $k>0,\ e\in\mathbb{Z}_p^{\times}$.
Thus we have proved the theorem.\qed
In this theorem we omitted the case of $a=0$.
However we can also treat this case.
In the case $\gamma=2$ and $a=0$, for example,
if we take 
\[
f_{2k}:=x_{2k}x_{2k-1},\ f_{2k-1}:=(x_{2k-1}x_{2k-2})^{-1}
\]
\eqref{discretemap} turns into the trivial linear mapping $f_{n+1}=f_n$ which has apparently good reduction modulo $\mathfrak{p}$.
Note that having an almost good reduction is equivalent to the integrability of the equation in these examples.

\section{Refined Almost Good Reduction}
Next, we introduce another generalization of the good reduction property, which can be used as a `refined' almost good reduction.
We decompose the domain $\mathbb{Q}_p^2$ of the system $\phi$ into three disjoint parts, so that we have the following three types of points\footnote{Another way to define $D_N$ is to take $D_N:=\Z_p^2\cap \phi^{-1}(\Z_p^2)$. The results are essentially the same in this way, however, the computation becomes more complicated.}:
\[
\mathbb{Q}_p^2=D_N\sqcup D_S\sqcup E,
\]
where
\begin{equation}
\left\{
\begin{array}{ll}
D_N&:=\{(x,y)\in\mathbb{Z}_p^2|\ \tilde{\phi}(\tilde{x},\tilde{y})\ \mbox{is well defined in}\ \mathbb{F}_p^2\},\\
D_S&:=\mathbb{Z}_p^2\setminus D_N,\\
E&:=\{(x,y)\in\mathbb{Q}_p^2|\ \tilde{x}=\infty\ \mbox{or}\ \tilde{y}=\infty\}.
\end{array}
\right.
\end{equation}
For example, in the case of $\phi=\Psi_2$ in the previous section,
\[
D_N=\left(\mathbb{Z}_p^{\times}\right)^2,\ \ D_S=\mathbb{Z}_p^2\setminus \left(\mathbb{Z}_p^{\times}\right)^2,\ \ E=\mathbb{Q}_p^2\setminus \mathbb{Z}_p^2.
\]
\begin{Definition}
The mapping $\phi:\ \mathbb{Q}_p^2\to(\P\mathbb{Q}_p)^2$ has the \textit{refined} AGR, if, for every point $(x,y)\in D_N$, there exists an integer $m>0$ such that
\[
\phi^m(x,y)\in D_N.
\]
Here the domain $D_N\subset \mathbb{Q}_p^2$ is defined as
\[
D_N:=\{(x,y)\in\mathbb{Z}_p^2|\ \tilde{\phi}(\tilde{x},\tilde{y})\ \mbox{is well defined in}\ \mathbb{F}_p^2\}.
\]
\end{Definition}
We call the domain $D_N$ the `normal' domain.
\begin{Proposition}
The system $\Psi_2$ has a refined AGR.
\end{Proposition}
\textbf{Proof}\;\;
First let us fix the initial condition $(x_0,\, x_{-1})\in D_N$ ($x_0,x_{-1}\in\mathbb{Z}_p^{\times}$). 

(i) If $\tilde{x}_0\neq \widetilde{-1/a}$, then
\[
x_1=\frac{ax_0+1}{x_0^2 x_{-1}}\in \mathbb{Z}_p^{\times}.
\]
Therefore we have $(x_1, x_0)\in D_N$.  We can take $m=1$ in this case.

(ii) If $\tilde{x}_0= \widetilde{-1/a}$, then we have $x_1\in p\mathbb{Z}_p$ ($\tilde{x}_1=0$). Thus $(x_1,x_0)\in D_S$. Continuing the iterations, we obtain $\tilde{x}_2=\infty$, $\tilde{x}_3=0$, $\tilde{x}_4=\widetilde{-1/a}$. Therefore,
\[
(x_2,x_1)\in E,\ (x_3,x_2)\in E,\ (x_4,x_3)\in D_S.
\]
Finally when $m=5$, we have $\tilde{x}_5=\tilde{x}_{-1}$, and
\[
(x_5,x_4)\in D_N.
\]
By an assumption that $x_{-1}\in \mathbb{Z}_p^{\times}$, we have $\tilde{x}_5\neq 0,\, \infty$.
\qed
We can prove by an argument similar to that in section \ref{AGRsection} that the mapping $\Psi_\gamma$ $(\gamma \ge 3)$ does not have a refine AGR. We can apply refine AGR to non-autonomous systems with minor modifications.
The AGR and the refined AGR are both effective as integrability detectors of dynamical systems over $\mathbb{F}_p$, as we will explain in the following sections. The refined AGR can be more suitable than AGR when we investigate in detail the behaviors around the singularities and zeros of the mapping over $\mathbb{Q}_p$.
On the other hand we have to note that refined AGR requires heavier computation than AGR does to be proved, in particular for non-autonomous systems. Also note that refined AGR and AGR are not equivalent, nor is one of them stronger/weaker than the other one.
In the case of the discrete Painlev\'{e} equations, we have basically the same results for both of the criteria. Therefore we will mainly explain the results regarding the AGR property in the following sections for simplicity. 
\section{Time evolution over finite fields}
We explain how to define the time evolution of discrete dynamical systems over finite fields. Of course, we cannot determine the time evolution solely from the information over finite fields, however, we can propose one reasonable way of evolution by applying the refined AGR.
Let $\phi$ be a dynamical system with refine AGR property.
Let us fix the initial condition $(v,w)\in\F_p^2$.

(i) In the case of $\pi^{-1}(v)\times \pi^{-1}(w)\subset D_N$:
We define $(x_0,x_{-1}):=(v,w)$. By the refined AGR property we have a positive integer $m$ such that
\[
\phi^m (x_0,x_{-1})\in D_N.
\]
We define $\phi^m(v,w):=(\pi(x_m),\pi(x_{m-1}))\in\F_p^2$.
By an assumption, we do not encounter indeterminacies in this calculation.
We also define the intermediate states as $\phi^j(v,w):=(\pi(x_j),\pi(x_{j-1}))\in (\P\F_p)^2$, for $j=1,2,\cdots, m-1$.
Since $\phi^m(v,w)\in D_N$ again, we can continue the time evolution.

(ii) In the case of $\pi^{-1}(v)\times \pi^{-1}(w)\not\subset D_N$:
We cannot define the time evolution by the refined AGR. In this case, we encounter the indeterminate points $\phi^i(v,w)$ for some $i>0$.
We can determine one path of evolution by considering $\phi(v',w')$
for some $v'\in\pi^{-1}(v)$ and $w'\in\pi^{-1}(w)$. However, we have an ambiguity with respect to the choice of the inverse image of $v,w$.
If the mapping $\phi$ has the ordinary AGR in section \ref{AGRsection}, it helps us to define the time evolution for a few steps, however, $\phi^m(v,w)$ is not necessarily in $D_N$.
When we need to know all the trajectories, we may return to the chapter \ref{sec3} and extend the space of initial conditions.
\section{Discrete Painlev\'{e} II equation over finite fields and its special solutions}
Now let us examine the dP\II \eqref{dP2} over $\Q_p$. 
We suppose that $p \ge 3$, and redefine the coefficients $\alpha_n$ and $\beta_n$ so that
they are periodic with period $p$:
\begin{align*}
\alpha_{i+mp}&:=\frac{(i\delta+z_0+a+n_\alpha p)}{2},\ \beta_{i+mp}:=\frac{(-i\delta-z_0+a+n_\beta p)}
{2},\\
(m\in\Z&,\  i\in \{0,1,2,\cdots,p-1\}),
\end{align*}
where the integer $n_\alpha$ ($n_\beta$) is chosen such that $0 \in \{\alpha_i\}_{i=0}^{p-1}$ $(0 \in \{\beta_i\}_{i=0}^{p-1})$.
As a result, we have $\tilde{\alpha}_{n}=\widetilde{\frac{n \delta +z_0+a}{2}}$, $\tilde{\beta}_{n}=\widetilde{\frac{-n \delta -z_0+a}{2}}$ and $|\alpha_n|_p,\ |\beta_n|_p\in\{0,1\}$ for any integer $n$.

\begin{Theorem}
Let $p\ge 7$.
Under the above assumptions, the dP\II equation has an almost good reduction modulo $\mathfrak{p}$ on $\mathcal{D}:=\Z_p^2\cap\phi_n^{-1}(\Q_p^2)=\{ (x,y) \in \Z_p^2\ |x \ne \pm 1\}$.
\label{PropdP2}
\end{Theorem}
\textbf{Proof}\;\; 
We put $(x_{n+1},y_{n+1})=\phi_n(x_n,y_n)=\left( \phi_n^{(x)}(x_n,y_n),\phi_n^{(y)} (x_n,y_n) \right)$. 

When $\tilde{x}_n \ne \pm 1$, we have
\[
\tilde{x}_{n+1}=\dfrac{\tilde{\alpha_n}}{1-\tilde{x}_n}+\dfrac{\tilde{\beta_n}}{1+\tilde{x}_n}-\tilde{y}_{n},
\quad \tilde{y}_{n+1}=\tilde{x}_n.
\]
Hence $\widetilde{\phi_n(x_n,y_n)}=\tilde{\phi}_n(\tilde{x}_n,\tilde{y}_n)$.

When $\tilde{x}_n=1$, we can write $x_n=1+p^k e$ $(k \in \Z_+,\ |e|_p=1)$. 
We have to consider four cases\footnote{Precisely speaking, there are some special cases for $p=3,\ 5$ where we have to consider the fact $\alpha_n=\alpha_{n+p}$ or $\beta_n=\beta_{n+p}$. In these cases the map does not have an almost good reduction on $\mathcal{D}$. They are treated later in this section.}:\\
\noindent
(i) For $\alpha_n= 0 $,
\[
\tilde{x}_{n+1}=\tilde{\phi}_n^{(x)}(\tilde{x}_n,\tilde{y}_n)=\widetilde{\left(\frac{\beta_n}{2}\right)}-\tilde{y}_n.
\]
Hence we have $\widetilde{\phi_n(x_n,y_n)}=\tilde{\phi}_n(\tilde{x}_n,\tilde{y}_n)$.\\
(ii) In the case $\alpha_n \neq 0$ and $\beta_{n+2} \neq 0$, 
\begin{align*}
x_{n+1}&=-\dfrac{(\alpha_n-\beta_n)(1+ep^k)+a}{ep^k(2+ep^k)}-y_n=-\dfrac{2\alpha_n+(\alpha_n-\beta_n)ep^k}{ep^k(2+ep^k)}-y_n,\\
x_{n+2}&=-\frac{\alpha_n^2+\mbox{polynomial of $O(p)$}}{\alpha_n^2+\mbox{polynomial of $O(p)$}},\\
x_{n+3}&=\dfrac{\{2\alpha_{n}y_n+2\delta \beta_{n+1}+(2-\delta)a \}\alpha_n^3 +\mbox{polynomial of $O(p)$}}{2\beta_{n+2}
\alpha_n^3 + \mbox{polynomial of $O(p)$} },
\end{align*}
Thus we have
\[
\tilde{x}_{n+3}=\frac{2\tilde{\alpha}_{n}\tilde{y}_n+2\delta \tilde{\beta}_{n+1}+(2-\delta)a}{2 \tilde{\beta}_{n+2}},
\quad \tilde{y}_{n+3}=-1,
\]
and $\widetilde{\phi_n^3(x_n,y_n)}=\widetilde{\phi_n^3}(\tilde{x}_n,\tilde{y}_n)$.\\
(iii) In the case $\alpha_n \neq 0$, $\beta_{n+2}= 0$ and $a \ne -\delta$, we have to calculate up to $x_{n+5}$.
After a lengthy calculation we find
\[
\tilde{x}_{n+4}=\widetilde{\phi_n^{5}}^{(y)}(1,\tilde{y}_n)=1,\;\mbox{and}\;
\tilde{x}_{n+5}=\widetilde{\phi_n^{5}}^{(x)}(1,\tilde{y}_n)=-\frac{a\delta-(a-\delta)\tilde{y}_n}{a+\delta},
\]  
and we obtain $\widetilde{\phi_n^5(x_n,y_n)}=\widetilde{\phi_n^5}(\tilde{x}_n,\tilde{y}_n)$.\\
(iv) Finally, in the case $\alpha_n \neq 0$, $\beta_{n+2}= 0$ and $a = -\delta$ we have to calculate up to $x_{n+7}$.
The result is
\[
\tilde{x}_{n+6}=\widetilde{\phi_n^{7}}^{(y)}(1,\tilde{y}_n)=-1,\;\mbox{and}\;
\tilde{x}_{n+7}=\widetilde{\phi_n^{7}}^{(x)}(1,\tilde{y}_n)=\frac{1+2\tilde{y}_n}{2},
\]
and we obtain $\widetilde{\phi_n^7(x_n,y_n)}=\widetilde{\phi_n^7}(\tilde{x}_n,\tilde{y}_n)$.
Hence we have proved that the dP\II equation has almost good reduction modulo $\mathfrak{p}$ at $\tilde{x}_n=1$.

We can proceed in the case $\tilde{x}_n=-1$ in an exactly similar manner and find;
(v) For $\beta_n= 0$,
we have $
\tilde{x}_{n+1}=\tilde{\phi}_n^{(x)}(\tilde{x}_n=-1,\tilde{y}_n)=\widetilde{\left(\frac{\alpha_n}{2}\right)}-\tilde{y}_n$.
Therefore we have $\widetilde{\phi_n(x_n,y_n)}=\widetilde{\phi_n}(\tilde{x}_n,\tilde{y}_n)$.\\
(vi) In the case $\beta_n \neq 0$ and $\alpha_{n+2} \neq 0$,
\[
\widetilde{\phi_n^3(x_n,y_n)}=\widetilde{\phi_n^3}(\tilde{x}_n=-1,\tilde{y}_n)
=\left(\frac{a(-2+\delta)-2\delta\alpha_{n+1}+2\beta_n \tilde{y}_n}{2 \alpha_{n+2}}, 1  \right).
\]
(vii) In the case $\beta_n \neq 0$, $\alpha_{n+2} = 0$ and $a \ne \delta$,
\[
\widetilde{\phi_n^5(x_n,y_n)}=\widetilde{\phi_n^5}(\tilde{x}_n=-1,\tilde{y}_n)
=\left(\frac{a \delta+(a+\delta)\tilde{y}_n}{a-\delta}, -1 \right).
\]
(viii) In the case $\beta_n \neq 0$, $\alpha_{n+2} = 0$ and $a = \delta$,
\[
\widetilde{\phi_n^7(x_n,y_n)}=\widetilde{\phi_n^7}(\tilde{x}_n=-1,\tilde{y}_n)=\left(\frac{-1+2\tilde{y}_n}{2}, 1 \right).\qed
\]
From this theorem, the evolution of the dP\II equation \eqref{dP2equation} over $\P\F_p$ can be constructed from the following seven cases which determine $u_{n+1},u_{n+2},\cdots$ from the initial values $u_{n-1}$ and $u_n$.
Note that we can assume that $u_{n-1} \ne \infty$ because all the cases in which the dependent variable $u_n$ becomes $\infty$ are included below\footnote{For $p \le 5$, there are some exceptional cases as shown in the proof of theorem \ref{PropdP2}.}. Here $a=\alpha_n+\beta_n$.
\begin{enumerate}
\item For $u_n \in \{2,3,...,p-2\}$, or $u_n=1$ and $\alpha_n = 0$, or $u_n=p-1$ and $\beta_n =0$,
\[
u_{n+1}=\dfrac{{\alpha}_n}{1-u_n}+\dfrac{{\beta}_n}{1+u_n}-u_{n-1}.\\
\]
\item For $u_n=1$, $\alpha_n \ne 0$ and $\beta_{n+2} \ne 0$,
\[
u_{n+1}=\infty,\  u_{n+2}=p-1,\  u_{n+3}=\frac{2\alpha_n u_{n-1}+2\delta\beta_{n+1}+(2-\delta)a}{2\beta_{n+2}}.
\]
\item For $u_n=1$, $\alpha_n \ne 0$, $\beta_{n+2} = 0$ and $a+\delta \ne 0$,
\begin{align*}
&u_{n+1}=\infty,\  u_{n+2}=p-1,\  u_{n+3}=\infty,\ u_{n+4}=1,\\
&\qquad u_{n+5}=-\frac{a\delta-(a-\delta){u}_{n-1}}{a+\delta}.
\end{align*}
\item For $u_n=1$, $\alpha_n \ne 0$, $\beta_{n+2} = 0$ and $a+\delta = 0$,
\begin{align*}
&u_{n+1}=\infty,\  u_{n+2}=p-1,\  u_{n+3}=\infty,\ u_{n+4}=1,\ u_{n+5}=\infty,\\
&\qquad u_{n+6}=p-1,\ u_{n+7}=\frac{1+2{u}_{n-1}}{2}.
\end{align*}
\item For $u_n=p-1$, $\beta_n \ne 0$ and $\alpha_{n+2} \ne 0$,
\[
u_{n+1}=\infty,\ u_{n+2}=1,\ u_{n+3}=\frac{a(-2+\delta)-2\delta\alpha_{n+1}+2\beta_n {u}_{n-1}}{2 \alpha_{n+2}}.
\]
\item For $u_n=p-1$, $\beta_n \ne 0$, $\alpha_{n+2} = 0$ and $a \ne \delta$,
\[
u_{n+1}=\infty,\ u_{n+2}=1,\ u_{n+3}=\infty,\ u_{n+4}=p-1,\ u_{n+5}=\frac{a \delta+(a+\delta){u}_{n-1}}{a-\delta}.
\]
\item For $u_n=p-1$, $\beta_n \ne 0$, $\alpha_{n+2} = 0$ and $a = \delta$,
\begin{align*} 
&u_{n+1}=\infty,\ u_{n+2}=1,\ u_{n+3}=\infty,\ u_{n+4}=p-1,\ u_{n+5}=\infty, \\
&\qquad u_{n+6}=1,\ u_{n+7}=\frac{-1+2{u}_{n-1}}{2}.
\end{align*}
\end{enumerate}
\subsection{Exceptional cases where $p=3$ and $p=5$.}
Now we study the exceptional cases: $p=3$ and $p=5$.
In these cases, the almost good reduction property does not hold for all points in $\mathcal{D}=\{(x,y)\in\Z_p^2|x\neq \pm 1\}$. The situations change depending on the value `$x\!\!\mod p^2$'.
Here $(x\!\!\mod p^2)$ for a $p$-adic integer $x\in\mathbb{Z}_p$ is defined as $x=x_0+x_1p$ from the $p$-adic expansion
\[
x=x_0+x_1p+x_2p^2+x_3p^3+\cdots,
\]
of $x$ where each $x_i\in\{0,1,2,\cdots, p-1\}$.
Let us first consider the case of $p=3$.
We explain the details via an example when $\delta=2$, $\alpha_0=1$ and $\beta_0=2$.
The dP\II equation in this case takes the following three forms periodically:
\begin{equation}
\left\{
\begin{array}{cl}
\phi_{n+3j}:&\; x_{n+1+3j}=-x_{n-1+3j}+\dfrac{-2}{1-x_{n+3j}}+\dfrac{2}{1+x_{n+3j}},\\
\phi_{n+1+3j}:&\;x_{n+2+3j}=-x_{n+3j}+\dfrac{-1}{1-x_{n+1+3j}}+\dfrac{1}{1+x_{n+1+3j}},\\
\phi_{n+2+3j}:&\;x_{n+3+3j}=-x_{n+1+3j},
\end{array}
\right.
\end{equation}
where $j$ is an integer.
Unfortunately, the dP\II equation over $\Z_3^2$ with $\delta=2$, $\alpha_0=1$ and $\beta_0=2$ does not have an almost good reduction. 
However, it has a somewhat weaker property than the almost good reduction on the following domain $\mathcal{D}$:
\[
\mathcal{D}=\{(x,y)\in\Z_3\times\Z_3|\ x\not\in \pm 1+9\Z_3 \}.
\]
\begin{Proposition}
Let $\mathcal{D}$ as above.
For every $(x,y)\in\mathcal{D}$, there exists a positive integer $m>0$ such that
\[
\widetilde{\phi_n^m(x_n,y_n)}=\tilde{\phi_n^m}\left((x_n\!\!\!\mod  9), \tilde{y}_n\right)
\]
holds. $($We will call this property `weak' almost good reduction.$)$

If $x \in 1+9 \Z_3$, then the solution modulo $\mathfrak{p}$ goes into the periodic orbit:
\[
(1,k)\mapsto(\infty,1)\mapsto(2,\infty)\mapsto(\infty,2)\mapsto(1,\infty)\mapsto(\infty,1)\cdots
\]
for $k=0,1,2$.

If $x \in -1+9 \Z_3$, then the solution goes into the periodic orbit:
\[
(2,k)\mapsto(\infty,2)\mapsto(1,\infty)\mapsto(\infty,1)\mapsto(2,\infty)\mapsto(\infty,2)
\]
for $k=0,1,2$.
\end{Proposition}
\textbf{Proof}\;\;

(i) If $\tilde{x}_n\neq \pm 1$ then we have $\widetilde{\phi_n(x_n,y_n)}=\widetilde{\phi_n}(\tilde{x}_n,\tilde{y}_n)$.

(ii) If $\tilde{x}_n=1$ then we have three cases to consider:

(ii-a) If ($\tilde{x}_{n-1}=1$ and $x_n\in 7+9\Z_3$) or ($\tilde{x}_{n-1}=2$ and $x_n\in 4+9\Z_3$) then,
\[
\widetilde{\phi_n^5(x_n,y_n)}=\widetilde{\phi_n^5}(x \!\!\!\mod 9,\ \tilde{y}_n)
=\left( 0, 1 \right).
\]

(ii-b) If ($\tilde{x}_{n-1}=0$ and $x_n\in 7+9\Z_3$) or ($\tilde{x}_{n-1}=1$ and $x_n\in 4+9\Z_3$) then,
\[
\widetilde{\phi_n^5(x_n,y_n)}=\widetilde{\phi_n^5}(x \!\!\!\mod 9,\ \tilde{y}_n)
=\left( 1, 1 \right).
\]

(ii-c) If ($\tilde{x}_{n-1}=2$ and $x_n\in 7+9\Z_3$) or ($\tilde{x}_{n-1}=0$ and $x_n\in 4+9\Z_3$) then,
\[
\widetilde{\phi_n^5(x_n,y_n)}=\widetilde{\phi_n^5}(x \!\!\! \mod 9,\ \tilde{y}_n)
=\left( 2, 1 \right).
\]

(ii-d) If $x_n\in 1+9\Z_3$ then, both the reduced mappings and the reduced coordinates return to the original position after iterating the mappings $12$ times
from the lemma \ref{caseD}.

(iii) If $\tilde{x}_n=-1$ then we have three points to consider: $(2,0)$, $(2,1)$ and $(2,2)$.
The proof is much the same as in the case of (ii).\qed

\begin{Lemma}\label{caseD}
For the initial value $(x_{n+1},x_n)\in\mathbb{Q}_3\times\mathbb{Q}_3$, with $x_n\in1+9\mathbb{Z}_3$ and $x_{n+1}\in\mathbb{Q}_3\setminus 3\mathbb{Z}_3$,
we have $x_{n+12}\in 1+9\mathbb{Z}_3$ and $x_{n+13}\in \mathbb{Q}_3\setminus 3\mathbb{Z}_3$ for
\[
\phi_{n+1}^{12}(x_{n+1},x_n)=(x_{n+13},x_{n+12}).
\]
\end{Lemma}
\textbf{Proof}\;\;
We can write $x_n=1+9m$ and $x_{n+1}=\dfrac{n}{3}$ with $m\in\mathbb{Z}_3$ and $n\in\mathbb{Q}_3\setminus 3\mathbb{Z}_3$.
By iterating the mappings we have
\begin{eqnarray*}
x_{n+12}&=&\frac{4n^6-8m n^7+4m^2n^8+O(3)}{4n^6-8m n^7+4m^2n^8+O(3)}\equiv 1,\\
x_{n+13}&=&\frac{16n^9-64mn^{10}-64m^3n^{12}+16m^4n^{13}+O(3)}{O(3)}\equiv \infty,
\end{eqnarray*}
where $O(3)$ denotes a polynomial whose coefficients are multiples of three.
Therefore we obtain $x_{n+12}\in 1+9\mathbb{Z}_3$ and $x_{n+13}\in \mathbb{Q}_3\setminus 3\mathbb{Z}_3$.
\qed

In the case of $p=5$, we have a similar result.
Let us consider the dP\II equations with the same parameters as in the case of $p=3$: $\delta=2$, $\alpha_0=1$ and $\beta_0=2$.
Then the dP\II equation is expressed as the following five maps.
\begin{equation}
\left\{
\begin{array}{cl}
\phi_{n+5j}:&\; x_{n+1+5j}=-x_{n-1+5j}+\dfrac{-4}{1-x_{n+5j}}+\dfrac{2}{1+x_{n+5j}},\\
\phi_{n+1+5j}:&\;x_{n+2+5j}=-x_{n+5j}+\dfrac{-3}{1-x_{n+1+5j}}+\dfrac{1}{1+x_{n+1+5j}},\\
\phi_{n+2+5j}:&\;x_{n+3+5j}=-x_{n+1+5j}+\dfrac{-2}{1-x_{n+2+5j}},\\
\phi_{n+3+5j}:&\;x_{n+4+5j}=-x_{n+2+5j}+\dfrac{-1}{1-x_{n+3+5j}}+\dfrac{-1}{1+x_{n+3+5j}},\\
\phi_{n+4+5j}:&\;x_{n+5+5j}=-x_{n+3+5j}+\dfrac{-2}{1+x_{n+4+5j}},
\end{array}
\right.
\end{equation}
where $j$ is an integer.
\begin{Proposition}
The dP\II equation above have `weak' almost good reduction on the following domain $\mathcal{D}$:
\[
\mathcal{D}=\{(x,y)\in\mathbb{Z}_5^2|x\neq +1+25\mathbb{Z}_5\}.
\]
If $x\in 1+25\mathbb{Z}_5$ then, the time evolution goes into a periodic orbit:
\[
(4,\infty)\to(\infty,4)\to(1,\infty)\to(\infty,1).
\]
\end{Proposition}
\textbf{Proof}\;\;

(i) If $\tilde{x}_n\neq \pm 1$ then we have $\widetilde{\phi_n(x_n,y_n)}=\widetilde{\phi_n}(\tilde{x}_n,\tilde{y}_n)$.

(ii) If $\tilde{x}_{n}=1$ then, the time evolution depends on $x_n\!\!\! \mod 25=1,6,11,16$ or $21$.
If $x_n\in 1+25\mathbb{Z}_5$ then, the orbit is periodic with a period $4$. We classify other four cases below.

(ii-a) If $x_n\in 6+25\mathbb{Z}_5$ then,
\[
\widetilde{\phi_n^7(x_n,y_n)}=\widetilde{\phi_n^7}(x_n\!\!\!\! \mod 25=6,\tilde{y}_n)
=\left( \widetilde{y_n - 1}, -1 \right).
\]

(ii-b) If $x_n\in 11+25\mathbb{Z}_5$ then,
\[
\widetilde{\phi_n^7(x_n,y_n)}=\widetilde{\phi_n^7}(x_n\!\!\!\! \mod 25=11,\tilde{y}_n)
=\left( \widetilde{y_n + 1}, -1 \right).
\]

(ii-c) If $x_n\in 16+25\mathbb{Z}_5$ then,
\[
\widetilde{\phi_n^7(x_n,y_n)}=\widetilde{\phi_n^7}(x_n\!\!\!\! \mod 25=16,\tilde{y}_n)
=\left( \widetilde{y_n}, -1 \right).
\]

(ii-d) If $x_n\in 21+25\mathbb{Z}_5$ then,
\[
\widetilde{\phi_n^7(x_n,y_n)}=\widetilde{\phi_n^7}(x_n\!\!\!\! \mod 25=21,\tilde{y}_n)
=\left( \widetilde{y_n + 2}, -1 \right).
\]

(iii) If $\tilde{x}_{n}=-1$ then,
\[
\widetilde{\phi_n^3(x_n,y_n)}=\widetilde{\phi_n^3}(\tilde{x}_n=1,\tilde{y}_n=1)
=\left( \widetilde{7-y_n}, 1 \right).
\]
In the case of (iii), the time evolution up to the third iteration does not depend on $x_n\!\!\! \mod 25$, but depends only on $(x_n\!\!\! \mod 5)=\tilde{x}_n$.
\qed
Note that in this case, the singularities are confined if $x\in -1+25\mathbb{Z}_5$, unlike the result in the case of $p=3$.
\subsection{Its Special solutions}
Next we consider special solutions to \eqref{dP2equation} over $\P\F_p$.
For the dP\II equation over $\C$, rational function solutions have already been obtained \cite{Kajiwara}.
Let $N$ be a positive integer and $\lambda \ne 0$  be a constant. Suppose that 
$a=-2(N+1)/ \lambda$, $\delta=z_0=2/\lambda$,
\[
L_k^{(\nu)}(\lambda):=\left\{ 
\begin{array}{cl}
\displaystyle
\sum_{r=0}^k(-1)^r
\begin{pmatrix}
k+\nu\\
k-r
\end{pmatrix} 
\dfrac{\lambda^r}{r!}&\quad(k \in \Z_{\ge 0}),\\
0 &\quad (k \in \Z_{<0}),
\end{array}
\right.
\]
and
\begin{equation}
\tau_N^n:=\det
\begin{vmatrix}
L_{N+1-2i+j}^{(n)}(\lambda)
\end{vmatrix}_{1\le i,j\le N}.
\label{Ltau}
\end{equation}
Then a rational function solution of the dP\II equation is given by
\begin{equation}
u_n=\frac{\tau_{N+1}^{n+1}\tau_{N}^{n-1}}{\tau_{N+1}^n\tau_N^n}-1.
\label{rationaldP2}
\end{equation}
If we deal with the terms in \eqref{Ltau} and \eqref{rationaldP2} by arithmetic operations over $\F_p$, 
we encounter terms such as $1/p$ or $p/p$ and \eqref{rationaldP2} is not well-defined.
However, from theorem \ref{PropdP2}, we find that \eqref{rationaldP2} gives a solution to the dP\II equation over $\P\F_q$ by
the reduction from $\Q (\subset \Q_p)$, as long as the solution avoids the points $(\tilde{\alpha}_n=0,\ u_n=1)$ and $(\tilde{\beta_n}=0,\ u_n=-1)$, which is equivalent to the solution satisfying
\begin{equation}
\tau_{N+1}^{-N-1} \tau_N^{-N-3}\not\equiv 0,\ \frac{\tau_{N+1}^{N+1}\tau_N^{N-1}}{\tau_{N+1}^N\tau_N^N}\not\equiv 2, \label{taucond}
\end{equation}
where the superscripts are considered modulo $p$. Note that $\tau_N^n\equiv \tau_N^{n+p}$ for all integers $N$ and $n$.
In the table below, we give several \textit{rational solutions to the dP\II equation} with $N=3$ and $\lambda=1$ over $\P\F_q$ for $q=3,5,7$ and $11$. We see that the period of the solution is $p$.
\begin{scriptsize}
\[
\begin{array}{|c|c|c|l|}
\hline
& & & \\[-2mm]
\raise3mm\hbox{$p$} & \raise3mm\hbox{$\tau_{N+1}^{-N-1}\tau_N^{-N-3}$} & \raise3mm\hbox{$\frac{\tau_{N+1}^{N+1}\tau_N^{N-1}}{\tau_{N+1}^N\tau_N^N}$}
&\enskip
\raise3mm\hbox{$\tilde{u}_1,\tilde{u}_2,\tilde{u}_3,\tilde{u}_4,\tilde{u}_5,\tilde{u}_6,\tilde{u}_7,\tilde{u}_8,\tilde{u}_9,\tilde{u}_{10},\ldots$} \\ \hline & & & \\[-3mm]
3 & \infty & \infty &\
\raise1mm\hbox{$\underbrace{1,2,\infty}_{\mbox{period
$3$}},1,2,\infty,1,2,\infty,1,2,\infty,1,2,\infty,\ldots$}
\\[5mm] \hline
& & & \\[-3mm]
5 & \infty & 4 &\
\raise1mm\hbox{$\underbrace{4,2,3,1,\infty}_{\mbox{period
$5$}},4,2,3,1,\infty,4,2,3,1,\infty,4,\ldots$} \\[5mm] \hline & & & \\[-3mm]
7 & \infty & 0 &\
\raise1mm\hbox{$\underbrace{1,\infty,6,5,1,\infty,6}_{\mbox{period
$7$}},1,\infty,6,5,1,\infty,6,1,\infty,\ldots$} \\[5mm] \hline & & & \\[-3mm]
11 & 0 & 7 &\
\raise1mm\hbox{$\underbrace{\infty,1,6,1,\infty,10,\infty,1,0,2,10}_{\mbox{period
$11$}},\infty,1,6,1,\ldots$} \\[5mm] \hline \end{array} \]
\end{scriptsize}
We see from the case of $p=11$ that we may have an appropriate solution even if the condition \eqref{taucond} is not satisfied, although this is not always true.
The dP\II equation has linearized solutions also for $\delta=2a$ \cite{Tamizhmani}. 
With our new method, we can obtain the corresponding solutions without difficulty.

\section{$q$-discrete Painlev\'{e} equations over finite fields}

\subsection{$q$-discrete Painlev\'{e} I equation}

One of the forms of the $q$-discrete analogs of the Painlev\'{e} I equation is as follows:
\begin{equation}
x_{n+1}x_{n-1}=\frac{aq^nx_n+b}{x_n^2}, \label{qp1eq}
\end{equation}
where $a$ and $b$ are parameters \cite{Sakai}.
We rewrite \eqref{qp1eq} for our convenience as a form of dynamical system with two variables:
\begin{equation}
\Phi_n: \left\{
\begin{array}{cl}
x_{n+1}&=\dfrac{aq^n x_n+b}{x_n^2 y_n},\\
y_{n+1}&=x_n.
\end{array}
\right.
\label{qP1}
\end{equation} 
We can prove the AGR property for this equation:
\begin{Theorem}
Suppose that $a, b, q$ are integers not divisible by $p$, then the mapping \eqref{qP1} has an almost good reduction   
modulo $\mathfrak{p}$ on the domain 
$\mathcal{D}:=\Z_p^2\cap\Phi_n^{-1}(\Q_p^2)=\{(x,y)\in \Z_p^2\ |x \ne 0, y\ne 0\}$.
\label{PropqP1}
\end{Theorem}
\textbf{proof}\;\;

Let $(x_{n+1},y_{n+1})=\Phi_n(x_n,y_n)$.
Just like we have done before, we have only to examine the cases $\tilde{x}_n=0$, and $\tilde{y}_n=0$.
We use the abbreviation $\tilde{q}=q, \tilde{a}=a, \tilde{b}=b$ for simplicity. By direct computation we obtain;

(i) If $\tilde{x}_n=0$ and $\tilde{y}_n\ne 0$, then
\[
\widetilde{\Phi_n^3(x_n,y_n)} = \widetilde{\Phi_n^3}(0,\tilde{y}_n)=\left(\frac{b^2}{a^2 q^2 \tilde{y}_n},0\right).
\]

(ii) If $\tilde{y}_n=0$ and $\tilde{x}_n\ne 0$, then
\[
\widetilde{\Phi_n^5(x_n,y_n)} = \widetilde{\Phi_n^5}(\tilde{x}_n,0)=\left(0, \frac{a^2 q^4}{b \tilde{x}_n}\right).
\]

(iii) If $\tilde{x}_n=0$ and $\tilde{y}_n= 0$, then
\[
\widetilde{\Phi_n^8(x_n,y_n)} = \widetilde{\Phi_n^8}(0,0)=\left( 0, 0\right).\qed
\]
The same is true for a refined AGR property.
\begin{Proposition}
Suppose that $a, b, q$ are integers not divisible by $p$, then the mapping \eqref{qP1} has a refined almost good reduction. Here the normal domain is
$D_N=(\mathbb{Z}_p^{\times})^2$.
\end{Proposition}
\textbf{Proof}\;\;
First let us fix the initial condition $(x_0,\, x_{-1})\in D_N$ ($x_0,x_{-1}\in\mathbb{Z}_p^{\times}$). 

(i) If $\tilde{x}_0\neq \pi(-b/a)$, then
\[
x_1=\frac{ax_0+1}{x_0^2 x_{-1}}\in \mathbb{Z}_p^{\times}.
\]
Therefore we have $(x_1, x_0)\in D_N$.  We can take $m=1$ in this case.

(ii) If $\tilde{x}_0= \pi(-b/a)$, then we have $x_1\in p\mathbb{Z}_p$ ($\tilde{x}_1=0$). Thus $(x_1,x_0)\in D_S$. Continuing the iterations, we obtain $\tilde{x}_2=\infty$, $\tilde{x}_3=0$ and $\tilde{x}_4=\displaystyle\pi\left(\frac{-b}{a q^4}\right)$. Therefore,
\[
(x_2,x_1)\in E,\ (x_3,x_2)\in E,\ (x_4,x_3)\in D_S.
\]
At the next step,
\[
\tilde{x}_5=\pi\left(\frac{q^6(a^3(q-q^5)+b^2 \tilde{x}_{-1})}{b^2}\right).
\]

(ii-1) If $\tilde{x}_{-1}\neq \pi\left(a^3(q^5-q)/b^2)\right)$ then
\[
(x_5,x_4)\in D_N.
\]

(ii-2) If $\tilde{x}_{-1}= \pi\left(a^3(q^5-q)/b^2)\right)$ then
we have to continue the iterations further until we obtain
\[
\tilde{x}_8=-\frac{b}{a q^8},\ \tilde{x}_9=-\frac{a^3(q^4-1)q^{19}}{b^2}.
\]
Here note that $\tilde{x}_9=0$ is equivalent to $q^4\equiv 1\mod \pi$, which is in turn equivalent to $\tilde{x}_{-1}=0$. Therefore by the assumption, we have $\tilde{x}_9\neq 0$.
Thus we have proved that
\[
(x_9,x_8)\in D_N.\qed
\]
%
%
\subsection{$q$-discrete Painlev\'{e} II equation}
%
%
We study the $q$-discrete analog of the Painlev\'{e} II equation ($q$P\II equation):
\begin{equation}
(z(q\tau)z(\tau)+1)(z(\tau)z(q^{-1}\tau)+1)=\frac{a \tau^2 z(\tau)}{\tau-z(\tau)},
\label{qP2eq}
\end{equation}
where $a$ and $q$ are parameters \cite{Kajiwaraetal}.
It is also convenient to rewrite \eqref{qP2eq} as
\begin{equation}
\Phi_n: \left\{
\begin{array}{cl}
x_{n+1}&=\dfrac{a(q^n\tau_0)^2x_n-(q^n\tau_0-x_n)(1+x_ny_n)}{x_n(q^n\tau_0-x_n)(x_ny_n+1)},\\
y_{n+1}&=x_n,
\end{array}
\right.
\label{qP2}
\end{equation}
where $\tau=q^n\tau_0$.
Similarly to the dP\II equation, we can prove the following theorem:
\begin{Theorem}
Suppose that $a, q, \tau_0$ are integers not divisible by $p$, then the mapping \eqref{qP2} has an almost good reduction   
modulo $\mathfrak{p}$ on the domain 
$\mathcal{D}^{(n)}:=\Z_p^2\cap\Phi_n^{-1}(\Q_p^2)=\{(x,y)\in \Z_p^2\ |x \ne 0, x \ne q^n\tau_0,\ xy+1 \ne 0\}$.
\label{PropqP2}
\end{Theorem}
\textbf{Proof}\;\; 
Let $(x_{n+1},y_{n+1})=\Phi_n(x_n,y_n)$. 
Just like the proof of theorem \ref{PropdP2}, we have only to examine the cases $\tilde{x}_n=0, \widetilde{q^n\tau_0}$ 
and $-\tilde{y}_n^{-1}$.
We use the abbreviation $\tilde{q}=q, \tilde{\tau}=\tau, \tilde{a}=a$ for simplicity. 
By direct computation, we obtain;\\
(i) If $\tilde{x}_n=0$ and $ -1+q^2-aq^2\tau^2+q^3\tau^2-q^2\tau \tilde{y}_n \ne 0$, 
\[
\widetilde{\Phi_n^3(x_n,y_n)} = \widetilde{\Phi_n^3}(\tilde{x}_n=0,\tilde{y}_n)=\left( \frac{ 1-q^2+aq^2\tau^2-q^3\tau^2-aq^4\tau^2+q^2\tau \tilde{y}_n}{q^2\tau( -1+q^2-aq^2\tau^2+q^3\tau^2-q^2\tau \tilde{y}_n ) }   , q^2\tau  \right).
\]
(ii) If $\tilde{x}_n=0$ and $ -1+q^2-aq^2\tau^2+q^3\tau^2-q^2\tau \tilde{y}_n = 0$, 
\[
\widetilde{\Phi_n^5(x_n,y_n)} = \widetilde{\Phi_n^5}(\tilde{x}_n=0,\tilde{y}_n) =\left( \frac{1-q^2+q^7\tau^2-aq^8\tau^2}{q^4\tau}, 0    \right).
\]
(iii) If $\tilde{x}_n=\tau$ and $1+\tau \tilde{y}_n\ne 0$,
\begin{align*}
&\widetilde{\Phi_n^3(x_n,y_n)} = \widetilde{\Phi_n^3}(\tilde{x}_n=\tau,\tilde{y}_n)\\
&\quad =\left(\frac{ 1-q^2+(a+q-aq^2)q^2\tau^2+(1-q^2)\tau\tilde{y}+(1-aq)q^3\tau^3\tilde{y}  }{q^2\tau(1+\tau\tilde{y}_n)}, 0 \right).
\end{align*}
(iv) If $\tilde{x}_n=\tau$ and $1+\tau \tilde{y_n}= 0$,
\[
\widetilde{\Phi_n^7(x_n,y_n)} = \widetilde{\Phi_n^7}(\tilde{x}_n=\tau,\tilde{y}_n)=\left(\frac{1}{aq^{12}\tau^3}, - aq^{12}\tau^3  \right).
\]
(v) If $\tilde{x_n}\tilde{y}_n+1=0$,
\[
\widetilde{\Phi_n^7(x_n,y_n)} = \widetilde{\Phi_n^7}(\tilde{x}_n=-\tilde{y}_n^{-1}, \tilde{y}_n)=\left(-\frac{1}{aq^{12}\tau^4\tilde{y}_n}, aq^{12}\tau^4\tilde{y}_n  \right).
\]
Thus we complete the proof. \qed
Note that the `refined' AGR is not properly defined for the $q$P\II equation, since we have the term $(q^n \tau_0-x_n)$ in the denominator of $x_{n+1}$, which prevents the definition of the normal domain $D_N$.
We can overcome this problem if we are to define the normal domain $D_N^{(n)}$ to be non-autonomous, however, the computation becomes heavier.
\subsection{Special solutions of $q$P\II equation}
From the previous theorem, we can define the time evolution of the $q$P\II equation explicitly just like the dP\II equation in the previous section.
We consider special solutions for $q$P\II equation \eqref{qP2eq} over $\P\F_p$. 
In \cite{HKW} it has been proved that \eqref{qP2eq} over $\C$ with $a=q^{2N+1}$ $(N \in \Z)$ is solved by
the functions given by
\begin{align}
z^{(N)} (\tau) &= 
\begin{cases}
\DIS \frac{g^{(N)} (\tau) g^{(N+1)} (q \tau)}{q^N g^{(N)} (q \tau) g^{(N+1)} (\tau)}
 & (N \ge 0) \\
\DIS \frac{g^{(N)} (\tau) g^{(N+1)} (q \tau)}{q^{N+1} g^{(N)} (q \tau) g^{(N+1)} (\tau)} & (N<0)
\end{cases}, \label{eq:gtoz} \\
g^{(N)} (\tau) &= 
\begin{cases}
\det |w(q^{-i+2j-1}\tau)|_{1\le i,j\le N} & (N>0) \\
1 & (N=0) \\
\det |w(q^{i-2j}\tau)|_{1\le i,j\le |N|}& (N<0)
\end{cases}, \label{eq:det_sol_g}
\end{align}
where $w(\tau)$ is a solution of the $q$-discrete Airy equation:
\begin{equation}
w(q\tau)-\tau w(\tau)+w(q^{-1}\tau)=0. 
\label{dAiryeq}
\end{equation}
As in the case of the dP\II equation, we can obtain the corresponding solutions 
to \eqref{eq:gtoz} over $\P\F_p$ by reduction modulo $\mathfrak{p}$ according to the theorem \ref{PropqP2}.
For that purpose, we have only to solve \eqref{dAiryeq} over $\Q_p$.
By elementary computation we obtain:
\begin{equation}
w(q^{n+1}\tau_0)=c_1P_{n}(\tau_0;q)+c_0P_{n-1}(q\tau_0;q),
\label{Airy:sol}
\end{equation}
where $c_0,\ c_1$ are arbitrary constants and $P_n(x;q)$ is defined by the tridiagonal determinant: 
\[
P_n(x;q):=
\left|
\begin{array}{ccccc}
qx&-1&&&\\
-1&q^2x&-1&&\hugesymbol{0}\\
&\ddots&\ddots&\ddots& \\
&&-1&q^{n-1}x&-1\\
\hugesymbol{0}&&&-1&q^{n}x
\end{array}
\right|.
\]
The function $P_n(x;q)$ is the polynomial of $n$th order in $x$,
\[
P_n(x;q)=\sum_{k=0}^{[n/2]}(-1)^k a_{n;k}(q)x^{n-2k},
\]
where $a_{n;k}(q)$ are polynomials in $q$.
If we let $i \ll j $ denotes $i<j-1$, and 
$
c(j_1,j_2,...,j_k):=\sum_{r=1}^k (2j_r+1),
$
then, we have
\begin{eqnarray*}
a_{n;0}&=&q^{n(n+1)/2},\\
a_{n;k}&=&\sum_{1\le j_1 \ll j_2 \ll \cdots \ll j_k \le n-1} q^{n(n+1)/2 -c(j_1,j_2,...,j_k)}.
\end{eqnarray*}
Therefore the solution of $q$P\II equation over $\P\F_p$ is obtained by reduction modulo $\mathfrak{p}$ from \eqref{eq:gtoz}, \eqref{eq:det_sol_g}
and \eqref{Airy:sol} over $\Q$ or $\Q_p$.


\subsection{$q$-discrete Painlev\'{e} III equation}
The $q$-discrete analog of the Painlev\'{e} III equation has the following form
\[
x_{n+1}x_{n-1}=\frac{ab(x_n-cq^n)(x_n-dq^n)}{(x_n-a)(x_n-b)},
\]
where $a,b,c,d$ and $q$ are parameters \cite{RGH}.
It is convenient to rewrite it as the following coupled form
\begin{equation}
\Phi_n: \left\{
\begin{array}{cl}
x_{n+1}&=\dfrac{ab(x_n-cq^n)(x_n-dq^n)}{y_n(x_n-a)(x_n-b)},\\
y_{n+1}&=x_n.
\end{array}
\right.
\label{qP3}
\end{equation}
\begin{Theorem}
Suppose that $a,b,c,d,q$ are parameters in $\{1,2,\cdots,p-1\}$ and that $a,b,c,d$ are distinct and we also suppose that $a+b\not \equiv (c+d)q^3$, then the mapping \eqref{qP3} has an almost good reduction   
modulo $\mathfrak{p}$ on the domain 
$\mathcal{D}:=\Z_p^2\cap \Phi_n^{-1}(\Q_p^2)=\{(x,y)\in \Z_p^2\ |x\neq a,b,\,y\neq 0 \}$.
\label{PropqP3}
\end{Theorem}
\textbf{Proof}\;\;
Let $(x_{n+1},y_{n+1})=\Phi_n(x_n,y_n)$.
In the case when $\tilde{x}_n\neq \tilde{a},\tilde{b}$ and $\tilde{y}_n\neq 0$, we have
\begin{equation}
\left\{
\begin{array}{cl}
\tilde{x}_{n+1}&=\dfrac{\tilde{a}\tilde{b}(\tilde{x}_n-\tilde{c}\tilde{q}^n)(\tilde{x}_n-\tilde{d}\tilde{q}^n)}{\tilde{y}_n(\tilde{x}_n-\tilde{a})(\tilde{x}_n-\tilde{b})},\\
\tilde{y}_{n+1}&=\tilde{x}_n.
\end{array}
\right.
\end{equation}
from the relation \eqref{prel}.
Hence $\widetilde{\Phi_n(x_n,y_n)}=\widetilde{\Phi_n}(\tilde{x}_n,\tilde{y}_n)$.
We have to examine the other cases. From here we sometimes abbreviate $\tilde{a}$ as $a$, $\tilde{b}$ as $b$ for simplicity.

(i) If $\tilde{x}_n=\tilde{a}$ and $(a-b)(a+b-cq-dq)\tilde{y}_n^t\not \equiv b(a-c)(a-d)$,
neither $\widetilde{\Phi_n}(\tilde{a},\tilde{y}_n)$ nor $\widetilde{\Phi_n^2}(\tilde{a},\tilde{y}_n)$ is well-defined. However,
$\widetilde{\Phi_n^3}(\tilde{a},\tilde{y}_n)$ is well-defined and we have,
\begin{eqnarray*}
&&\widetilde{\Phi_n^3(x_n,y_n)} = \widetilde{\Phi_n^3}(\tilde{x}_n=\tilde{a},\tilde{y}_n)\\
&=&\left(\frac{a(b-cq^2)(b-dq^2)\tilde{y}_n}{b(a-c)(a-d)-(a-b)(a+b-cq-dq)\tilde{y}_n},b\right).
\end{eqnarray*}

(ii) If $\tilde{x}_n=\tilde{a}$ and $(a-b)(a+b-cq-dq)\tilde{y}_n^t\equiv b(a-c)(a-d)$, none of $\widetilde{\Phi_n^i}(\tilde{a},\tilde{y}_n)$ is well-defined for $i=1,2,3,4$. However, $\widetilde{\Phi_n^5}(\tilde{a},\tilde{y}_n)$ is well-defined and we have,
\[
\widetilde{\Phi_n^5(x_n,y_n)} = \widetilde{\Phi_n^5}(\tilde{x}_n=\tilde{a},\tilde{y}_n)=\left(\frac{b(a-cq^4)(a-dq^4)}{(a-b)(a+b-cq^3-dq^3)},a\right).
\]

(iii) If $\tilde{x}_n=\tilde{b}$ and $(a-b)(a+b-cq-dq)\tilde{y}_n^t \not \equiv -a(b-c)(b-d)$,
\begin{eqnarray*}
&&\widetilde{\Phi_n^3(x_n,y_n)} = \widetilde{\Phi_n^3}(\tilde{x}_n=\tilde{b},\tilde{y}_n)\\
&=&\left(\frac{b(a-cq^2)(a-dq^2)\tilde{y}_n}{a(b-c)(b-d)+(a-b)(a+b-cq-dq)\tilde{y}_n},a\right).
\end{eqnarray*}

(iv) If $\tilde{x}_n=\tilde{b}$ and $(a-b)(a+b-cq-dq)\tilde{y}_n^t \equiv -a(b-c)(b-d)$, we have,
\[
\widetilde{\Phi_n^5(x_n,y_n)} = \widetilde{\Phi_n^5}(\tilde{x}_n=\tilde{b},\tilde{y}_n)=\left(-\frac{a(b-cq^4)(b-dq^4)}{(a-b)(a+b-cq^3-dq^3)},b\right).
\]

(v) If $\tilde{y}_n=0$ and $\tilde{x}_n\not = 0$,
\[
\widetilde{\Phi_n^3(x_n,y_n)} = \widetilde{\Phi_n^3}(\tilde{x}_n,\tilde{y}_n=0)=\left(0,\frac{ab}{\tilde{x}_n}\right).
\]

(vi) If $\tilde{y}_n=0$ and $\tilde{x}_n = 0$,
\[
\widetilde{\Phi_n^4(x_n,y_n)} = \widetilde{\Phi_n^4}(\tilde{x}_n=0,\tilde{y}_n=0)=\left(0,0\right).\qed
\]

\subsection{$q$-discrete Painlev\'{e} IV equation}
The $q$-discrete analog of the Painlev\'{e} IV equation has the following form:
\[
(x_{n+1}x_n-1)(x_nx_{n-1}-1)=\frac{aq^{2n}(x_n^2+1)+bq^{2n}x_n}{cx_n+dq^n},
\]
where $a,b,c,d$ and $q$ are parameters \cite{RGH, RG}.
It can be rewritten as  follows:
\begin{equation}
\Phi_n: \left\{
\begin{array}{cl}
x_{n+1}&=\dfrac{\tau^2(ax_n^2+bx_n+a)+(x_ny_n-1)(x_n+\tau)}{x_n(x_ny_n-1)(x_n+\tau)},\\
y_{n+1}&=x_n,
\end{array}
\right.
\label{qP4}
\end{equation}
where $\tau=q^n\tau_0$.
Here we took $\tau_0=\frac{d}{c}$ and redefined $a,b$ as $\frac{ac}{d^2}\to a$ and $\frac{bc}{d^2}\to b$.
\begin{Theorem}
Suppose that $|a|_p=|b|_p=|q|_p=|\tau_0|_p=1$, then the mapping \eqref{qP4} has an almost good reduction   
modulo $\mathfrak{p}$ on the domain 
$\mathcal{D}^{(n)}:=\Z_p^2 \cap \Phi_n^{-1}(\Q_p^2)=\{(x,y)\in \Z_p^2\ |x\neq 0,\ xy\neq 1\ x\neq -\tau \}$, on the condition that
$aq^2\tau_0\neq 1$ and $aq^4\tau_0\neq 1$.
\label{PropqP4}
\end{Theorem}
\textbf{Proof}\;\;
In the proof we use the abbreviation as $\tilde{a}\to a,\ \tilde{b}\to b,\tilde{\tau}_0\to\tau_0$.

(i) If $\tilde{x}_n=0$ and $1+q^3\tau_0^2+q^2(-1-b\tau_0^2+\tau_0 y_n+a\tau_0-a\tau_0^2 y_n)\not\equiv 0$,
\begin{eqnarray*}
&&\widetilde{\Phi_n^3(x_n,y_n)} = \widetilde{\Phi_n^3}(\tilde{x}_n=0,\tilde{y}_n)\\
&=&\left(\frac{-1-q^3\tau_0^2-bq^4\tau_0^2+aq^6\tau_0^3+q^2(1+b\tau_0^2-\tau_0 y_n+a\tau_0^2 y_n)}{q^2\tau_0\{1+q^3\tau_0^2+q^2(-1-b\tau_0^2+\tau_0 y_n+a\tau_0-a\tau_0^2 y_n)\}},-q^2\tau_0\right).
\end{eqnarray*}

(ii) If  $\tilde{x}_n=0$ and $1+q^3\tau_0^2+q^2(-1-b\tau_0^2+\tau_0 y_n+a\tau_0-a\tau_0^2 y_n)\equiv 0$,
\[
\widetilde{\Phi_n^5(x_n,y_n)} = \widetilde{\Phi_n^5}(\tilde{x}_n=0,\tilde{y}_n)=\left(\frac{-1+q^2+aq^4\tau_0+q^7\tau_0^2-bq^8\tau_0^2}{q^4\tau_0(-1+aq^4\tau_0)},0\right),
\]
where we assumed that $aq^4\tau_0\neq 1$.

(iii) If $\tilde{x}_n=-q^n\tau_0$ and $\tilde{y}_n\neq -\tau_0^{-1}$,
\begin{eqnarray*}
\widetilde{\Phi_n^3(x_n,y_n)}& =& \widetilde{\Phi_n^3}(\tilde{x}_n=-q^n\tau_0,\tilde{y}_n)\\
&=&\left(\frac{-1-\tau_0 y_n+(q^3-bq^4)\tau_0^2(1+\tau y_n)+C}{q^2\tau_0(-1+aq^2\tau_0)(1+\tau_0 y_n)},0\right),
\end{eqnarray*}
where
\[
C=q^2\{1+b\tau_0^2+\tau_0 y_n+a\tau_0^2(-\tau_0+y_n)\}.
\]
Here we assumed $aq^2\tau_0\neq 1$.

(iv) If $\tilde{x}_n=-q^n\tau_0$ and $\tilde{y}_n= -\tau_0^{-1}$,
\[
\widetilde{\Phi_n^5(x_n,y_n)} = \widetilde{\Phi_n^5}(\tilde{x}_n=-q^n\tau_0,\tilde{y}_n=-\tau_0^{-1})=\left(-\frac{1}{aq^6\tau_0^2},-aq^6\tau_0^2\right).
\]

(v) If $\tilde{x}_n\tilde{y}_n=1$,
\[
\widetilde{\Phi_n^5(x_n,y_n)} = \widetilde{\Phi_n^5}\left(\tilde{x}_n=\frac{1}{\tilde{y}_n},\tilde{y}_n\right)=\left(\frac{1}{aq^6\tau_0^3\tilde{y}_n},aq^6\tau_0^3\tilde{y}_n\right).\qed
\]

\subsection{$q$-discrete Painlev\'{e} V equation}
The $q$-discrete analog of the Painlev\'{e} V equation has the following form:
\[
(x_{n+1}x_n-1)(x_nx_{n-1}-1)=\frac{abq^n(x_n-c)(x_n-1/c)(x_n-d)(x_n-1/d)}{(x_n-aq^n)(x_n-bq^n)},
\]
where $a,b,c,d$ and $q$ are parameters \cite{RGH}.
It can be rewritten as the following form:
\begin{equation}
\Phi_n: \left\{
\begin{array}{cl}
x_{n+1}&=\dfrac{1}{x_n}\left(\dfrac{abq^n(x_n-c)(x_n-1/c)(x_n-d)(x_n-1/d)}{(x_n-aq^n)(x_n-bq^n)(x_ny_n-1)}+1\right),\\
y_{n+1}&=x_n.
\end{array}
\right.
\label{qP5}
\end{equation}
\begin{Theorem}
Suppose that $a,b,c,d,q$ are in $\{1,2,\cdots, p-1\}$ and $a,b,c,d,c^{-1},d^{-1}$ are distinct from each other, then the mapping \eqref{qP5} has almost good reduction   
modulo $\mathfrak{p}$ on the domain 
$\mathcal{D}^{(n)}:=\Z_p^2\cap \Phi_n^{-1}(\Q_p^2)=\{(x,y)\in \Z_p^2\ | x\neq aq^n,bq^n,\ xy\neq 1\}$.
\label{PropqP5}
\end{Theorem}
\textbf{Proof}\;\;
The calculation is extremely lengthy and we need about 13 gigabytes of memory. We deal with $n=0$ for simplicity. (Since ord$_p(q)=0$, the same argument applies to other cases.)

(i) If $\tilde{x}_n=a$,
\[
\widetilde{\Phi_n^3(x_n,y_n)} = \widetilde{\Phi_n^3}(\tilde{x}_n=a,\tilde{y}_n)=\left(\frac{1}{bq},bq\right).
\]

(ii) If $\tilde{x}_n=b$,
\[
\widetilde{\Phi_n^3(x_n,y_n)} = \widetilde{\Phi_n^3}(\tilde{x}_n=b,\tilde{y}_n)=\left(\frac{1}{aq},aq\right).
\]

(iii) If $\tilde{x}_n\tilde{y}_n=1$,
\[
\widetilde{\Phi_n^3(x_n,y_n)} = \widetilde{\Phi_n^3}\left(\tilde{x}_n,\tilde{y}_n=\frac{1}{\tilde{x}_n}\right)=\left(\frac{1}{abq\tilde{y}_n},abq\tilde{y}_n\right).\qed
\]
\section{Hietarinta-Viallet equation}
The Hietarinta-Viallet equation \cite{HV} is the following difference equation: 
\begin{equation}
x_{n+1}+x_{n-1}=x_n+\frac{a}{x_n^2}, \label{HV1}
\end{equation}
with $a$ as a parameter.
The equation \eqref{HV1} passes the singularity confinement test \cite{Grammaticosetal}, which is a notable test for integrability of equations, but yet is not integrable in the sense that its algebraic entropy is positive and that the orbits display chaotic behaviors.
We prove that the AGR is satisfied for this Hietarinta-Viallet equation.
We again rewrite \eqref{HV1} as the following coupled form:
\begin{equation}
\Phi_n: \left\{
\begin{array}{cl}
x_{n+1}&=x_n+\dfrac{a}{x_n^2}-y_n,\\
y_{n+1}&=x_n.
\end{array}
\right.
\label{HV}
\end{equation}
\begin{Theorem}
Suppose that $|a|_p=1$, then the mapping \eqref{HV} has an almost good reduction   
modulo $\mathfrak{p}$ on the domain
$\mathcal{D}:=\Z_p^2\cap \Phi_n^{-1}(\Q_p^2)=\{(x,y)\in \Z_p^2\ |x\neq 0\}$.
\label{PropHV}
\end{Theorem}
\textbf{Proof}\;\;
If $\tilde{x}_n=0$,
\[
\widetilde{\Phi_n^4(x_n,y_n)} = \widetilde{\Phi_n^4}(\tilde{x}_n=0,\tilde{y}_n)=(\tilde{y}_n,0).\qed
\]
\begin{Proposition}
Suppose that $|a|_p=1$, then the mapping \eqref{HV} has a refined almost good reduction.
Here the normal domain is $D_N=\mathbb{Z}_p^{\times}\times \Z_p$. Other domains are defined as
$D_S=p\Z_p \times \Z_p$, $E=\Q_p^2\setminus \Z_p^2$.
\end{Proposition}
\textbf{Proof}\;\;
First let us fix the initial condition $(x_0,x_{-1})\in D_N$, $(x_0\in\Z_p^{\times})$.

(i) If $\pi(a/x_0^2+x_0)\neq \tilde{x}_{-1}$ then, we have $\tilde{x}_1\neq 0$. Thus
\[
(x_1,x_0)\in D_N.
\]

(ii) If $\pi(a/x_0^2+x_0)= \tilde{x}_{-1}$ then, we have $x_1\in p\Z_p\ (\tilde{x}_1=0)$, therefore
\[
(x_1,x_0)\in D_S.
\]
By iterating further, we obtain the followings:
$\tilde{x}_2=\tilde{x}_3=\infty$, $\tilde{x}_4=0$, $\tilde{x}_5=\tilde{x}_0$.
Therefore,
\[
(x_2,x_1)\in E,\ (x_3,x_2)\in E,\ (x_4,x_3)\in E,\ (x_5,x_4)\in D_N.\qed
\]
Therefore we learn that the AGR and refined AGR work similarly to the singularity confinement test in distinguishing the integrable systems from the non-integrable ones.
In fact, the AGR and refined AGR can be seen as an arithmetic analog of the singularity confinement test.

\section{The $p$-adic singularity confinement}

The above approach is closely related to the singularity confinement method which is an effective test to judge the integrability of the given equations \cite{Grammaticosetal}.
In the proof of the theorem \ref{PropdP2}, we have taken
\[
x_n=1+e p^k\ \ (e\in\mathbb{Z}_p^{\times},\ k>0)
\]
instead of taking $x_n=1+\epsilon$ and showed that the limit
\[
\lim_{|e p^k|_p \to 0}(x_{n+m}, x_{n+m+1})
\] 
is well defined for some positive integer $m$.
Here $ep^k\ (k>0)$ is an alternative in $\Q_p$ for the infinitesimal parameter $\epsilon>0$ in the singularity confinement test in $\C$. Note that $p^k\ (k>0)$ is a `small' number in terms of the $p$-adic metric $(|p^k|_p=p^{-k})$. In fact, in most cases, we may just replace $\epsilon$ for $p$ in order to test the $p$-adic singularity confinement.
From this observation and previous theorems, we postulate that having almost good reduction in arithmetic mappings is similar to passing the singularity confinement test.

\section{Relation to the `Diophantine integrability'}
Lastly we discuss a relationship between the systems over  finite fields and the algebraic entropies of the systems.
Let $\phi$ be a difference equation and let the degree of the map $\phi$ be $d>0$. We define the degree of the iterates $\phi^n$ as $\deg (\phi^n)=d_n$.
The na\"{i}ve composition suggests $d_n=d^n$, however, common factors can be eliminated, lowering the degree of the iterates. Algebraic entropy $E$ of $\phi$ is the following well-defined quantity \cite{BV}.
\[
E:=\lim_{n\to\infty}\frac{1}{n}\log d_n\; (\ge 0).
\]
The existence of $E\ge 0$
We can postulate from a lot of examples that the mapping $\phi$ is
integrable if and only if $E=0$, that is, $d_n$ has a polynomial growth.
We can construct an arithmetic analog of the algebraic entropy which has first been introduced in \cite{Halburd}.
If we consider the map with rational numbers as coefficients, and choose initial values to be rational numbers, then we have $x_n\in \mathbb{Q}$ for all $n\in\mathbb{Z}_{>0}$. The arithmetic complexity of rational numbers can be expressed by the height function $H(x)$:
\[
H(x)=\max\{|u|,|v|\},
\]
where $x=\frac{u}{v}$ and $u$ and $v$ are integers without common factors. ($H(0)=0$.)
The map $\phi$ is said to be `Diophantine integrable' if and only if $\log H(x_n)$ grows as slowly as some polynomial.
Thus we define the arithmetic analog of algebraic entropy, which may be called as a `Diophantine entropy' as
\[
\epsilon:=\lim_{n\to\infty} \frac{1}{n}\log\left(\log H(x_n)\right).
\]
Precisely speaking, the value $\epsilon$ depends on the choice of initial data of the systems, however, we conjecture that the value $\epsilon$ is independent of that choice for most of the initial conditions. We conjecture that for most of the dynamical systems with rational numbers as coefficients, two values $E$ and $\epsilon$ is the same.
We have the following two conjectures from numerical observations:

\paragraph{(i)} The Hietarinta-Viallet equation \eqref{HV1} has $\epsilon=\log\left(\frac{3+\sqrt{5}}{2}\right)$, which is exactly equal to the original algebraic entropy $E=\log\left(\frac{3+\sqrt{5}}{2}\right)$ obtained in \cite{Takenawa,HV,BV}.

\paragraph{(ii)} In the case of the equation \eqref{discretemap}, $\epsilon=\log 3>0$ for $\gamma=3$, while, for $\gamma=1,2$, we have $\epsilon=0$ and $\log H(x_n)$ has a polynomial growth of second degree for generic initial conditions.

Therefore, in these cases, the Diophantine entropy $\epsilon$ motivated by the Diophantine integrability is expected to be equivalent to the (original) algebraic entropy $E$.
We do not explain the proof of these conjectures, some part of which is incomplete. We give some numerical examples which support (i) and (ii).

\paragraph{(i)} Let us suppose that $p=3$, $(x_0,x_1)=(1,3)$ and that the parameter $a=1$ in the Hietarinta-Viallet equation \eqref{HV1}. Then
\begin{eqnarray*}
\{\log_3 H(x_n)\}_{n=0}^\infty&=&\{0,1,2,7,19,50,132,347,911,2385,6245,\\
&& 16352,42811,112082,293434,768221,\cdots\}.
\end{eqnarray*}
Here we only displayed the integer part of the values.
We can see numerically that
\[
\left. \frac{\log_3 H(x_{n+1})}{\log_3 H(x_n)}\right|_{n\to\infty}\sim 2.61804 \sim \frac{3+\sqrt{5}}{2}.
\]

\paragraph{(ii)} In the case of the equation \eqref{discretemap}, we give two examples.
First let us suppose that $p=3$, $\gamma=3$, $(x_0,x_1)=(1,3)$, and that the parameter in the equation \eqref{discretemap} is $a=2$. Then,
\[
\{\log_3 H(x_n)\}_{n=0}^\infty=\{0,1,3,8,26,79,236,711,2133,6400,19201,\cdots\}.
\]
Therefore we see numerically that
\[
\frac{\log_3 H(x_{n+1})}{\log_3 H(x_n)} \sim 3,\ \ \epsilon\sim \log 3.
\]
On the other hand, if $\gamma=1$, we have
\[
\{\log_3 H(x_n)\}_{n=0}^\infty=\{0,1,1,1,2,3,3,4,5,7,7,8,9,12,13,14,16,18,20,\cdots\}.
\]
Therefore we see numerically that
\[
\frac{\log_3 H(x_{n+1})}{\log_3 H(x_n)} \sim 1,\ \ \epsilon\sim 0.
\]
The rate of growth of $\log_3 H(x_n)$ is quadratic:
if we estimate $x_0,\cdots,x_{100}$ using a cubic polynomial, we obtain
\[
\log_3 H(x_n)\sim 0.120+0.185n+0.0454n^2+5\times 10^{-7} n^3,
\]
which indicates a quadratic growth.

In the case of original algebraic entropy $E$, we can rigorously obtain the recurrence relation for the sequence $d_n$ of degrees of rational functions with several methods. However, in the case of `Diophantine entropy', it is not easy in many cases to exactly estimate the elimination of common factor $p$ between the numerator and the denominator.
This idea is essentially equivalent to studying the growth of the number of digits of the numerator (or denominator) of $x_n\in\mathbb{Q}$ when expressed as $p$-adic expansions. Therefore the procedure can be seen as an analog of algebraic entropy of a system over a finite field $\mathbb{F}_p$.
As a technique of the numerical simulations, instead of the height $H(x_n)$, we can also use only the denominator $r_n$ or the numerator $s_n$ of $x_n=s_n/r_n$: i.e.,
both of the values
\[
\lim_{n\to \infty}\frac{1}{n}\log\left(\log | s_n|\right),\; \lim_{n\to \infty}\frac{1}{n}\log\left(\log | r_n |\right)
\]
should give the same value as $\epsilon$ due to a result by Silverman \cite{Silvermanduke}.
The biggest benefit of this `Diophantine' approach might be that the time of computation is greatly reduced by using rational numbers instead of the using formal variables. This allows us to obtain a conjecture for the integrability, and a conjecture for the value of algebraic entropy with comparably short time.
In 2014, after the thesis is submitted, a series of generalized versions of the Hietarinta-Viallet equation
\[
x_{n+1}=-x_{n-1}+x_n+\frac{1}{x_n^k},
\]
where $k\ge 2$ is under investigation by the author and his collaborators.
They numerically computed an approximation to the Diophantine entropy $\epsilon$ of this system for $k=2,3,4,5,\cdots$ and conjectured the exact values of algebraic entropy $E$ from these approximations.
We have found that the situation depends on the parity of the integer $k$.
This topic will be dealt with in other papers.

\section{Systems over the extended fields}
In the preceding subsections we have successfully defined the dynamical systems over the finite field $\mathbb{F}_p$ through the extensions to / reductions from the field of $p$-adic numbers $\mathbb{Q}_p$.
In this subsection we generalize this result to the systems over a larger finite field $\mathbb{F}_{p^m}$ where $m>1$, and then study the ways of reduction to some finite field from the field of complex values $\mathbb{C}$.
Since a field extension $L$ of the degree $m$ over $\mathbb{Q}_p$ is a simple extension, there exist an element $\alpha\in L$ such that $L=\mathbb{Q}_p(\alpha)$. The reduction from $L$ to the set $\mathbb{F}_p(\alpha)\cup \{\infty\}$ is defined naturally using the reduction map \eqref{padicreductionmap} in the previous sections:
\begin{eqnarray*}
L=\mathbb{Q}_p(\alpha)&\ni& x_0+x_1\alpha+x_2\alpha^2+\cdots+x_{m-1}\alpha^{m-1}\\
&\mapsto & \tilde{x}_0+\tilde{x}_1\alpha+\tilde{x}_2\alpha^2+\cdots+\tilde{x}_{m-1}\alpha^{m-1} \in \mathbb{F}_p(\alpha)\cup\{\infty\}.
\end{eqnarray*}
For example let us define dynamical systems over $\mathbb{F}_{p^2}$ and discuss the properties of the reductions.
Let $\alpha$ be a generator of $\mathbb{F}_{p^2}$ over $\mathbb{F}_p$.
Then we have $\mathbb{F}_{p^2}=\F_p(\alpha)$ and $\alpha\in\F_{p^2}\setminus \F_p$ and $a:=\alpha^2\in\mathbb{F}_p$.
\begin{Lemma}\label{lem411}
The field $\Q_p(\alpha)$ is the extension field of $\Q_p$ of degree two.
\end{Lemma}
\textbf{Proof}\;\;
We can see $a$ as an element of $\Z_p^{\times}$. Since $a\!\!\mod p\Z_p\in \F_p^{\times}$ is not a square element in $\F_p^{\times}$, $a$ is not a square in $\Q_p^{\times}$ either. Therefore $\alpha$ in not in $\Q_p$.\qed
We define the reduction map $\pi_2$ from $\Q_p(\alpha)$ to $\F_{p^2}$ as follows:
\begin{equation}
\pi_2:\;\;\Q_p(\alpha)\ni x+y\alpha\; (x,y\in\mathbb{Q}_p)\longmapsto \left\{
\begin{array}{cl}
\tilde{x}+\tilde{y}\alpha \in\F_p(\alpha)& (x,y\in\Z_p)\\
\infty & (\mbox{otherwise})
\end{array}
\right.
\end{equation} 
Note that $\pi_2|_{\Z_p[\alpha]}$ is a ring homomorphism.
We define the almost good reduction in a similar manner to the case of systems over $\F_p$.
\begin{Definition}
A non-autonomous rational system $\phi_n$: $(\Q_p(\alpha))^2 \to (\Q_p(\alpha))^2$ $(n \in \Z)$  has an almost good reduction modulo $\pi_2$ on the domain $\mathcal{D} \subseteq (\Z_p[\alpha])^2$, if there
exists a positive integer $m_{S;n}$ for any $S=(x,y) \in \mathcal{D}$ and time step $n$ such that
\begin{equation}
\pi_2({\phi_n^{m_{S;n}}(x,y)})=\widetilde{\phi_n^{m_{S;n}}}(\pi_2({x}),\pi_2({y})),
\label{AGRsquare}
\end{equation}
where $\phi_n^m :=\phi_{n+m-1} \circ \phi_{n+m-2} \circ \cdots \circ \phi_n$.
\end{Definition}
Next we apply these results to the field $\mathbb{C}$.
Note that we already have a method to obtain cellular automata from the discrete systems via extended ultra-discretization \cite{TYajima}.
We take a different approach, which is based on the arithmetic of $p$-adic numbers. We use without proof the following fact in the number theory.
\begin{Lemma}
The field $\mathbb{Q}_p$ has a square root of $-1$ if and only if $p\equiv 1\!\mod 4$.
\end{Lemma}
From this fact we consider the following two cases:
\begin{itemize}
\item If $p=2$ or $p\equiv 3\!\mod 4$, then the lemma \ref{lem411} holds for $\alpha=\sqrt{-1}$.
Thus we obtain the following reduction mapping $\pi_{\mathbb{C}}$:
\begin{equation}
\pi_{\mathbb{C}}:\;\;\Q_p(\sqrt{-1})\ni x+\sqrt{-1}y\;\longmapsto \left\{
\begin{array}{cl}
\tilde{x}+\sqrt{-1}\tilde{y} \in\F_{p^2}& (x,y\in\Z_p)\\
\infty & (\mbox{otherwise})
\end{array}
\right.
\end{equation}
\item If $p\equiv 1\mod 4$, on the other hand, $\mathbb{Q}_p(\sqrt{-1})=\mathbb{Q}_p$ holds.
Therefore, the reduction mapping $\pi_{\mathbb{C}}$ takes values in $\P\F_p=\mathbb{F}_p\cup\{\infty\}$.
\end{itemize}
The values of the form $x+\sqrt{-1}y\in\mathbb{C}$, $(x,y\in\mathbb{Q})$ can be reduced to either $\mathbb{F}_p\cup\{\infty\}$ or $\mathbb{F}_{p^2}\cup\{\infty\}$.
Note that we cannot apply this method if $x,y$ are not rational numbers.
By using this approach to the equations with complex variables such as a discrete version of the nonlinear Schr\"{o}dinger equation (dNLS) and a discrete sine-Gordon equation, we expect to obtain the cellular automata related to the equations. One of the future problems is to investigate the cellular automata (ultra-discrete) analogs of the breather solutions of dNLS.


\chapter{Two-dimensional systems over finite fields} \label{sec5}
In chapter \ref{sec3}, we have successfully determined the time evolution of the discrete Painlev\'{e} equations through the construction of their space of initial conditions by blowing-up twice at each of the singular points so that the mapping becomes bijective.
However, for a general nonlinear equation, explicit construction of the space of initial conditions over a finite field is not so straightforward (for example see \cite{Takenawa} or consider the higher dimensional lattice systems). Therefore it does not help us to obtain the explicit solutions.
In this section we study the soliton equations evolving as a two-dimensional lattice over finite fields by following the discussions made in \cite{KMT}.

\section{Discrete KdV equation over the field of rational functions}\label{section51}

Let us consider the discrete KdV equation \eqref{dKdV1}
over a finite field $\F_q$ where $q=p^m$, $p$ is a prime number and $m\in\Z_{+}$.
Let us reproduce the discrete KdV equation here:
\begin{equation*}
\frac{1}{x_{n+1}^{t+1}}-\frac{1}{x_n^t}+\frac{\delta}{1+\delta}\left(x_n^{t+1}-x_{n+1}^t \right)=0.
\end{equation*}
Here $n,t \in \Z$ and $\delta$ is a parameter. If we take
\[
\frac{1}{y_n^t}:=(1+\delta)\frac{1}{x_n^{t+1}}-\delta x_n^t
\]
we obtain equivalent coupled equations
\begin{equation}
\left\{
\begin{array}{cl}
x_n^{t+1}&=\dfrac{(1+\delta)y_n^t}{1+\delta x_n^ty_n^t},\vspace{2mm} \\
y_{n+1}^{t}&=\dfrac{(1+\delta x_n^ty_n^t)x_n^t}{1+\delta}.
\end{array}
\right.
\label{dKdV2}
\end{equation}
Clearly \eqref{dKdV2} does not determine the time evolution when $1+\delta x_n^t y_n^t\equiv 0$.
Over a field of characteristic 0 such as $\C$, the time evolution of $(x_n^t,y_n^t)$ will not hit this exceptional line 
for generic initial conditions, but on the contrary, the evolution comes to this exceptional line in many cases over a finite field as a division by $0$ appears.
The mapping, $(x_n^t,y_n^t) \mapsto (x_n^{t+1}, y_{n+1}^t)$, is lifted to an automorphism of the surface $\tilde{X}$,
where $\tilde{X}$ is obtained from $\P^1 \times \P^1$ by blowing up twice at $(0,\infty)$ and $(\infty, 0)$ respectively:
\begin{align*}
\tilde{X}&=\A_{(0,\infty)} \cup \A_{(\infty,0)},\\
\A_{(0,\infty)}&:=\left\{ \left(\left(x, y^{-1}\right), [\xi:\eta], [u:v]\right)  \Big| \ x \eta=y^{-1} \xi,\  \eta u = y^{-1} (\eta+\delta \xi) v \right\}\\
&\subset \A^2 \times \P^1\times\P^1, \\
\A_{(\infty,0)}&:=\left\{ \left(\left(x^{-1}, y\right), [\xi:\eta], [u:v]\right)  \Big|\ x^{-1}\eta=y\xi,\ (\eta + \delta \xi) u = y \eta v\right\}\\
&\subset \A^2 \times \P^1\times\P^1,
\end{align*}
where $[a:b]$ denotes a set of homogeneous coordinates for $\P^1$.
To define the time evolution of the system with $N$ lattice points from \eqref{dKdV2}, however, we have to consider the mapping 
\[
(y_1^t;x_1^t,x_2^t,...,x_N^t) \longmapsto  (x_1^{t+1},x_2^{t+1},...,x_N^{t+1};y_{N+1}^t).
\]
Since there seems no reasonable decomposition of $\tilde{X}$ into a direct product of two independent spaces, successive use of \eqref{dKdV2} becomes impossible.
Note that if we blow down $\tilde{X}$ to $\P^1 \times \P^1$, the information of the initial values is lost in general.
If we intend to construct an automorphism of a space of initial conditions, it will be inevitable to start from $\P^{N+1}$ and blow-up to some huge manifold, which is beyond the scope of the present paper.
There should be so many exceptional hyperplanes in the space of initial conditions if it does exist, and it is practically impossible to check all the ``singular'' patterns in the na\"{i}ve extension of the singularity confinement test. Another difficulty is that, in high dimensional lattice systems, we cannot properly impose the boundary conditions to be compatible with the extension of the spaces.
These difficulties seem to be some of the reasons why the singularity confinement method has not been used for construction of integrable partial difference equations or judgment for their integrability, though some attempts have been proposed in the bilinear form \cite{RGS}.
On the other hand, when we fix the initial condition for a partial difference equation, the number of singular patterns is restricted in general and we have only to enlarge the domain so that the mapping becomes well defined.
This is the strategy that we will adopt in this section.

Suppose that $x_1^0=6,\ x_2^0=5,\ y_1^0=2,\ y_1^1=2$, then 
we have 
\[
x_1^{1}=4/13 \equiv 3,\quad y_2^{0}=78/2 \equiv 4 \mod 7.
\]
With further calculation we have
\[
x_1^2=4/7 \equiv 4/0,\quad y_2^1=21/2 \equiv 0,\quad x_2^1=8/21 \equiv 1/0.
\]
Since $4/0$ and $1/0$ are not defined over $\F_7$, we now extend $\F_7$ to $\P\F_7$ and 
take $\dfrac{j}{0}\equiv \infty$ for $j\in\{1,2,3,4,5,6\}$.
However, at the next time step, we have
\[
x_2^2=\frac{2 \cdot 0}{1+ \infty \cdot 0},\qquad y_3^1=\frac{(1+ \infty \cdot 0)\cdot \infty}{2}
\]
and reach a deadlock.

The first idea to overcome this problem is to consider the equation over the field of rational functions \cite{KMT}.
We try the following two procedures:

(I) we keep $\delta$ as a parameter for the same initial condition, and obtain as a system over $\F_7(\delta)$,
\begin{align*}
x_1^{1}&=\frac{2(1+\delta)}{1+5\delta},\quad y_2^{0}=\frac{6(1+5\delta)}{1+\delta},\\
x_2^1&=\frac{6(1+\delta)(1+5\delta)}{1+3\delta+3\delta^2},
\quad y_2^1=\frac{2(1+2\delta+4\delta^2)}{(1+5\delta)^2},\quad
x_1^2=\frac{2(1+\delta)(1+5\delta)}{1+2\delta+4\delta^2},\\
x_2^2&=\frac{4(1+\delta)(2+\delta)(3+2\delta)}{(1+5\delta)(5+5\delta+2\delta^2)},\quad y_3^1=\frac{2(5+5\delta+2\delta^2)}{(2+\delta)^2}.
\end{align*}

\begin{figure}
\centering
\includegraphics[width=10cm,bb=80 250 500 750]{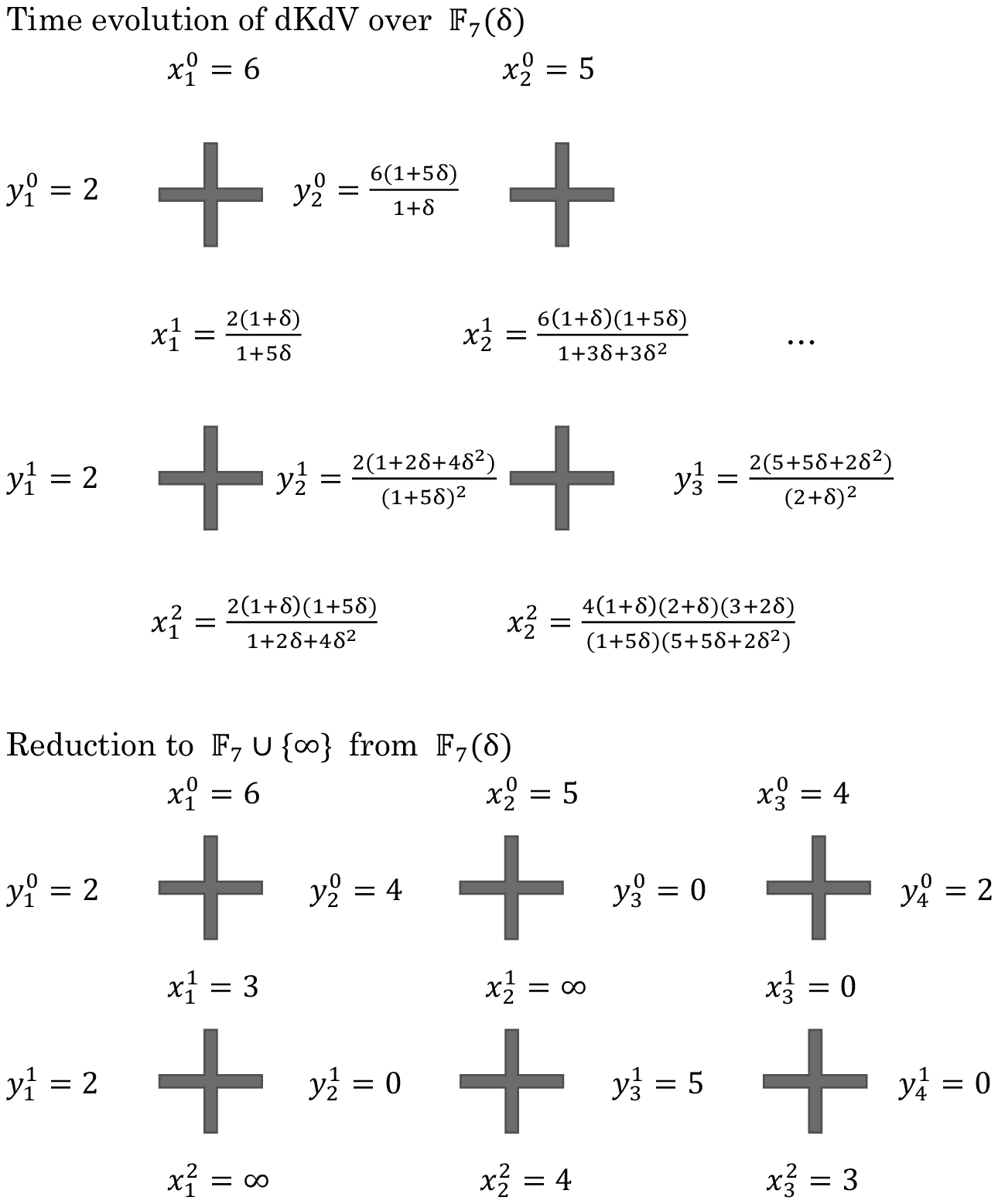}
\caption{An example of the time evolution of the coupled discrete KdV equation \eqref{dKdV2} over $\P\F_7$ where $\delta=1$.}
\label{figure1}
\end{figure}

\begin{figure}
\centering
\includegraphics[width=12cm,bb=80 540 500 720]{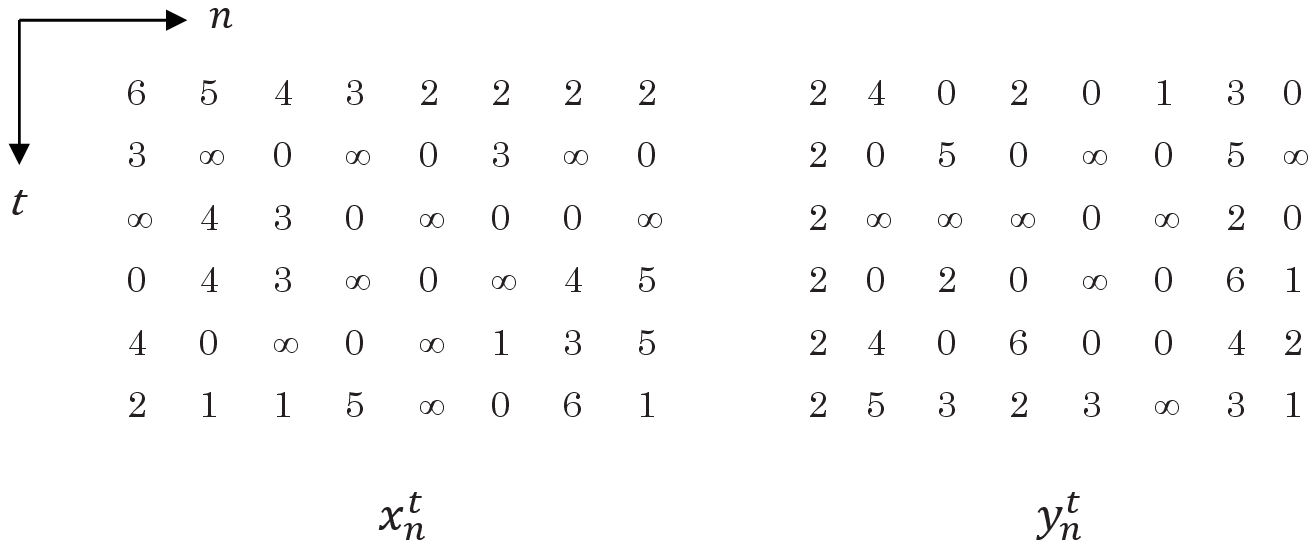}
\caption{The time evolution pattern of $x_n^t$ (left) and $y_n^t$ (right) of the discrete KdV equation \eqref{dKdV2} over $\P\F_7$ where $\delta=1$.}
\label{figure2}
\end{figure}

(II) Then we put $\delta=1$ to have a system over $\P\F_7$ as
\begin{align*}
x_1^{1}&=3,\quad y_2^{0}=4,\quad x_2^1=72/7\equiv \infty,
\quad y_2^1=14/36\equiv 0,\quad x_1^2=24/7 \equiv \infty,\\
x_2^2&=120/72 \equiv 4,\quad y_3^1=24/9 \equiv 5.
\end{align*}
Thus all the values are uniquely determined over $\P\F_7$.
Figures \ref{figure1} and \ref{figure2} show a time evolution pattern of the discrete KdV equation \eqref{dKdV2} over $\P\F_7$ for the initial conditions $x_1^0=6,\ x_2^0=5,\ x_3^0=4,\ x_4^0=3,\  x_j^0=2\ (j\ge 5)$ and $y_1^t=2\ (t\ge 0)$.

This example suggests that the equation \eqref{dKdV2} should be understood as evolving over the field $\F_q(\delta)$, the rational function field with indeterminate $\delta$ over $\F_q$. 
To obtain the time evolution pattern over $\P\F_q$, we have to substitute $\delta$ with a suitable value $\delta_0 \in\F_q$ ($\delta_0=1$ in the example above).
This substitution can be expressed as the following reduction map:
\begin{equation}\label{reductionmapping}
\F_q(\delta)^{\times}\rightarrow \P\F_q:\ (\delta-\delta_0)^s\frac{g(\delta-\delta_0)}{f(\delta-\delta_0)}\mapsto \left\{
\begin{array} {cl}
0  & (s>0),\vspace{1mm}\\
\infty & (s<0),\vspace{1mm}\\
g(0)/f(0) & (s=0),
\end{array}
\right.
\end{equation}
where $s\in \Z$, $f(h),\ g(h)\in\F_q[h]$ are co-prime polynomials and $f(0)\neq 0, g(0)\neq 0$.
With this prescription, we know that $0/0$ does not appear and we can uniquely determine the time evolution for generic initial conditions defined over $\F_q$.
Of course we can also overcome the indeterminacy by using the filed of $p$-adic numbers as we have done in previous sections. This approach is introduced in section \ref{padicdkdv}.  
\section{Soliton solutions of the (generalized) discrete KdV equations over the field of rational functions}\label{sectiondkdvsoliton}
%
%
First we consider the $N$-soliton solutions to \eqref{dKdV1} over $\F_q$.
It is well-known that the $N$-soliton solution is given as
\begin{equation}\label{dkdvsoliton}
\begin{array}{cl}
x_n^t&=\dfrac{\sigma_n^t\sigma_{n+1}^{t-1}}{\sigma_{n+1}^t\sigma_n^{t-1}},\\
\sigma_n^t&:=\det_{1\le i,j\le N}\left( \delta_{ij}+\dfrac{\gamma_i}{l_i+l_j-1}\left(\dfrac{1-l_i}{l_i}\right)^t
\left(\dfrac{l_i+\delta}{1+\delta-l_i}\right)^n\right)
\end{array}
\end{equation}
where $\gamma_i,\ l_i$ $(i=1,2,...,N)$ are arbitrary parameters but $l_i \ne l_j$ for $i \ne j$.
When $l_i,\ \gamma_i$ are chosen in $\F_q$, $x_n^t$ becomes a rational function in $\F_q(\delta)$. 
Hence we obtain soliton solutions over $\P\F_q$ by substituting $\delta$ with a value in $\F_q$.

\begin{figure}
\centering
\includegraphics[width=12cm, bb=60 350 520 730]{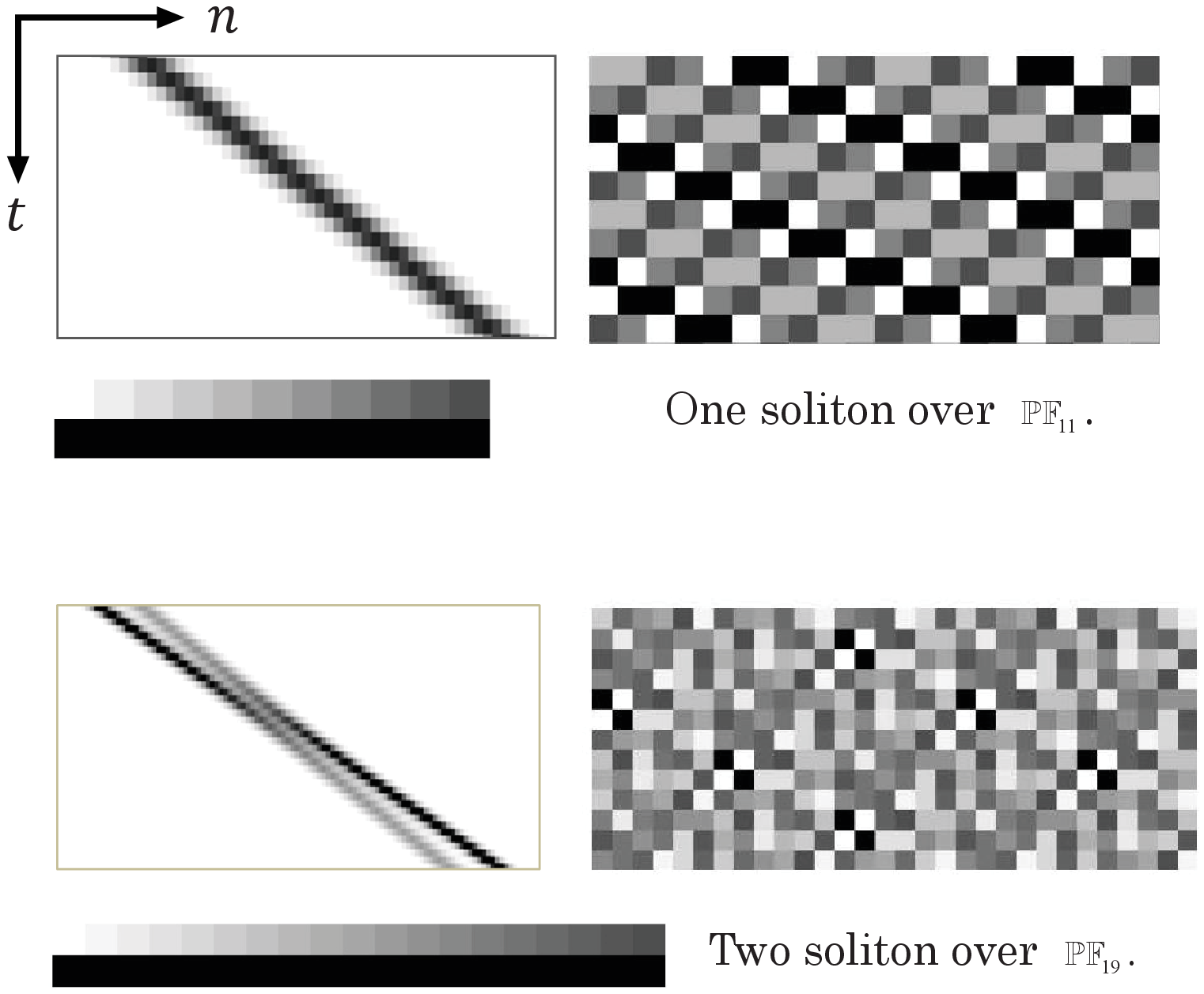}
\caption{The soliton solutions of the discrete KdV equation \eqref{dKdV1} over $\R$ (left) and finite fields (right).
The top one is the one-soliton over $\P\F_{11}$ where $\delta=7,\ \gamma_1=2,\ l_1=9$.
The bottom one is the two-soliton solution over $\P\F_{19}$ where $\delta=8,\ \gamma_1=15,\ l_1=2,\ \gamma_2=9,\ l_2=4$. 
Elements of the finite fields $\P\F_{p}$ are represented on the following grayscale: from $0$ (white) to $p-1$ (gray) and $\infty$ (black).}
\label{figure4}
\end{figure}

The figure \ref{figure4} shows one and two soliton solutions for the discrete KdV equation \eqref{dKdV1} over the finite fields $\P\F_{11}$ and $\P\F_{19}$. Here we have chosen the values $(1-l_i)/l_i$ and $(l_i+\delta)/(1+\delta-l_i)$ so that their reduction by the reduction map  \eqref{reductionmapping} is neither $0$ nor $\infty$. In this case, the reduced soliton solutions exhibit periodicity with periods $q-1$ as in figure \ref{figure4}.
The corresponding time evolutionary patterns over the field $\R$ are also presented for comparison.
Note that, if some of the reduced values of $(1-l_i)/l_i$ or $(l_i+\delta)/(1+\delta-l_i)$ take $0$ or $\infty$, the reduced soliton solutions do not exhibit periodicity in general: they might become stationary, vanish after a few time steps, or look like the normal solitary waves.
These phenomena are described in detail in section \ref{padicdkdv}.

Next we consider the generalized form of the discrete KdV equation.
We introduce the following discrete integrable system:
\begin{equation}
\left\{
\begin{array}{cl}
x_n^{t+1}&=\dfrac{\left\{(1-\beta)+\beta x_n^ty_n^t\right\}y_n^t}{(1-\alpha)+\alpha x_n^ty_n^t}\vspace{2mm},\\
y_{n+1}^{t}&=\dfrac{\left\{(1-\alpha)+\alpha x_n^ty_n^t\right\}x_n^t}{(1-\beta)+\beta x_n^ty_n^t},
\end{array}
\right.
\label{YBdKdV}
\end{equation}
with arbitrary parameters $\alpha$ and $\beta$.
This is a natural and important generalization of the discrete KdV equation, partly because it becomes the generalized version of the BBS called `Box Ball Systems with a Carrier' (BBSC) through ultra-discretization. The parameter $\beta$ corresponds to the capacity of the box, and $\alpha$ to the capacity of the carrier. The equation \eqref{YBdKdV} are known to have soliton solutions whose speeds and widths are intuitively understood from the BBSC \cite{KMTJPSJ}.
We consider soliton solutions to the generalized discrete KdV equation \eqref{YBdKdV}.
Note that 
by putting $u_n^t:=\alpha x_n^t,\ v_n^t:=\beta y_n^t$, we obtain
\[
\left\{
\begin{array}{cl}
u_n^{t+1}&=\dfrac{(\alpha(1-\beta)+u_n^tv_n^t)v_n^t}{\beta(1-\alpha)+u_n^t v_n^t},\vspace{2mm} \\
v_{n+1}^{t}&=\dfrac{(\beta(1-\alpha)+u_n^tv_n^t)u_n^t}{\alpha(1-\beta)+ u_n^t v_n^t}.
\end{array}
\right.
\]
Hence \eqref{YBdKdV} is
essentially equivalent to the `consistency of the discrete potential KdV equation around a $3$-cube' \cite{Tongasetal}: $(u,v) \to (u',v')$, as
\begin{eqnarray*}
u'=vP,\ v'=uP^{-1},\ P=\dfrac{a+uv}{b+uv}.
\label{YBMAP}
\end{eqnarray*}
The map is also obtained from discrete BKP equation \cite{KakeiNimmoWillox}. 
We will obtain $N$-soliton solutions to \eqref{YBdKdV} from the $N$-soliton solutions to the discrete KP equation
by a reduction similar to the one adopted in \cite{KakeiNimmoWillox}.

Let us consider the four-component discrete KP equation:
\begin{align}
&(a_1-b)\tau_{l_1t}\tau_n+(b-c)\tau_{l_1}\tau_{tn}+(c-a_1)\tau_{l_1n}\tau_t=0,
\label{eq1}\\
&(a_2-b)\tau_{l_2t}\tau_n+(b-c)\tau_{l_2}\tau_{tn}+(c-a_2)\tau_{l_2n}\tau_t=0.
\label{eq2}
\end{align}
Here $\tau=\tau(l_1,l_2,t,n)$ is the $\tau$-function of integer variables $(l_1,l_2,t,n) \in \Z^4$, and $a_1$, $a_2$, $b$ and $c$ are arbitrary parameters.
We express the shift operations by subscripts:
$\tau \equiv \tau(l_1,l_2,t,n),\  \tau_{l_1} \equiv \tau(l_1+1,l_2,t,n),\ \tau_{l_1t} \equiv \tau(l_1+1,l_2,t+1,n) \ $
and so on.
If we shift $l_1 \to l_1+1$ in \eqref{eq2}, we have
\begin{equation}
(a_2-b)\tau_{l_1l_2t}\tau_{l_1n}+(b-c)\tau_{l_1l_2}\tau_{l_1tn}+(c-a_2)\tau_{l_1l_2n}\tau_{l_1t}=0.\label{kptaushift}
\end{equation}
Then, by imposing the reduction condition:
\begin{equation}
\tau_{l_1l_2}=\tau,
\label{reduction}
\end{equation}
the equation \eqref{kptaushift} turns to
\[
(a_2-b)\tau_{t}\tau_{l_1n}+(b-c)\tau\tau_{l_1tn}+(c-a_2)\tau_{n}\tau_{l_1t}=0.
\]
Hence, putting $f:=\tau,\ g:=\tau_{l_1}$, we obtain
\begin{align*}
&(a_1-b)g_{t}f_n+(b-c)gf_{tn}+(c-a_1)g_{n}f_t=0,
 \\
&(a_2-b)f_{t}g_{n}+(b-c)fg_{tn}+(c-a_2)f_{n}g_{t}=0,
\end{align*}
and
\begin{equation*}
\frac{fg_{tn}}{g f_{tn}}=\frac{(a_2-b)f_{t}g_{n}+(c-a_2) f_{n}g_{t}}{(a_1-b)g_{t}f_n+(c-a_1) g_{n}f_t}=\frac{(c-a_2)+(a_2-b)\frac{f_{t}g_{n}}{f_{n}g_{t}}}{(a_1-b)+(c-a_1) \frac{g_{n}f_t}{g_{t}f_n}}.
\end{equation*}
Now we denote 
\begin{equation}
x_n^t:=\frac{f g_n}{g f_n},\qquad y_n^t:=\frac{g f_t}{f g_t}.
\label{xy}
\end{equation}
From the equality
\[
x_n^{t+1}y_{n+1}^t=x_n^ty_n^t=\frac{f_tg_n}{f_ng_t}, \quad
\frac{x_n^{t+1}}{y_n^t}=\frac{f g_{tn}}{g f_{tn}},
\]
we find that $x_n^t,\ y_n^t$ defined in \eqref{xy} satisfy the equation \eqref{YBdKdV} by defining $\alpha:=(c-a_1)/(c-b),\ \beta:=(a_2-b)(c-b)$.

The $N$-soliton solution to \eqref{eq1} and \eqref{eq2} is known as
\begin{equation*}
\tau=\det_{1\le i,j \le N}\left[ \delta_{ij}+\frac{\gamma_i}{p_i-q_j}\left(\frac{q_i-a_1}{p_i-a_1 }\right)^{l_1}
\left(\frac{q_i-a_2}{p_i-a_2 }\right)^{l_2}\left(\frac{q_i-b}{p_i-b }\right)^{t}\left(\frac{q_i-c}{p_i-c }\right)^{n} \right],
\end{equation*}
where $\{p_i,q_i\}_{i=1}^N$ are distinct parameters from each other and $\{ \gamma_i \}_{i=1}^N$ are
arbitrary parameters \cite{DJKM}.
The reduction condition \eqref{reduction} gives the constraint, 
\[
\left(\frac{a_1-p_i}{a_1-q_i}\right)\left(\frac{a_2-p_i}{a_2-q_i}\right)=1,
\]
to the parameters $\{p_i,\, q_i\}$.
Since $p_i \ne q_i$, the restriction is equivalent to $p_i+q_i=a_1+a_2$.
By rewriting $\dfrac{p_i-a_1}{c-b}  \rightarrow  p_i$,
 $\dfrac{\gamma_i}{c-b} \rightarrow  \gamma_i  $, defining $\Delta := \dfrac{a_1-a_2}{c-b}$ and taking $l_1=l_2$ we have
\begin{align}
f&=\det_{1\le i,j \le N}\left[ \delta_{ij}+\frac{\gamma_i}{p_i+p_j+\Delta }\left(\frac{-p_i+\beta}{p_i+1-\alpha }\right)^{t} 
\left(\frac{p_i+1-\beta}{-p_i+\alpha}\right)^{n} 
\right], \label{fform}\\
g&=\det_{1\le i,j \le N}\left[ \delta_{ij}+\frac{\gamma_i}{p_i+p_j+\Delta }\frac{-\Delta -p_i}{p_i}\left(\frac{-p_i+\beta}{p_i+1-\alpha }\right)^{t} 
\left(\frac{p_i+1-\beta}{-p_i+\alpha}\right)^{n}
\right]. \label{gform}
\end{align}
Thus we obtain the $N$-soliton solution of \eqref{YBdKdV} by \eqref{xy}, \eqref{fform} and \eqref{gform}.

\begin{figure}
\centering
\includegraphics[width=12cm, bb=50 430 550 740]{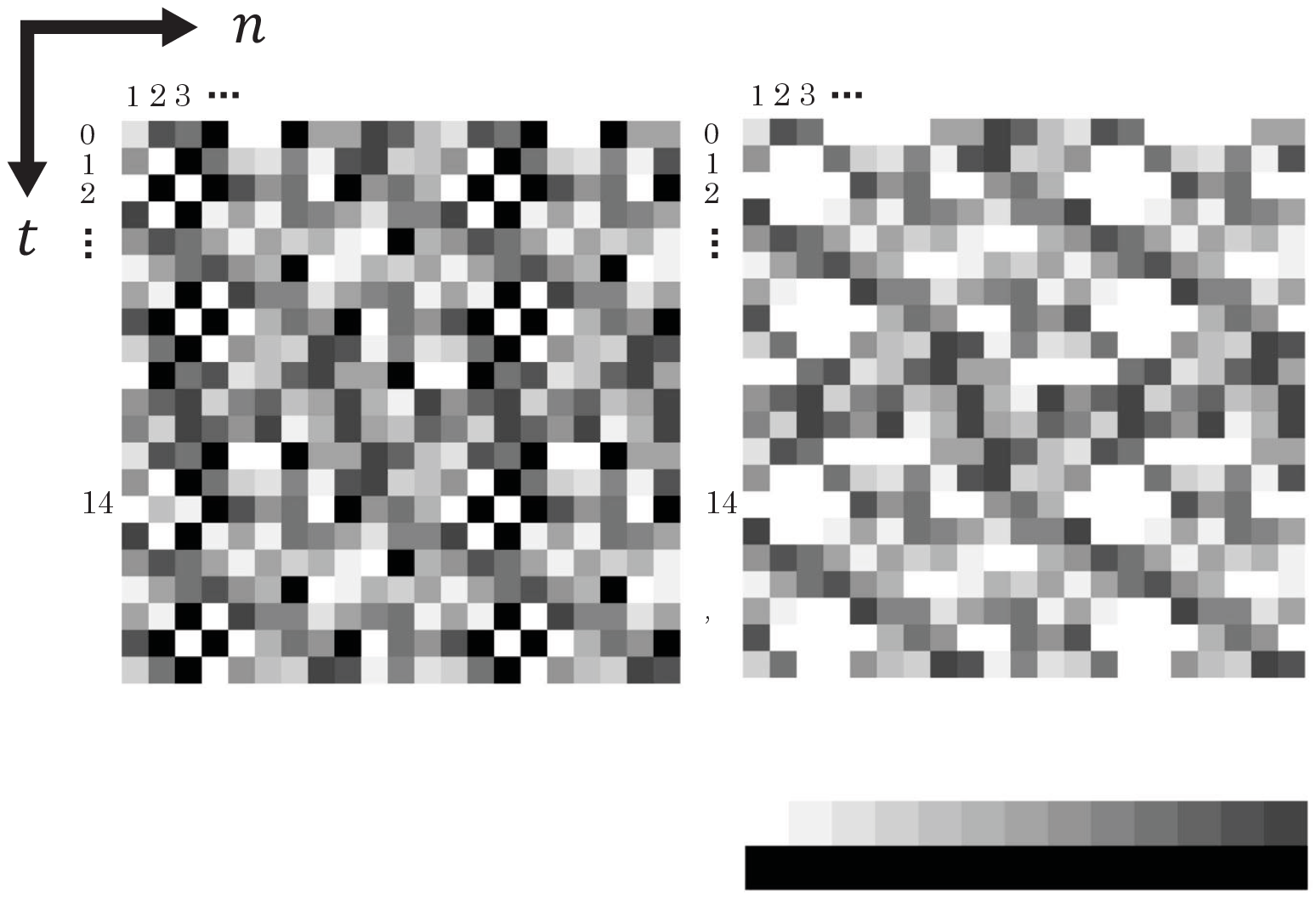}
\caption{The two-soliton solution of the generalized discrete KdV equation \eqref{YBdKdV} over $\P\F_{13}$ calculated in two different ways (left: $\tilde{x}_n^t$, right: $\hat{x}_n^t$), where $\alpha=14/15,\,\beta=5/6,\, r_1=-1/6,\,l_1=2/15,\, r_2=-1/30,\, l_2=1/30$. Elements of $\P\F_{13}$ are represented on the grayscale: from $0$ (white) to $12$ (gray) and $\infty$ (black).}
\label{figure8}
\end{figure}

Although the generalized discrete KdV equation has more than one parameters $\alpha$ and $\beta$, we can do the same approach of using the field of rational function as in the case of \eqref{dKdV2}.
If we want to consider the equation at $\alpha=a, \beta=b$ $(a, b \in \F_q)$, then we substitute $\alpha=a+\epsilon,\ \beta=b+\epsilon$  using a new parameter $\epsilon$, which will be considered as a variable. Then we can construct soliton solutions in $\F_q(\epsilon)$ by a reduction for suitable values of
$\{p_i,\ \gamma_i\}$ and $\Delta$. The reduced solutions defined in $\P\F_q$ are obtained by putting $\epsilon=0$ and are expressed as $\tilde{f},\ \tilde{g},\ \tilde{x}_n^t$ and $\tilde{y}_n^t$.
Lastly, let us comment on the periodicity of the soliton solutions over $\P\F_q$.
We have
\begin{eqnarray*}
\tilde{f}(n+q-1,\,t)&=&\tilde{f}(n,\,t+q-1)=\tilde{f}(n,\,t),\\
\tilde{g}(n+q-1,\,t)&=&\tilde{g}(n,\,t+q-1)=\tilde{g}(n,\,t),
\end{eqnarray*}  
for all $t,\,n\in\Z$ since we have $a^{q-1}\equiv 1$ for all $a \in \F^{\times}_q$.
Thus the functions $\tilde{f}$ and $\tilde{g}$ have periods $q-1$ over $\F_q$.
However we cannot conclude that $\tilde{x}_n^t$ and $\tilde{y}_n^t$ are also periodic with periods $q-1$, unlike the case in the discrete KdV equation.
The values of $\tilde{x}_n^t$ may not be periodic when $\tilde{f}(n,t)\tilde{g}(n+1,t)=0$ and $\tilde{g}(n,t)\tilde{f}(n+1,t)=0$ (See \eqref{xy}).
First we write $f(n,t)g(n+1,t)$ and $g(n,t)f(n+1,t)$ as follows:
\begin{eqnarray*}
f(n,t)g(n+1,t)&=&\epsilon^l k(\epsilon),\\
g(n,t)f(n+1,t)&=&\epsilon^m h(\epsilon),
\end{eqnarray*}
where $l,\ m\in\Z,\ h(0)\neq 0,\ k(0)\neq 0$ and $k(\epsilon),\ h(\epsilon)\in \F_q[\epsilon]$.
We also write $f(n+q-1,t)g(n+q,t)=\epsilon^{l'} k'(\epsilon),\; g(n+q-1,t)f(n+q,t)=\epsilon^{m'} h'(\epsilon)$ in the same manner.
Let us write down the reduction map again:
\[
\tilde{x}_n^t=
\left\{
\begin{array} {cl}
k(0)/h(0)  & (l=m),\vspace{1mm}\\
0 & (l>m),\\
\infty & (l<m).
\end{array}
\right.
\]
In the case when $\tilde{f}(n,t)\tilde{g}(n+1,t)=0$ and $\tilde{g}(n,t)\tilde{f}(n+1,t)=0$, $x_n^t=\dfrac{f(n,t)g(n+1,t)}{g(n,t)f(n+1,t)}\in\F_q(\epsilon)$ and $x_n^{t+q-1}=\dfrac{f(n+q-1,t)g(n+q,t)}{g(n+q-1,t)f(n+q,t)}\in\F_q(\epsilon)$ may have different reductions with respect to $\epsilon$, since $l'$ is not necessarily equal to $l$, and neither is $m'$ equal to $m$.
The left part of figure \ref{figure8} shows a gray-tone  plot of a two-soliton solution. In some points $\tilde
{x}_n^t$ does not have period 12 (for example $x_2^2\neq x_2^{14}$) while almost all other points do have this periodicity.
If we want to recover full periodicity, there is another reduction to obtain the reduced variables from $x_n^t$ and $y_n^t$.
This time, we define another reduction to the finite field $\hat{x}_n^t$ as
\[
\hat{x}_n^t=
\left\{
\begin{array} {cl}
k(0)/h(0)  & (l=0,\ m=0),\vspace{1mm}\\
0 & (\mbox{otherwise}).
\end{array}
\right.
\]
The right part of figure \ref{figure8} shows the same two-soliton solution as in the left part but calculated with this new method. We see that all points have periods 12.
It is important to determine how to reduce values in $\P\F_q(\epsilon)$ to values in $\P\F_q$, depending on the properties one wishes the soliton solutions to possess.

\section{Discrete KdV equation over the field of $p$-adic numbers}\label{padicdkdv}
\begin{figure}
\centering
\includegraphics[width=12cm, bb=50 350 540 750]{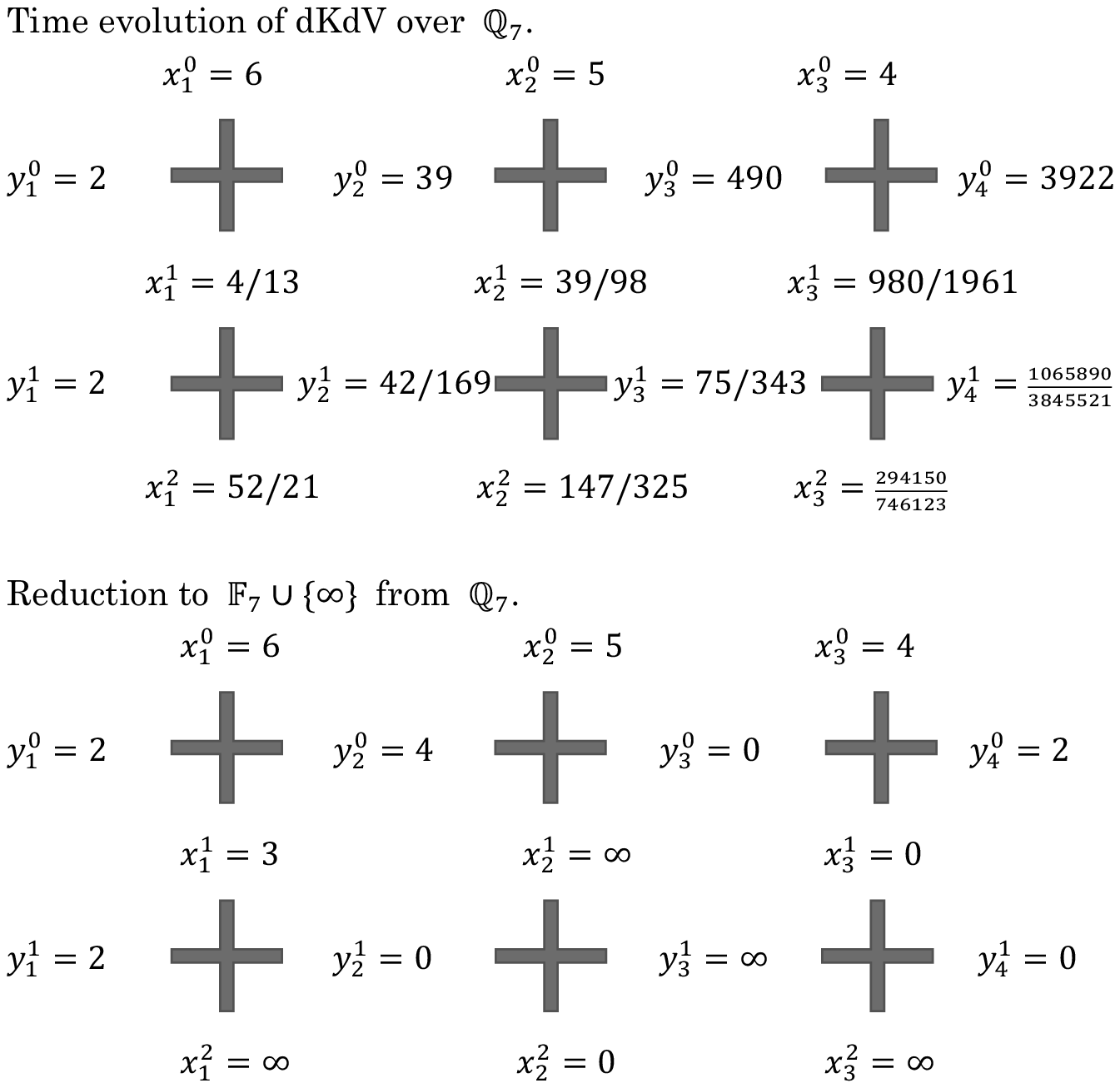}
\caption{The time evolution of the discrete KdV equation over the field $\Q_7$ and its reduction to $\P\F_7$.}
\label{figure9}
\end{figure}

\begin{figure}
\centering
\includegraphics[width=12cm, bb=70 350 550 750]{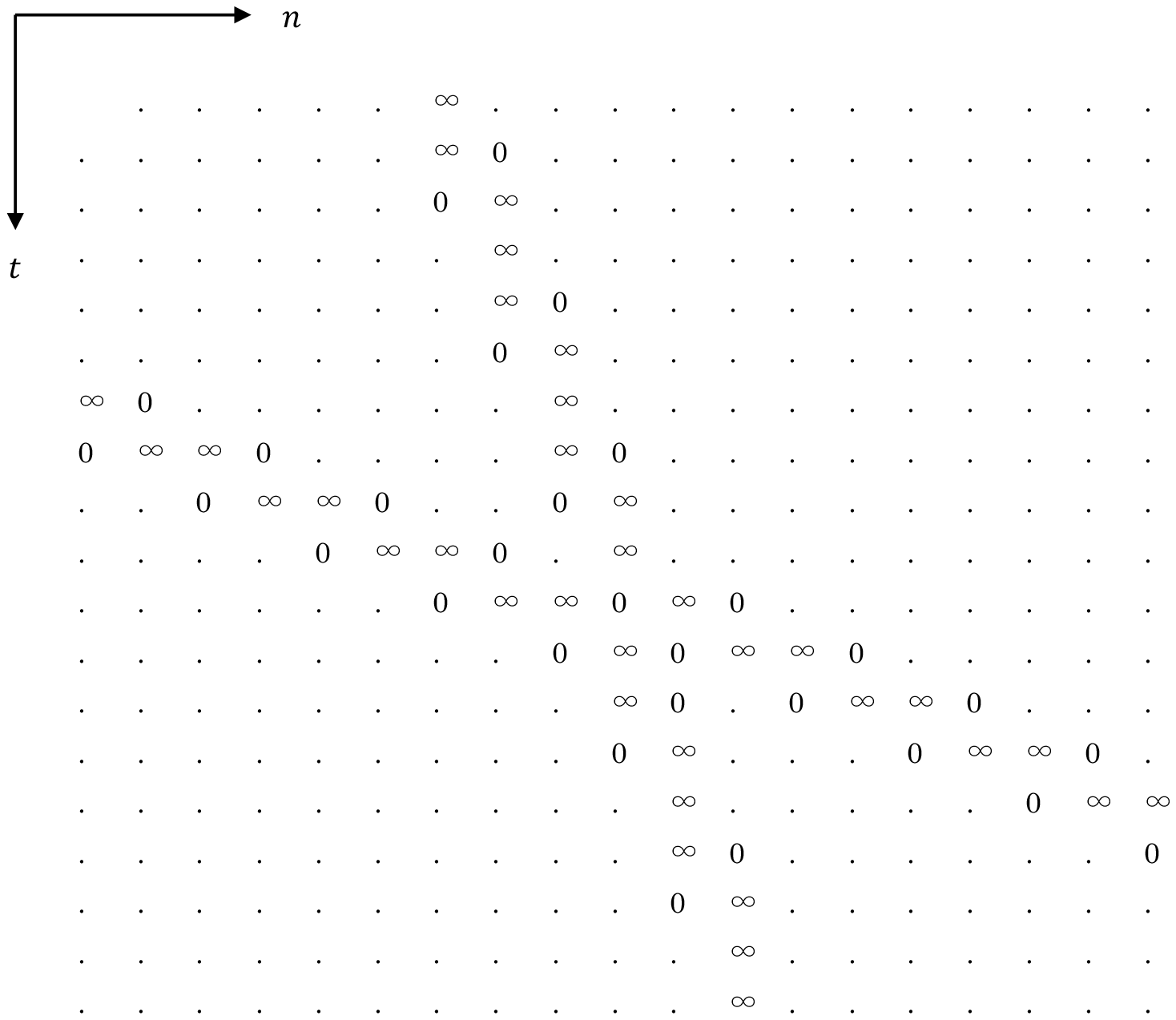}
\caption{The plot of the $2$-soliton solution of the discrete KdV equation over the field $\tilde{x}_n^t \in \P\F_2$ where $\delta=2$, $\gamma_1=\gamma_2=1$ and $l_1=5,l_2=6$. The dot `.' denotes that $\tilde{x}_n^t=1$. The range is $-9\le t,n\le 9$.}
\label{figure11}
\end{figure}

\begin{figure}
\centering
\includegraphics[width=12cm, bb=70 350 550 750]{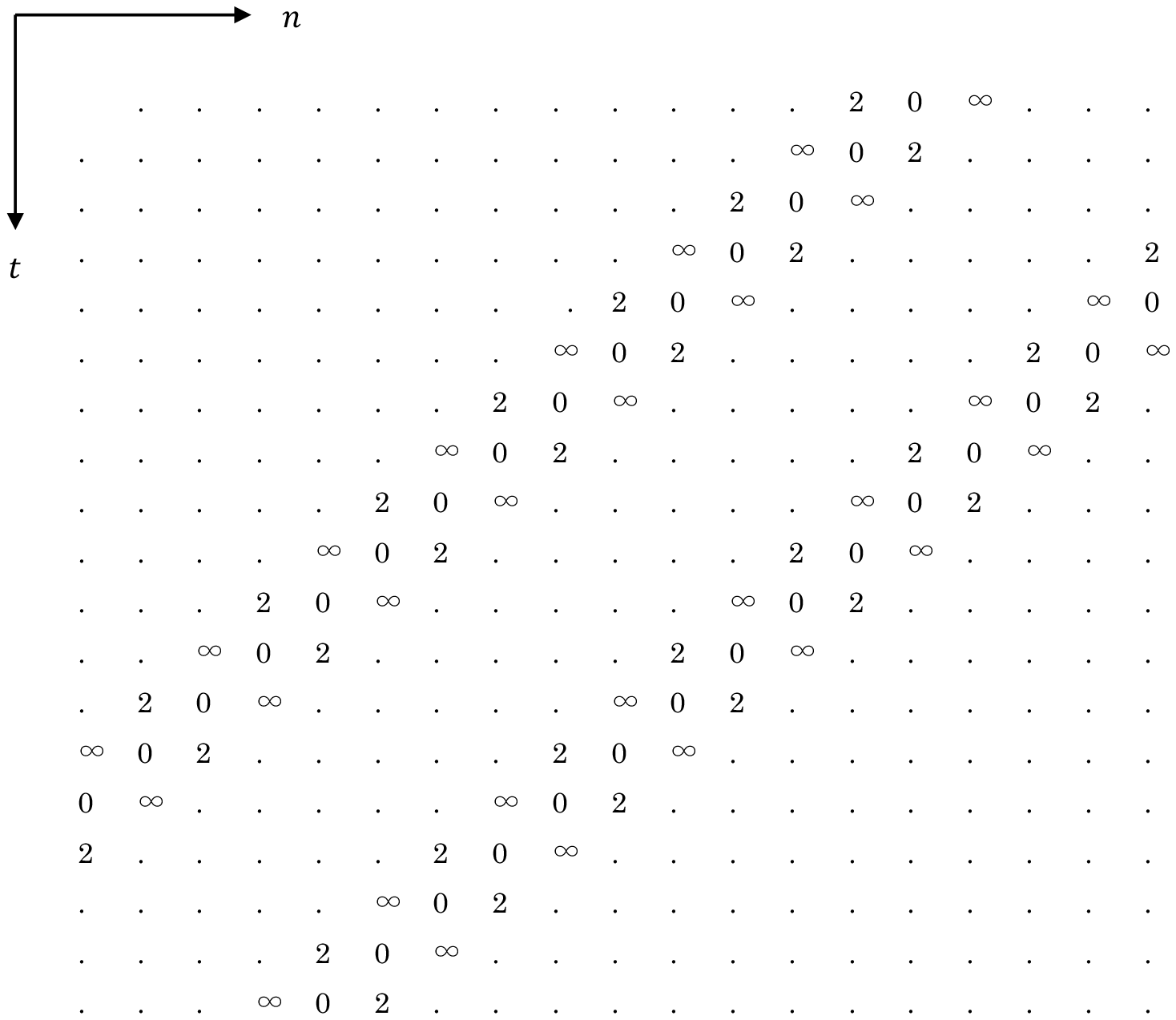}
\caption{The plot of the $2$-soliton solution of the discrete KdV equation over the field
$\tilde{x}_n^t \in \P\F_3$ where $\delta=2$, $\gamma_1=\gamma_2=1$ and $l_1=6,l_2=13$. The dot `.' denotes that $\tilde{x}_n^t=1$.
The range is $-9\le t,n\le 9$.}
\label{figure12}
\end{figure}

Instead of dealing with the systems over the field of rational functions, we can consider them over the field of $p$-adic numbers just like we have done for discrete Painlev\'{e} equations.
The calculation of soliton solutions and its reduction to the finite field can be done exactly the same as in section \ref{section51}. 
We can define the time evolution of the discrete KdV equation over the field of $p$-adic numbers $\Q_p$, and then obtain the time evolution of the equation over $\P\F_p$ by reducing it.
One of the good thing about this approach is the efficiency in numerical calculations. One of the weakness is that we have to limit ourselves to $m=1$ of $q=p^m$. (This is not a problem if we consider the extended field of $\Q_p$.)
The example with the same initial conditions as in figure \ref{figure1} is presented in figure \ref{figure9}.
The values $x_2^2=0$ and $x_3^1=\infty$ in the figure \ref{figure9} are different from those ($x_2^2=4,\ x_3^1=5$) reduced from the field $\F_p(\delta)$ in figure \ref{figure1}. The two systems do not present the same singularities for the same initial conditions in general because of the structural difference in the addition between the fields $\Q_p$ and $\F_p(\delta)$. However, the overall appearance of soliton solutions are unchanged.
We add some examples of the soliton solutions we have missed in the section \ref{sectiondkdvsoliton}.
We describe the behavior of solutions of the discrete KdV equation \eqref{dkdvsoliton}.  We consider the $p$-adic valuations of the parameter values $X:=(1-l_i)/l_i$ and $Y:=(l_i+\delta)/(1+\delta-l_i)$.
First, note that if we take $X,Y$ satisfying $v_p(X)=0$ and $v_p(Y)=0$ then the soliton solutions of the discrete KdV equation \eqref{dKdV2} over $\P\F_p$ is always periodic with respect to $t$ and $n$ with a period $p-1$ just like in figure \ref{figure4}.
Second, if we have $v_p(X)\neq 0$ or $v_p(Y)\neq 0$ then, at least one of the reductions of $X,Y$ by the reduction map  \eqref{padicreductionmap} are either $0$ or $\infty$. We have two cases:

(i) If $v_p(X)\neq 0$ and $v_p(Y)\neq 0$, then the soliton solution over $\P \mathbb{F}_p$ looks similar to that over the field $\mathbb{R}$.
The solitary waves which include the value $\infty$ move both to the left and to the right, over the background arrays of $1$'s. We introduce two examples where $p=2$ and $p=3$ respectively.
The first example concerns the $2$-soliton solution over $\P\F_2$ of the form $x_n^t=(\sigma_n^t \sigma_{n+1}^{t-1})/(\sigma_{n+1}^t \sigma_n^{t-1})$, where
\[
\sigma_n^t=\left(
\begin{array}{cl}
1+\frac{1}{9}\left(-\frac{4}{5}\right)^t \left(-\frac{7}{2}\right)^n & \frac{1}{10}\left(-\frac{4}{5}\right)^t \left(-\frac{8}{3}\right)^n\\
\frac{1}{10}\left(-\frac{5}{6}\right)^t \left(-\frac{7}{2}\right)^n & 1+\frac{1}{11}\left(-\frac{5}{6}\right)^t \left(-\frac{8}{3}\right)^n
\end{array}
\right).
\]
Note that all values concerning the speed of solitons ($v_2(-4/5)$, $v_2(-7/2)$, $v_2(-5/6)$, $v_2(-8/3)$) are non-zero.
The second one is the $2$-soliton solution over $\P\F_3$, written as
\[
\sigma_n^t=\left(
\begin{array}{cl}
1+\frac{1}{11}\left(-\frac{5}{6}\right)^t \left(-\frac{8}{3}\right)^n & \frac{1}{18}\left(-\frac{5}{6}\right)^t \left(-\frac{3}{2}\right)^n\\
\frac{1}{18}\left(-\frac{12}{13}\right)^t \left(-\frac{8}{3}\right)^n & 1+\frac{1}{25}\left(-\frac{12}{13}\right)^t \left(-\frac{3}{2}\right)^n
\end{array}
\right).
\]
See the figures \ref{figure11} and \ref{figure12} for their solitonic shapes.

\begin{figure}
\centering
\includegraphics[width=9cm, bb=60 460 450 740]{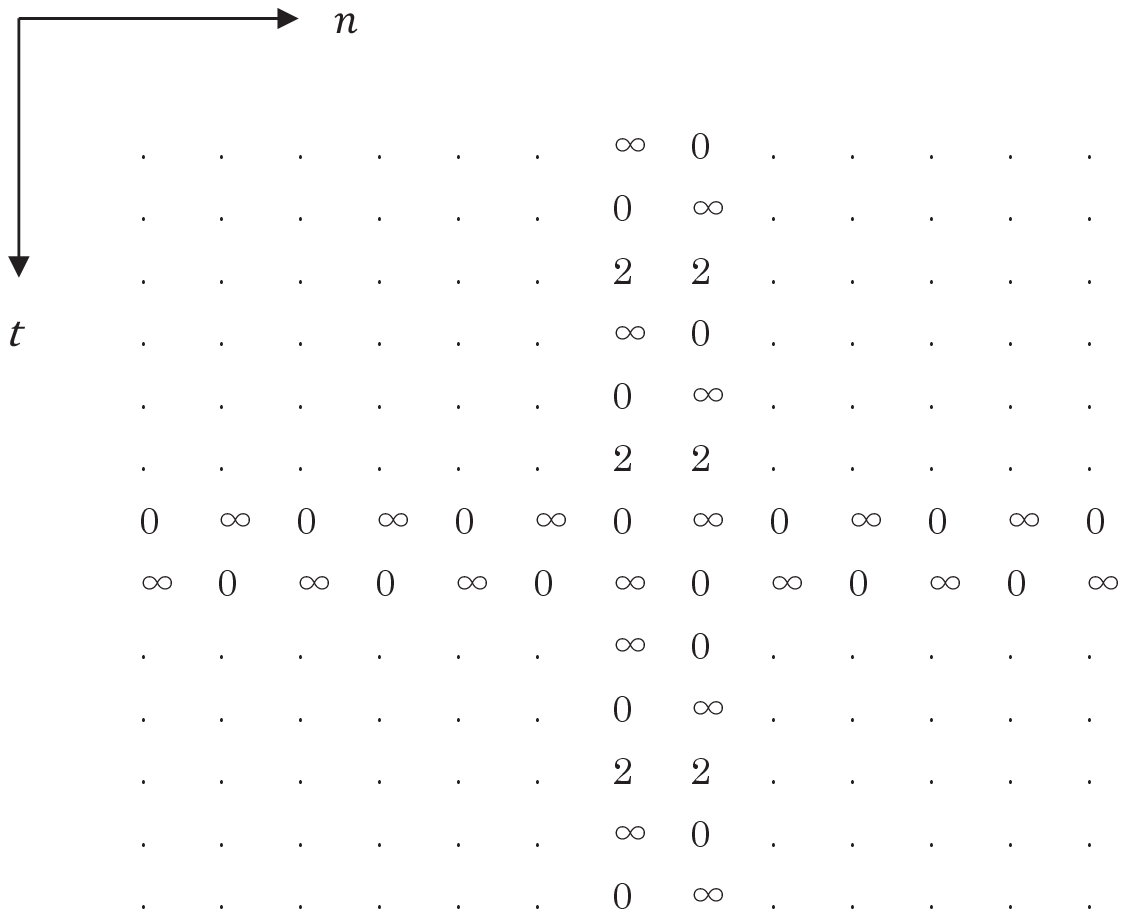}
\caption{The plot of the $2$-soliton solution of the discrete KdV equation over the field
$\tilde{x}_n^t \in \P\F_3$ where $\delta=1$, $\gamma_1=\gamma_2=1$ and $l_1=3,l_2=8$. The dot `.' denotes $\tilde{x}_n^t=1$.
The range is $-6\le t,n\le 6$.}
\label{figure13}
\end{figure}

(ii) If just one of the values $v_p(X)$ and $v_p(Y)$ is zero, the speed of solitary waves is either $0$ or $\infty$. In the figure \ref{figure13}, we present the example of the $2$-soliton solution over $\P\F_3$ with the speed of solitons $0$ and $\infty$ respectively.

We can prove that  the case (i) occurs if and only if $\tilde{\delta}=0,-1$.
\begin{Proposition}
We obtain $v_p(X)\neq 0$ and $v_p(Y)\neq 0$ for some $l_i$
if and only if the parameter $\delta\in\mathbb{Q}_p$ satisfies $\pi(\delta)=0$ or $\pi(\delta)=1$.
The solitary waves over $\mathbb{P}\mathbb{F}_p$ go to the right if $\pi(\delta)=0$, and to the left if $\pi(\delta)=-1$.
\end{Proposition}
\textbf{Proof}\;\;
Let us rewrite $l_i\to l$.
We have $v_p(X)\neq 0$ if and only if
\begin{equation}
v_p(l)=0\ \mbox{and}\ \pi(l)=1,\label{Xcase1}
\end{equation}
or
\begin{equation}
v_p(l)>0 \label{Xcase2}.
\end{equation}
In the case of \eqref{Xcase1},
\[
v_p(Y)=\left\{
\begin{array}{cl}
\infty & (v_p(\delta)>0),\\
(1+\delta )/ \delta & (v_p(\delta)=0),\\
1 & (v_p(\delta)<0).
\end{array}
\right.
\]
Therefore, $v_p(Y)\neq 0$ if and only if ($v_p(\delta)=0$ and $\pi(\delta)=-1$) or $v_p(\delta)>0$.

In the case of \eqref{Xcase2},
\[
v_p(Y)=\left\{
\begin{array}{cl}
0 & (v_p(\delta)>0),\\
\delta /(1+\delta) & (v_p(\delta)=0),\\
1 & (v_p(\delta)<0).
\end{array}
\right.
\]
Therefore, $v_p(Y)\neq 0$ if and only if ($v_p(\delta)=0$ and $\pi(\delta)=-1$) or $v_p(\delta)>0$.
\qed
Note that if $\pi(\delta)=0$ or $-1$, the discrete KdV equation \eqref{dKdV1} is reduced to the linear difference equations
\begin{equation}
x_{n+1}^{t+1}\equiv x_n^t,\ \mbox{or},\ x_n^{t+1}\equiv x_{n+1}^t, \label{reduced1}
\end{equation}
respectively. These equations have trivial waves with speed $\pm 1$, and do not have soliton solutions.
What we are observing in this section is the reduction of the soliton solution of the discrete KdV equation \eqref{dKdV1} over $\mathbb{Q}_p$, \textit{not} the solutions of the `reduced' discrete KdV equations \eqref{reduced1}. Through our methods, we successfully extract the solitonic structure of solutions over the finite field.

\subsection{Relation to the cellular automata}
\begin{figure}
\centering
\includegraphics[width=12cm, bb=70 485 530 740]{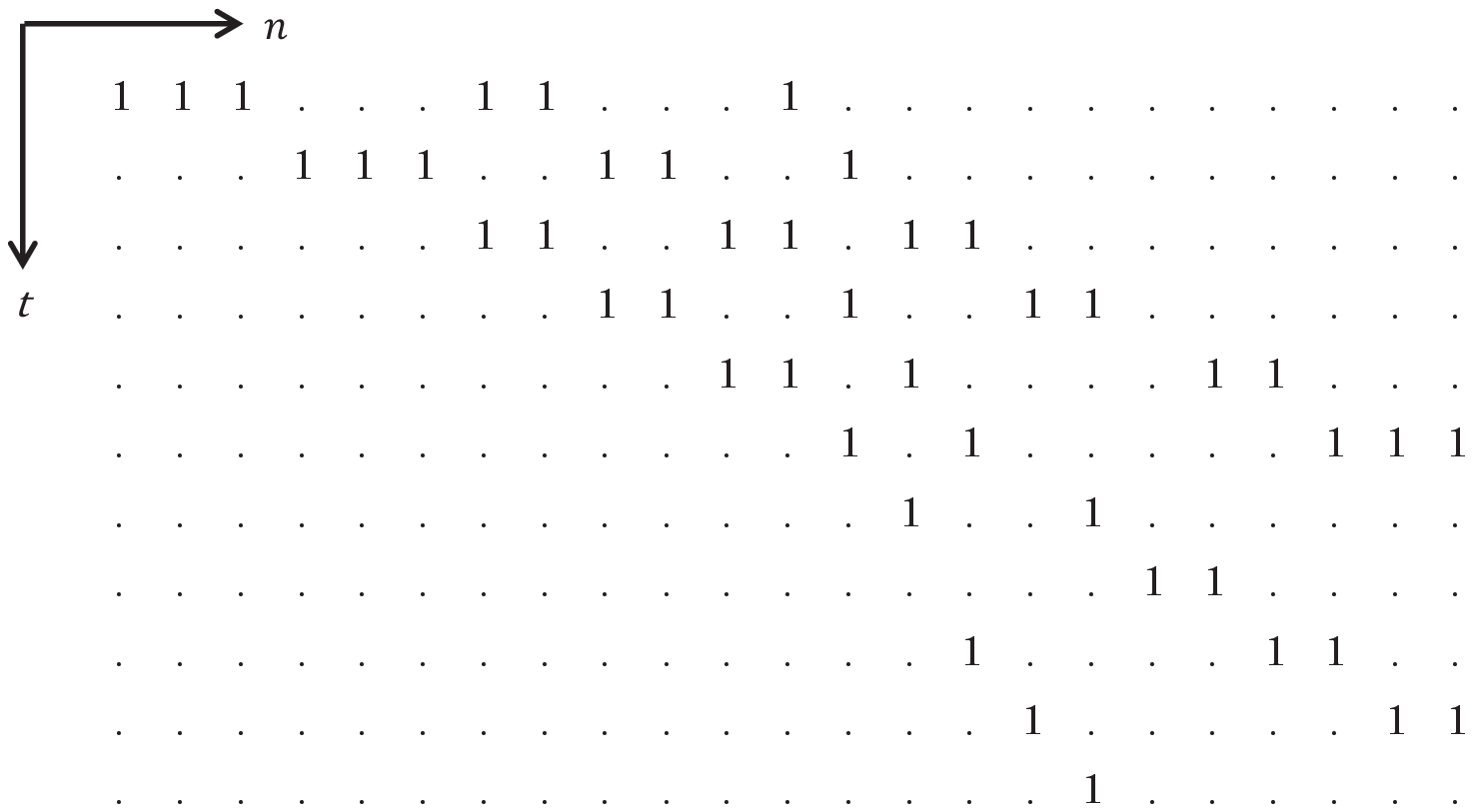}
\caption{Plot of the `roundoff' $p$-adic valuations $\hat{x}_n^t$ of the discrete KdV equation \eqref{dKdV2}. Here the dot `.' represents the point where $\hat{x}_n^t=0$.}
\label{figure15}
\end{figure}

In this section, we study how the discrete KdV equation over $\mathbb{Q}_p$ is related to the Box and Ball System (BBS) by taking the $p$-adic valuations. The BBS is a famous cellular automaton obtained by taking the ultra-discrete limit of the discrete KdV equation (section \ref{udintegrable}).
Let us fix $\delta=p^m,(m>0)$ for the discrete KdV equation \eqref{dKdV2}.
We define the new system $\{ \hat{x}_n^t\}$ from $\{x_n^t \}$ as
\[
\hat{x}_n^t:=-\mbox{Round} \left( \frac{v_p (x_n^t)}{m}\right),
\]
where Round$(x)\in\mathbb{Z}$ denotes the closest integer to $x\in\mathbb{R}$.
Then the system $\hat{x}_n^t$ goes to the time evolution of the BBS in the limit $m\to\infty$, or $p\to\infty$.
Note that if $x_n^t=p^{-m}$ then we have $\hat{x}_n^t=1$, and that if $x_n^t=1$ we have $\hat{x}_n^t=0$.
Here is an example where $p=2$ and $m=10$. We start from the initial condition of the equation \eqref{dKdV2}:
\[
\{x_n^0\}=\{\cdots,1,1,p^{-m},p^{-m},p^{-m},1,1,1,p^{-m},p^{-m},1,1,1,p^{-m},1,1,\cdots\}.
\]
Then the evolution of $\hat{x}_n^t$ is as in figure \ref{figure15}.
The time evolution obtained here is almost the same as the three soliton interaction of BBS in figure \ref{figure2}.
The block of $p^{-m}$'s in the initial condition of \eqref{dKdV2} is a $p$-adic analog of a BBS soliton (array of $1$'s) in the system $\hat{x}_n^t$.
The underlying fact is that the ultra-discrete limit is a super-exponential estimate, whose $p$-adic analog is taking $p$-adic valuations of the variables.

\section*{Concluding remarks and future problems}
We studied the discrete dynamical systems over finite fields. We investigated how to define them without indeterminacies, how to judge their integrability by a simple test similar to the singularity confinement method, and how to obtain the special solutions of them, in particular the solitary wave solutions.
In the first part of the paper, we constructed the space of initial conditions for discrete Painlev\'{e} equations defined over a finite field via the application of the Sakai theory.
In particular, we defined the time evolution graph for the discrete Painlev\'{e} II equation over finite field $\F_{q}$.
We have found out that, in case of the systems over the finite field, the space of initial conditions can be minimized compared to those obtained through Sakai theory, because of the discrete topology of the space.
The second part concerns the extension of the value spaces to local fields, in particular, to the field of $p$-adic numbers.
Our idea is to define the equations over the field of $p$-adic numbers $\mathbb{Q}_p$ and then reduce them to the finite field $\F_p$.
We generalized good reduction in order to be applied to integrable mappings, in particular, to the discrete Painlev\'{e} equations. We called this generalized notion an `almost good reduction' (AGR). It has been proved that AGR is satisfied for discrete Painlev\'{e} II equation and for $q$-discrete Painlev\'{e} I, II, III, IV and V equations. We have found out that AGR was satisfied for the Hietarinta-Viallet equation, and hence was an integrability detector which worked as an arithmetic analog of the singularity confinement test.
In the third part, we applied our methods to the two-dimensional lattice systems, in particular, to the discrete KdV equation and its generalized equation.
We obtained the solitary wave solutions defined over finite fields and showed that they have periods $q-1$ in generic cases. Other special solitary wave solutions which only appear over finite fields have also been presented and their properties have been studied.
One of the future problems is to construct a theory to solve the initial value problems over the non-archimedean valued fields. We also wish to study further the properties of the reduction modulo a prime of the higher dimensional lattice integrable equations.
In this paper, we have not dealt with the theory of continuous integrable equations over the field of $p$-adic numbers and over the finite fields. For example, $p$-adic soliton theory has been investigated by G. W. Anderson \cite{Anderson}. The continuous Painlev\'{e} equations over finite fields have been studied in terms of their symmetric solutions by K. Okamoto and S. Okumura. We also wish to study the relation of our methods to these approaches.

\section*{Acknowledgments}
The author would particularly like to thank his advisor, Professor Tetsuji Tokihiro for 
generous support and advice throughout his Ph.D course studies.
He has greatly benefited from Professors Jun Mada and K. M. Tamizhmani who collaborated in the research and jointly published several papers. He would like to thank Professor Ralph Willox
for carefully reading the papers and making insightful suggestions.
He would like to thank Professors Shinsuke Iwao, Nalini Joshi, Saburo Kakei, Shigeo Kusuoka, Kenichi Maruno, Yousuke Ohyama, Hidetaka Sakai, Junkichi Satsuma, Junichi Shiraishi, for valuable discussions and comments.
This work is partially supported by Grant-in-Aid for JSPS Fellows 24-1379.

\end{document}